\documentclass[11pt]{article}
\usepackage{geometry}
\usepackage{textcomp,booktabs}
\usepackage[usenames,dvipsnames]{color}
\usepackage{colortbl}
\usepackage{epsfig,graphicx}
\usepackage{subfig}
\usepackage{float}
\usepackage{tikz,pgfplots}
\usepgfplotslibrary{fillbetween}
\usepackage{setspace,multirow}
\usepackage{amsmath,amsthm,amsfonts}
\usepackage{amssymb,enumerate}
\usepackage[noend]{algpseudocode}
\usepackage{setspace,paralist,color}
\usepackage{bm,verbatim}
\usepackage{lineno}
\usepackage{multicol}
\usepackage{rotating}
\usepackage{MnSymbol}
\usepackage{threeparttable}
\usepackage[titletoc]{appendix}
\usepackage{titlesec}
\usepackage{longtable}
\usepackage{makecell}
\usepackage[round]{natbib}
\usepackage[colorlinks,citecolor=black,linkcolor=black]{hyperref}
\usepackage{algorithm}

\definecolor{mygray}{gray}{.9}
\definecolor{mypink}{rgb}{.99,.91,.95}
\definecolor{mycyan}{cmyk}{.3,0,0,0}
\allowdisplaybreaks

\newtheorem{Th}{\bf Theorem}
\newtheorem{Rem}{\bf Remark}

\newtheorem{Pro}{Proposition}

\newtheorem{Cor}{\bf Corollary}
\newtheorem{Ex}{Example}

\newtheorem{Assump}{Assumption}
\newtheorem{Con}{Condition}

\newtheorem*{examplecore}{Example \continueref (continued)}

\def\E{\mathbb{E}}

\def\log{\hbox{log}}

\def\T{{ \mathrm{\scriptscriptstyle T} }}

\graphicspath{{image/}}

\title{Model Selection and Averaging: Signal Strength, Ordering, and Oracle Risks}
\author{Guangyuan Cui$^{a}$, and Yuhong Yang$^{a,b\thanks{Corresponding author: yyangsc@mail.tsinghua.edu.cn}}$ 
		\\{ \footnotesize $^{a}$ Yau Mathematical Sciences Center, Tsinghua University, Beijing, China}\\
		{\footnotesize $^{b}$ Beijing Institute of Mathematical Sciences and Applications (BIMSA), Beijing, China}
	}
\date{}

\begin{document}
	
	\maketitle
	\begin{abstract}
		Existing oracle risk comparisons of model selection (MS) and model averaging (MA) have primarily distinguished slowly and rapidly decaying ordered signal-importance sequences, leaving the comparison beyond these idealized but practically unrepresentative decay conditions insufficiently characterized. We revisit this problem for a complete nested collection of linear regression models with independent homoscedastic errors. Without prescribing a coefficient decay pattern, we establish necessary and sufficient conditions for the relative oracle gain of MA over MS to be asymptotically substantial or negligible. Under complete ordering of the coefficients, the determining quantity is the share of oracle MS risk contributed by components near the noise level. This characterization explains how the strengths and numbers of strong, moderate, and weak signals jointly affect the comparison. Under an arbitrary fixed ordering, a common pool-adjacent-violators partition represents both oracle risks and yields an analogous characterization through block averages and sizes. For application, we develop a consistent diagnostic under sub-Gaussian errors and suitable growth conditions, together with finite-sample confidence intervals under Gaussian errors. The Gaussian framework also provides confidence sets for a certain Rashomon set. Simulations and a diamond-price application illustrate the theoretical findings and the empirical assessment of the potential MA advantage.
	\end{abstract}
	\noindent%
	
	\section{Introduction}\label{Section 1}
		
		\subsection{Model selection and model averaging}
		
		Due to the pervasive model uncertainty, regression analyses often involve multiple candidate models that differ in their covariates or complexity. A central challenge is how to use these models effectively for estimation and prediction when the best candidate is unknown. Model selection (MS) addresses this problem by choosing one candidate model, whereas model averaging (MA) combines estimators from multiple candidates. Common MS procedures compare candidate models through measures of fit, complexity, or predictive accuracy. AIC \citep{Akaike1973} and Mallows' $C_p$ \citep{mallows1973} correct in-sample fit for model complexity to assess predictive performance, while BIC \citep{Schwarz1978} uses a complexity penalty that increases with sample size from a Bayesian perspective. Cross-validation \citep{Allen1974,stone1974} evaluates candidate models on observations excluded from estimation and selects a model according to the resulting prediction errors. In comparison, frequentist model averaging determines data-dependent weights for combining candidate estimators \citep{buckland1997,hjort2003claeskens,Claeskens2003}. Adaptive regression by mixing \citep{Yang2001} uses sample splitting to assess candidate procedures and construct combination weights, while \citet{yang2004} derives risk bounds for combined forecasting procedures. Mallows model averaging \citep{hansen2007} chooses weights by minimizing a Mallows criterion that estimates the squared error of the averaged estimator. Subsequent work has extended averaging procedures to more general model collections and error structures \citep{wan2010,hansen2012,ando2014,Zhang2016,yu2025unified,Peng2024}.
		
		MS provides a single, potentially parsimonious model that facilitates interpretation of the regression relationship, whereas MA can reduce sensitivity to model choice and improve prediction through contributions from different candidate estimators. Their relative performance has been examined from both empirical and theoretical perspectives. The simulations of \citet{yang2003} associate the benefits of averaging with greater uncertainty in model choice, while \citet{Yuan2005} develop a measure of estimation instability and use it to guide the choice between selection and averaging. Beyond these empirical comparisons, \citet{yang2004} derives forecast risk bounds that quantify both the potential gains from combination and the cost of estimating combination weights. Aggregation theory provides further evidence of the limitations of selection: \citet{Juditsky2008} show that procedures restricted to a single candidate can be suboptimal in excess risk relative to the best candidate, whereas suitable convex combinations achieve sharper oracle bounds. More recently, \citet{LeClarke2022} establish conditions under which several averaging methods, including Mallows model averaging, Bayesian model averaging \citep{Madigan1994,draper1995,clyde1999}, and stacking \citep{Wolpert1992,Breiman1996stacked}, are asymptotically no worse than their constituent predictors under squared prediction error. These studies demonstrate advantages of averaging under different assumptions and performance criteria, but the magnitude of the improvement remains less understood and is essential to the comparison. For nested regression models, this leads to the question of when the oracle MA risk improves over the oracle MS risk by a nonvanishing proportion, and which features of the regression signal determine that improvement.
		
		\subsection{Asymptotic efficiency}
		
		Asymptotic efficiency concerns whether a procedure approaches the best risk or loss available within its own class \citep{shibata1983,li1987,Andrews1991,hansen2007}. However, the literature solely focuses on verifying asymptotic optimality within the respective domains of MS and MA in isolation, but they lack a direct comparison between the two approaches.
		
		\citet{Peng2022} directly investigated this gap for nested models based on orthonormal series expansions. Under their regularity conditions, slowly decaying ordered coefficients yield a substantial relative MA improvement, whereas rapidly decaying coefficients yield asymptotically equivalent oracle risks. \citet{xu2022} extended the comparison to more general nested linear models, allowing nonorthogonal designs, grouped predictors, heteroscedastic and autocorrelated errors, and zero coefficients. Their results also show that the size of the candidate model set affects the possible improvement. These studies establish important regimes in which MA offers a substantial or negligible advantage, leave several important questions open for further investigation.
		
		\begin{itemize}
			\item \textit{Which features of the signal sequence determine the relative gain?} The existing characterizations rely on particular decay conditions for the ordered coefficient magnitudes. These conditions distinguish broad classes of substantial and negligible improvements, but they do not directly identify the feature that determines the comparison when different parts of the coefficient sequence exhibit different decay patterns. A general characterization can help connect the coefficient sequence to the relative risk improvement without restrictively prescribing its overall decay pattern.
			
			\item \textit{How does ordering affect the comparison?} The comparison results in the literature typically assume that the coefficient magnitudes are non-increasing along the candidate model sequence. This ordering depends on unknown signal coefficients and is generally unavailable in applications. For nested models constructed according to a prespecified covariate ordering, the coefficient magnitudes may be arbitrarily unordered, and the effect of this ordering on the relative advantage of MA requires a corresponding characterization.
			
			\item \textit{How can the potential gain be assessed from data?} The oracle risks depend on unknown signal coefficients, so their theoretical comparison does not by itself provide a practical assessment of the potential gain. Moreover, asymptotic conclusions alone do not quantify the uncertainty of such an assessment at a given sample size. A practical analysis requires both a consistent diagnostic for the relative oracle gain and a confidence procedure that accounts for its finite-sample uncertainty.
		\end{itemize}
		
		\subsection{Our contributions}
		
		We study a complete nested set of fixed-design linear regression models with independent homoscedastic errors, allowing the design and regression coefficients to vary with the sample size. Our central question is how the magnitudes and numbers of strong, moderate, and weak signal components, together with their ordering along the nested path, determine the relative oracle gain from MA over MS. We address this question through theoretical risk comparisons and develop practical diagnostics and inference procedures to assess the gain from observed data. Our main contributions are as follows.
		
		\begin{itemize}
			\item \textit{A general characterization under complete ordering.} We establish necessary and sufficient conditions for the relative oracle gain to be asymptotically substantial or negligible, without prescribing a specific coefficient decay pattern. Our argument decomposes the exact risk difference and identifies the proportion of oracle MS risk contributed by components whose squared coefficients lie near the noise level as the determining quantity. This characterization accommodates polynomial, exponential, and mixed decay patterns.
			
			\item \textit{A common oracle-risk representation under arbitrary fixed ordering.} We extend the comparison to arbitrary fixed orderings, for which coordinatewise optimization may conflict with the nested model structure. We use the existing pool-adjacent-violators algorithm (PAVA) representation of the oracle MA risk \citep{Peng2024shrinkage} and show that the same partition also yields the oracle MS risk. This common representation extends the characterization to risk contributions from blocks whose average squared coefficients lie near the noise level. We further derive a sharp finite-sample upper bound on the relative gain.
			
			\item \textit{A consistent practical diagnostic for the relative gain.} We construct a diagnostic from the minima of a projected Mallows criterion over the MS and MA classes. The criterion corrects the variance contribution in the observed squared coefficients. To account for optimization, we control its approximation error uniformly over the admissible weights and establish consistency under sub-Gaussian errors and suitable growth conditions. We also provide a feasible version for unknown error variance and specify the accuracy required of the variance estimator.
			
			\item \textit{Finite-sample inference for the relative gain.} A consistent point diagnostic does not by itself quantify finite-sample uncertainty. Under Gaussian errors, we construct confidence intervals and associated threshold tests with finite-sample guarantees. The key step is to obtain simultaneous bounds on tail signal strengths and propagate them through both oracle optimizations, with the sharp gain bound providing an additional restriction. A feasible construction accommodates unknown variance through the full-model residual estimator when the full model contains the regression mean and has positive residual degrees of freedom.
			
			\item \textit{Near-optimal models and the advantage of averaging.} We characterize the Rashomon set of models whose risks lie within a specified tolerance of the oracle MS risk. This set forms an interval along a completely ordered nested path but may contain gaps under arbitrary ordering. Its span and cardinality, after normalization by the oracle MS risk in noise units, provide lower bounds on the relative MA gain. 
		\end{itemize}
		
		The remainder of the paper is organized as follows. Section~\ref{Section 2} introduces the model setup. Section~\ref{Section 3} establishes the oracle comparison under complete ordering and illustrates the roles of different signal strengths. Section~\ref{Section 4} develops the common PAVA representation, the comparison under arbitrary ordering, and the effects of ordering on both oracle risks. Section~\ref{Section 6} presents the practical diagnostic and confidence procedures.  Section~\ref{sec:rashomon} studies Rashomon sets and inference for their membership and cardinality. Sections~\ref{sec:simulations} and~\ref{sec:diamonds} report the simulations and the diamond-price application. Section~\ref{Section 8} concludes.

	\section{Model set-up}\label{Section 2}
	We consider the following data-generating process (DGP),
	\begin{equation}\label{Con 9}
		{y_i}={\mu_i}+{\epsilon_i}=\sum_{j=1}^{p_n}\beta_jx_{ij}+\epsilon_i,
	\end{equation}
	where $\boldsymbol{\mu}=(\mu_1,\ldots,\mu_n)^{\T}$ is the unknown true regression function, $\boldsymbol{y}=(y_1,\ldots,y_n)^{\T}$ is the response, $\boldsymbol{x_i}=(x_{i1},\ldots,x_{ip_n})^{\T}$ represents a $p_n$-dimensional covariate vector with the corresponding coefficients $\boldsymbol{\beta}=(\beta_1,\ldots,\beta_{p_n})^{\T}$, and $\boldsymbol{\epsilon}=(\epsilon_1,\ldots,\epsilon_n)^{\T}$ consists of error terms with $\E(\boldsymbol{\epsilon})=\mathbf{0}_{p_n\times 1}$ and $\mathrm{Cov}(\boldsymbol\epsilon)=\boldsymbol\Omega$. Here, $\boldsymbol{x}_i$'s are considered as non-random variables.
	We allow a triangular-array formulation in which the nonrandom design vectors $\{\boldsymbol x_i\}_{i=1}^n$, the original regression coefficient vector $\boldsymbol\beta$, and the regression mean vector $\boldsymbol\mu$ may all depend on the sample size $n$. For notational simplicity, we suppress this additional $n$-index throughout.
	
	We first fix an ordering of the covariates in \eqref{Con 9}. Let $\boldsymbol{X}_m$ denote the $n\times m$ design matrix containing the first $m$ covariates, and assume that $\boldsymbol{X}_{p_n}$ has full column rank, so that $p_n\leq n$. This ordering defines the complete nested model collection $\mathcal{M}_A=\{\{1,\ldots,m\}:1\leq m\leq p_n\}$, where the $m$th model uses the first $m$ covariates. For each model, define the projection matrix
		\begin{equation*}
			\boldsymbol{P}_{m\mid\mathcal{M}_A}
			=\boldsymbol{X}_m\left(\boldsymbol{X}_m^{\T}\boldsymbol{X}_m\right)^{-1}\boldsymbol{X}_m^{\T},
			\quad m=1,\ldots,p_n,
		\end{equation*}
		and set $\boldsymbol{P}_{0\mid\mathcal{M}_A}=\mathbf{0}_{n\times n}$.
		
		Following \citet{xu2022} and \cite{Peng2024}, we express the DGP in an orthonormal basis that preserves this nested structure. Since each successive model adds one linearly independent covariate, $\boldsymbol{P}_{j\mid\mathcal{M}_A}-\boldsymbol{P}_{(j-1)\mid\mathcal{M}_A}$ is a rank-one orthogonal projection. Let $\boldsymbol{\phi}_j$ be a unit eigenvector corresponding to its eigenvalue one, so that
		\begin{equation*}
			\boldsymbol{P}_{j\mid\mathcal{M}_A}
			-\boldsymbol{P}_{(j-1)\mid\mathcal{M}_A}
			=\boldsymbol{\phi}_j\boldsymbol{\phi}_j^{\T},
			\quad j=1,\ldots,p_n.
		\end{equation*}
		The resulting vectors $\{\boldsymbol{\phi}_1,\ldots,\boldsymbol{\phi}_{p_n}\}$ form an orthonormal basis for the column space of $\boldsymbol{X}_{p_n}$. Moreover, the first $m$ basis vectors span the column space of $\boldsymbol{X}_m$ for every $m=1,\ldots,p_n$. Equivalently, these vectors can be obtained from a QR factorization of $\boldsymbol{X}_{p_n}$ without column pivoting.
		
		For this basis, define
		\begin{equation}\label{Con 11}
			\theta_j=\frac{\boldsymbol{\phi}_j^{\T}\boldsymbol{\mu}}{\sqrt{n}},
			\quad j=1,\ldots,p_n.
		\end{equation}
		Since $\boldsymbol{\mu}$ belongs to the column space of $\boldsymbol{X}_{p_n}$ under \eqref{Con 9}, the DGP can be written as
		\begin{equation}\label{Con 10}
			\boldsymbol{y}
			=\sqrt{n}\sum_{j=1}^{p_n}\theta_j\boldsymbol{\phi}_j
			+\boldsymbol{\epsilon}.
		\end{equation}
		The sequence $\{\theta_j^2\}_{j=1}^{p_n}$ is determined by the given covariate ordering and the regression mean, and need not be non-increasing. Thus, the construction of $\mathcal{M}_A$ and its projection matrices does not require prior knowledge of these signal strengths. We refer to $\{|\theta_j|\}_{j=1}^{p_n}$ as the covariate-importance sequence for this ordering, as discussed in Section~\ref{Section 1}. Our theoretical analysis is based on the representation in \eqref{Con 10}. We estimate $\theta_j$ by $\hat\theta_j=\boldsymbol{\phi}_j^{\T}\boldsymbol{y}/\sqrt{n}$. 
		In particular, $\theta_j$ denotes $\theta_{n,j}$ and may vary with $n$ even for a fixed index $j$. Thus, the coefficient vectors $\{\theta_j\}_{j=1}^{p_n}$ need not be truncations of a single infinite coefficient sequence that is independent of $n$.
		
		Because the orthonormal representation preserves every candidate model space, the least squares estimator under the $m$th model in $\mathcal{M}_A$ is
		\begin{equation*}
			\hat{\boldsymbol{\mu}}_{m\mid\mathcal{M}_A}
			=\boldsymbol{P}_{m\mid\mathcal{M}_A}\boldsymbol{y}
			=\sum_{j=1}^{m}\boldsymbol{\phi}_j\boldsymbol{\phi}_j^{\T}\boldsymbol{y}
			=\sqrt{n}\sum_{j=1}^{m}\hat\theta_j\boldsymbol{\phi}_j.
		\end{equation*}
		Let $\boldsymbol{w}=(w_1,\ldots,w_{p_n})^{\T}$ belong to the weight space $\mathcal{W}_n=\{\boldsymbol{w}\in[0,1]^{p_n}:\sum_{m=1}^{p_n}w_m=1\}$. The corresponding model averaging estimator is
		\begin{equation*}
			\hat{\boldsymbol{\mu}}_{\boldsymbol{w}\mid\mathcal{M}_A}
			=\sum_{m=1}^{p_n}w_m\hat{\boldsymbol{\mu}}_{m\mid\mathcal{M}_A}.
		\end{equation*}
		For a fixed model index $m$ and a fixed weight vector $\boldsymbol{w}$, we measure estimation accuracy by the normalized squared-error risks
		\begin{align*}
			R_n(m\mid\mathcal{M}_A,\boldsymbol{\mu})
			=\frac{1}{n}\E\left\|\hat{\boldsymbol{\mu}}_{m\mid\mathcal{M}_A}-\boldsymbol{\mu}\right\|^2,\quad\text{and}\quad
			R_n(\boldsymbol{w}\mid\mathcal{M}_A,\boldsymbol{\mu})
			=\frac{1}{n}\E\left\|\hat{\boldsymbol{\mu}}_{\boldsymbol{w}\mid\mathcal{M}_A}-\boldsymbol{\mu}\right\|^2.
	\end{align*}
	
	
	Next, we consider another way of constructing candidate models. First, we sort the values of $\{|\theta_j|\}_{j=1}^{p_n}$ in descending order and correspondingly rearrange the original basis functions $\left\{\boldsymbol{\phi}_1, \ldots, \boldsymbol{\phi}_{p_n}\right\}$ into $\left\{\boldsymbol{\phi}_{(1)}, \ldots, \boldsymbol{\phi}_{{(p_n)}}\right\}$. Let $|\theta|_{(1)}, \ldots,|\theta|_{\left(p_n\right)}$ denote the decreasing rearrangement of $\left|\theta_1\right|, \ldots,\left|\theta_{p_n}\right|$, and let $\mathcal{M}_N$ denote the nested model set constructed analogously to $\mathcal{M}_A$, but based on the ordered sequence $|\theta|_{(1)}, \ldots, |\theta|_{(p_n)}$. Specifically, let $\mathcal{I}_m$ be a subset of $\{1,\ldots,p_n\}$, and then $\mathcal{M}_N=\{S_m:m=1,\ldots,p_n\}$ with $S_m=\mathop{\arg\max}_{\mathcal{I}\subset\{1,\ldots,p_n\}}\sum_{|\mathcal{I}=m|,j\in\mathcal{I}}|\theta|_{(j)}^2$. Define $\hat{\boldsymbol{\mu}}_{m\mid\mathcal{M}_N}$ as the least squares estimator for $\boldsymbol{\mu}$ under the $m$th model in $\mathcal{M}_N$, and it is given by 
	\begin{equation*}
		\hat{\boldsymbol{\mu}}_{m\mid\mathcal{M}_N}=\sqrt{n}\sum_{j=1}^{m}\hat\theta_{(j)}\boldsymbol{\phi}_{(j)},
	\end{equation*}
	and the averaging estimator is $\hat{\boldsymbol\mu}_{\boldsymbol{w}\mid\mathcal{M}_N}=\sum_{m=1}^{p_n}w_m\hat{\boldsymbol{\mu}}_{m\mid\mathcal{M}_N}$. The risks of these two estimators, denoted as $R_n(m\mid{\mathcal{M}_N},\boldsymbol{\mu})$ and $R_n(\boldsymbol{w}\mid{\mathcal{M}_N},\boldsymbol{\mu})$, can be similarly defined.
	
		\begin{Rem}[Connection between $\theta_j^2$ and partial correlation]\label{rem:signal-partial-correlation}
			The quantity $\theta_j^2$ measures the additional signal explained when the $j$th covariate enters the given nested model sequence. Let $\boldsymbol X_{\cdot j}$ denote the $j$th column of $\boldsymbol X_{p_n}$, and define
			\[
			\boldsymbol z_j=(\boldsymbol I_n-\boldsymbol P_{(j-1)\mid\mathcal M_A})\boldsymbol X_{\cdot j},
			\quad
			\boldsymbol r_{j-1}=(\boldsymbol I_n-\boldsymbol P_{(j-1)\mid\mathcal M_A})\boldsymbol\mu.
			\]
			These vectors are the parts of the $j$th covariate and the regression mean that remain after projection onto the first $j-1$ covariates. Full column rank ensures $\boldsymbol z_j\ne\boldsymbol0$. Since $\boldsymbol\phi_j=\pm\boldsymbol z_j/\|\boldsymbol z_j\|$ and $\boldsymbol z_j^{\T}\boldsymbol\mu=\boldsymbol z_j^{\T}\boldsymbol r_{j-1}$, for $\boldsymbol r_{j-1}\ne\boldsymbol0$ we obtain
			\begin{equation}\label{eq:signal-partial-correlation}
				\theta_j^2
				=
				\frac{\|\boldsymbol r_{j-1}\|^2}{n}
				\left(
				\frac{\boldsymbol z_j^{\T}\boldsymbol r_{j-1}}
				{\|\boldsymbol z_j\|\|\boldsymbol r_{j-1}\|}
				\right)^2.
			\end{equation}
			If $\boldsymbol r_{j-1}=\boldsymbol0$, then $\theta_j^2=0$.
			
			The term in parentheses in \eqref{eq:signal-partial-correlation} is the cosine of the angle between the two residual vectors. When these residuals are centered, as occurs if the preceding model includes an intercept or all design columns are centered, it is the partial correlation between $\boldsymbol X_{\cdot j}$ and $\boldsymbol\mu$ after adjustment for the first $j-1$ covariates. Thus, $\theta_j^2$ equals the previously unexplained signal energy $\|\boldsymbol r_{j-1}\|^2/n$ multiplied by the fraction captured by the newly added direction. This interpretation depends on the covariate ordering in $\mathcal{M}_A$. 
	\end{Rem}
	
	\section{Improvability under complete ordering of $\{|\theta_j|\}_{j=1}^{p_n}$}\label{Section 3}
		
		This section compares the oracle risks of MS and MA when the coefficient magnitudes are non-increasing, as specified in Assumption~\ref{Assump 6}, and the errors satisfy Assumption~\ref{Assump 5}, with $\sigma^2>0$. Under this ordering, $\theta_{(j)}^2=\theta_j^2$, and the risk functions under $\mathcal M_A$ and $\mathcal M_N$ coincide. We therefore use $\mathcal M_N$ and omit the ordering parentheses throughout this section.
		
		For $\boldsymbol w\in\mathcal W_n$, define the transformed weights $\gamma_j=\sum_{m=j}^{p_n}w_m$, $j=1,\ldots,p_n$. Their feasible set is
		\[
		\Gamma_n=\left\{\boldsymbol\gamma:1=\gamma_1\geq\gamma_2\geq\cdots\geq\gamma_{p_n}\geq0\right\}.
		\]
		The MS and MA risks have the bias--variance representations
		\begin{align*}
			R_n(m\mid\mathcal M_N,\boldsymbol\mu)
			&=\frac{m\sigma^2}{n}+\sum_{j=m+1}^{p_n}\theta_j^2,\\
			R_n(\boldsymbol w\mid\mathcal M_N,\boldsymbol\mu)
			&=\sum_{j=1}^{p_n}(1-\gamma_j)^2\theta_j^2+\frac{\sigma^2}{n}\sum_{j=1}^{p_n}\gamma_j^2;
		\end{align*}
		see \citet{Peng2022}. These representations also hold under an arbitrary fixed ordering, with $\mathcal M_N$ replaced by $\mathcal M_A$. Complete ordering is used below to obtain explicit expressions for the oracle risks.
		
		Let $m_n^*$ and $\boldsymbol w^*$ minimize these risks over $m\in\{1,\ldots,p_n\}$ and $\boldsymbol w\in\mathcal W_n$, respectively. For MS, an optimal cutoff can be chosen as
		\[
		m_n^*=1+\#\left\{j\in\{2,\ldots,p_n\}:\theta_j^2>\frac{\sigma^2}{n}\right\}.
		\]
		Additional components at the threshold $\theta_j^2=\sigma^2/n$ may be included without changing the risk. Thus, with $d_j=\min(\theta_j^2,\sigma^2/n)$ for $j=2,\ldots,p_n$, the oracle MS risk is
		\begin{equation}\label{Con 32}
			R_n(m_n^*\mid\mathcal M_N,\boldsymbol\mu)
			=\frac{\sigma^2}{n}+\sum_{j=2}^{p_n}d_j.
		\end{equation}
		
		For MA, the coordinatewise minimizers $\theta_j^2/(\theta_j^2+\sigma^2/n)$, $j=2,\ldots,p_n$, are non-increasing and therefore satisfy the constraints in $\Gamma_n$ when combined with $\gamma_1=1$. Consequently, the optimal transformed weights are
		\begin{equation}\label{Con 24}
			\gamma_1^*=1,
			\quad
			\gamma_j^*=\frac{\theta_j^2}{\theta_j^2+\sigma^2/n},
			\quad j=2,\ldots,p_n,
		\end{equation}
		and the oracle MA risk is
		\begin{equation}\label{Con 22}
			R_n(\boldsymbol w^*\mid\mathcal M_N,\boldsymbol\mu)
			=\frac{\sigma^2}{n}+\sum_{j=2}^{p_n}\frac{\theta_j^2\sigma^2}{n\theta_j^2+\sigma^2}.
		\end{equation}
		The common term $\sigma^2/n$ in \eqref{Con 32} and \eqref{Con 22} arises because the first component is included without shrinkage in every candidate estimator.
		
		Existing comparisons relate these oracle risks to the decay of the ordered coefficients. In particular, \citet{Peng2022} considered the following two conditions.
		
		\begin{Con}(Slowly decaying ordered coefficients)\label{Polynomial decaying rate}
			There exist constants $\kappa>1$ and $0<\delta\leq\nu<1$, with $\kappa\nu^2<1$, such that
			\[
			\delta\leq\frac{|\theta_{\lfloor\kappa j\rfloor}|}{|\theta_j|}\leq\nu
			\]
			for all sufficiently large $j$.
		\end{Con}
		
		\begin{Con}(Fast decaying ordered coefficients)\label{Exponential decaying rate}
			For every constant $\kappa>1$,
			\[
			\lim_{j\to\infty}\frac{|\theta_{\lfloor\kappa j\rfloor}|}{|\theta_j|}=0.
			\]
		\end{Con}
		
		Under their regularity conditions, \citet{Peng2022} established an asymptotically substantial improvement,
		\begin{equation}\label{Con 1}
			R_n(m_n^*\mid\mathcal M_N,\boldsymbol\mu)-R_n(\boldsymbol w^*\mid\mathcal M_N,\boldsymbol\mu)
			\asymp R_n(m_n^*\mid\mathcal M_N,\boldsymbol\mu),
		\end{equation}
		under Condition~\ref{Polynomial decaying rate}, and an asymptotically negligible improvement,
		\begin{equation}\label{Con 2}
			R_n(m_n^*\mid\mathcal M_N,\boldsymbol\mu)-R_n(\boldsymbol w^*\mid\mathcal M_N,\boldsymbol\mu)
			=o\!\left(R_n(m_n^*\mid\mathcal M_N,\boldsymbol\mu)\right),
		\end{equation}
		under Condition~\ref{Exponential decaying rate}.
		
		These specified conditions do not cover general ordered coefficient sequences. Different segments may have different decay patterns, or the sequence may contain groups of strong, moderate, and weak signals. We therefore seek necessary and sufficient conditions for \eqref{Con 1} and \eqref{Con 2} that do not prescribe a common decay pattern. Our characterization uses the share of oracle MS risk contributed by coefficients near the noise level $\sigma^2/n$.
		
		\subsection{Main results}
		
		The noise level $\sigma^2/n$ provides a reference for signal strength. The relations $\theta_j^2/(\sigma^2/n)\to\infty$, $\theta_j^2\asymp\sigma^2/n$, and $\theta_j^2/(\sigma^2/n)\to0$ describe strong, moderate, and weak signals, respectively. To quantify the contribution of signals near the noise level, for any fixed $c>1$, define
			\[
			B_n(c)=\left\{j\in\{2,\ldots,p_n\}:\frac{\sigma^2}{cn}\leq\theta_j^2\leq\frac{c\sigma^2}{n}\right\}.
			\]
			The set $B_n(c)$ contains components whose $\theta_j^2$'s lie within a factor $c$ of $\sigma^2/n$. Since $d_j=\min(\theta_j^2,\sigma^2/n)$, their total contribution to the oracle MS risk satisfies
			\[
			\frac{\sigma^2}{cn}|B_n(c)|
			\leq
			\sum_{j\in B_n(c)}d_j
			\leq
			\frac{\sigma^2}{n}|B_n(c)|.
			\]
			Thus, for fixed $c$, this contribution is bounded above and below by fixed multiples of the number $|B_n(c)|$ times the noise level $\sigma^2/n$.
			
			We normalize this contribution by the oracle MS risk and define
			\begin{equation}\label{Con 14}
				f_n(c)
				=\frac{\sum_{j\in B_n(c)}d_j}{R_n(m_n^*\mid\mathcal M_N,\boldsymbol\mu)}
				=\frac{\sum_{j\in B_n(c)}d_j}{\sigma^2/n+\sum_{j=2}^{p_n}d_j}
				\in[0,1].
			\end{equation}
			The quantity $f_n(c)$ therefore measures the share of oracle MS risk contributed by these moderate signals. Its magnitude depends on both their number $|B_n(c)|$ and the total oracle MS risk, so a diverging number alone does not ensure a non-vanishing risk share. The first component is excluded from $B_n(c)$ because it contributes the same risk $\sigma^2/n$ to MS and MA.
	
	\begin{Assump}\label{Assump 1}
		The regression mean vector $\boldsymbol\mu$ satisfies $\mathop{\lim \sup}_n \|\boldsymbol{\mu}\|^2/n<\infty$.
	\end{Assump}
	\begin{Assump}\label{Assump 6}
		The transformed coefficients $\{|\theta_j|\}_{j=1}^{p_n}$ form a non-increasing sequence.
	\end{Assump}
	\begin{Assump}\label{Assump 5}
		The error terms $\{\epsilon_i\}_{i=1}^n$ are independent and homoscedastic with $\E(\epsilon_i)=0$ and $\E(\epsilon_i^2)=\sigma^2$.
	\end{Assump}
	
	Assumption \ref{Assump 1} imposes a uniform bound on the average signal strength. By the orthogonal representation in \eqref{Con 10},
	\[
	\frac{\|\mu\|^2}{n}
	=
	\sum_{j=1}^{p_n}\theta_j^2
	=
	O(1),
	\]
	which places the regression problems in a conventional \(L_2\)-bounded signal class that is commonly considered in the model-averaging literature \citep{hansen2007,hansen2008,Peng2022}. Note that the oracle comparisons and the inference results below do not otherwise rely on Assumption \ref{Assump 1}, but we retain it as a standing condition to define the signal class and facilitate comparison with the existing literature.
	Assumption \ref{Assump 6} implies that the transformed coefficients are ordered from the most important to the least important \citep{Peng2022,xu2022,Peng2024,Lin2025nonstationary}. Under Assumption \ref{Assump 6},
	\[
	\theta_j^2
	\leq
	\frac{1}{j}\sum_{k=1}^{j}\theta_k^2
	\leq
	\frac{C}{j}
	\]
	for all sufficiently large \(n\) and \(j\leq p_n\). This is a global restriction on the magnitude of $\theta_j^2$. 
	Assumption \ref{Assump 5} further assumes the homoscedasticity of $\{\epsilon_i\}_{i=1}^n$, which is a very standard condition.
	
	The following theorem gives necessary and sufficient conditions for
	the two risk comparisons in \eqref{Con 1} and \eqref{Con 2}.
	
	\begin{Th}\label{Theorem 1}
		Under Assumptions~\ref{Assump 1}--\ref{Assump 5},
		(i) the MA improvement is asymptotically substantial, as in
		\eqref{Con 1}, if and only if there exists a fixed $c>1$ such that
		$\liminf_{n\to\infty}f_n(c)>0$;
		(ii) the MA improvement is asymptotically negligible, as in
		\eqref{Con 2}, if and only if there exists a sequence $c_n\to\infty$
		such that $f_n(c_n)\to0$.
	\end{Th}
	
	Theorem~\ref{Theorem 1} characterizes the relative advantage of oracle MA through the share of oracle MS risk contributed by coefficients near the noise level, without imposing a common decay pattern on the entire coefficient sequence. Part~(i) requires a non-vanishing risk share within some fixed multiplicative neighborhood of $\sigma^2/n$.
	Part~(ii) requires this share to vanish over some expanding
	neighborhood. The relevant quantity is the risk share $f_n(c)$;
	the number of coefficients in the neighborhood alone does not
	determine the comparison.
	
	\begin{Rem}\label{Remark 1}
		The condition in Theorem~\ref{Theorem 1}(ii) is equivalent to
		$f_n(c)\to0$ for every fixed $c>1$. Indeed, if $c_n\to\infty$ and
		$f_n(c_n)\to0$, then, for each fixed $c>1$, eventually $c\leq c_n$,
		so $0\leq f_n(c)\leq f_n(c_n)\to0$.
		Conversely, the bounds established in the proof of
		Theorem~\ref{Theorem 1} show that $f_n(c)\to0$ for every fixed $c>1$
		implies \eqref{Con 2}, and hence the existence of such a sequence
		$c_n$. The theorem requires only one suitable diverging sequence,
		not every diverging sequence. If a sequence $c_n$ satisfies the
		condition, then any sequence $\tilde c_n\to\infty$ with
		$1<\tilde c_n\leq c_n$ eventually also satisfies it, because
		$f_n(\tilde c_n)\leq f_n(c_n)$.
		\end{Rem}
	
	The same comparison can be expressed through an interval that need not be centered at $\sigma^2/n$. For any fixed constants $0<a<b$, define
	\begin{equation}\label{Con 36}
		B_n(a,b)=\{j\in\{2,\ldots,p_n\}:a\leq n\theta_j^2\leq b\}
	\end{equation}
	and
	\[
	f_n(a,b)=\frac{\sum_{j\in B_n(a,b)}d_j}
	{R_n(m_n^*\mid\mathcal M_N,\boldsymbol\mu)}.
	\]
	These definitions also apply to positive sequences $a_n<b_n$.
	
	\begin{Cor}\label{Corollary 1}
		Suppose that Assumptions \ref{Assump 1}--\ref{Assump 5} hold. Then we have \\ 
		(i) if there exists \(\eta>0\), such that $\liminf_{n\to\infty} f_n(a,b)\ge\eta$ for some fixed constants \(0<a<b\), then \eqref{Con 1} is satisfied;\\ 
		(ii) if there exist two positive sequences $a_n$ and $b_n$ with \(a_n\to0\) and $b_n\to\infty$, such that \(f_n(a_n,b_n)\to 0\) with respect to $n$, then \eqref{Con 2} is satisfied.
	\end{Cor}
	Corollary \ref{Corollary 1} allows a more flexible construction of $B_n(c)$, which further benefits the verification of the results \eqref{Con 1} and \eqref{Con 2}. Moreover, Corollary \ref{Corollary 1}(i) shows that if $\sum_{j\in B_n(a,b)}d_j$ from any subset $B_n(a,b)$ share the same order as that of $R_n(m_{n}^*\mid\mathcal{M}_N,\boldsymbol{\mu})$, then \eqref{Con 1}  is satisfied. 
	Finally, Corollary \ref{Corollary 1} can be similarly extended to the case of arbitrary ordering of $\{|\theta_j|\}_{j=1}^{p_n}$ in Sections \ref{Section 4}, but for simplicity, we will not provide further details.
	
	We next consider the risk functions evaluated at the model index $\hat m_n\in\{1,\ldots,p_n\}$ and weight vector $\hat{\boldsymbol w}\in\mathcal W_n$. Here $R_n(\hat m_n\mid\mathcal M_N,\boldsymbol\mu)$ and $R_n(\hat{\boldsymbol w}\mid\mathcal M_N,\boldsymbol\mu)$ denote the fixed-model and fixed-weight risk functions evaluated at these random choices; they do not denote the expected losses of the resulting data-dependent estimators.
	
	\begin{Cor}\label{Corollary 2}
		Suppose that Assumptions~\ref{Assump 1}--\ref{Assump 5} hold and
		\[
		\frac{R_n(\hat m_n\mid\mathcal M_N,\boldsymbol\mu)}
		{R_n(m_n^*\mid\mathcal M_N,\boldsymbol\mu)}
		\to_p1,
		\quad
		\frac{R_n(\hat{\boldsymbol w}\mid\mathcal M_N,\boldsymbol\mu)}
		{R_n(\boldsymbol w^*\mid\mathcal M_N,\boldsymbol\mu)}
		\to_p1.
		\]
		Then the following conclusions hold.
		
		\textup{(i)} If there exists a fixed $c>1$ such that $\liminf_{n\to\infty}f_n(c)>0$, then there exists a constant $\eta>0$ such that
		\[
		\Pr\left\{
		\frac{R_n(\hat m_n\mid\mathcal M_N,\boldsymbol\mu)-R_n(\hat{\boldsymbol w}\mid\mathcal M_N,\boldsymbol\mu)}
		{R_n(\hat m_n\mid\mathcal M_N,\boldsymbol\mu)}
		\geq\eta
		\right\}\longrightarrow1.
		\]
		
		\textup{(ii)} If there exists a sequence $c_n\to\infty$ such that $f_n(c_n)\to0$, then
		\[
		R_n(\hat m_n\mid\mathcal M_N,\boldsymbol\mu)-R_n(\hat{\boldsymbol w}\mid\mathcal M_N,\boldsymbol\mu)
		=o_p\!\left\{R_n(\hat m_n\mid\mathcal M_N,\boldsymbol\mu)\right\}.
		\]
	\end{Cor}
	
	Corollary~\ref{Corollary 2} extends the oracle comparison to these risk functions under the stated convergence assumptions. The same argument applies under an arbitrary fixed ordering using Theorem~\ref{Theorem 2}.
	
	\subsection{Signal composition and the relative oracle gain}\label{sec:signal-composition}
		This subsection explains how the strong, moderate and weak components affect $f_n(c)$. Through representative examples, we show how the associated quantities determine whether the relative oracle gain $G_{n,N}$ is substantial or negligible.
		
\begin{figure}[H]
			\centering
			\begin{tikzpicture}[x=1cm,y=1cm,font=\small,text=black]
				\fill[blue!8] (0,0.23) rectangle (2.1,0.93);
				\fill[blue!15] (4.4,0.23) rectangle (8.6,0.93);
				\fill[blue!24] (10.9,0.23) rectangle (13,0.93);
				\draw[fill=gray!8,draw=gray!60,densely dashed,line width=0.45pt]
				(2.1,0.23) rectangle (4.4,0.93);
				\draw[fill=gray!8,draw=gray!60,densely dashed,line width=0.45pt]
				(8.6,0.23) rectangle (10.9,0.93);
				\node at (1.05,0.58) {Weak};
				\node at (3.25,0.58) {Transition};
				\node at (6.5,0.58) {Moderate};
				\node at (9.75,0.58) {Transition};
				\node at (11.95,0.58) {Strong};
				
				\draw[->,blue,line width=0.65pt] (0,0) -- (13.3,0);
				\foreach \x in {0,2.1,4.4,6.5,8.6,10.9}
				\draw[blue,line width=0.55pt] (\x,-0.10) -- (\x,0.12);
				\node[anchor=north] at (0,-0.16) {$0$};
				\node[anchor=north] at (2.1,-0.16) {$v_n\to0$};
				\node[anchor=north] at (4.4,-0.16) {$c_1$};
				\node[anchor=north] at (6.5,-0.16) {$1$};
				\node[anchor=north] at (8.6,-0.16) {$c_2$};
				\node[anchor=north] at (10.9,-0.16) {$u_n\to\infty$};
				\node[anchor=north] at (13.18,-0.16) {$\infty$};
				\node[anchor=north] at (6.5,-0.70) {$n\theta_j^2/\sigma^2$};
			\end{tikzpicture}
			\caption{Strong, moderate, and weak signals on the normalized scale ($n\theta_j^2/\sigma^2$). The two transition intervals lie between the three groups; $c_1,c_2$ are fixed, while $v_n\to0$ and $u_n\to\infty$.}
			\label{fig:signal-strength-partition}
		\end{figure}
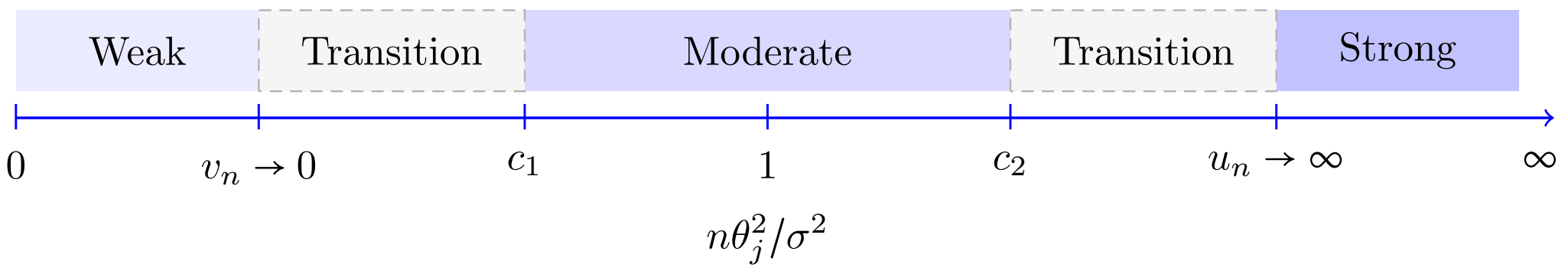

		Throughout, $\sigma^2>0$ is fixed, and the coefficients may depend on $n$. We distinguish strong, moderate, and weak signals through four thresholds. Let $0<c_1<1<c_2<\infty$ be fixed, and let $u_n$ and $v_n$ be positive sequences tending to infinity and zero, respectively. Both limits may be approached arbitrarily slowly. For all sufficiently large $n$, we have $0<v_n<c_1<1<c_2<u_n$. For the components indexed by $j=2,\ldots,p_n$, define
		\[
		\begin{aligned}
			\mathcal S_n
			&=\left\{j\in\{2,\ldots,p_n\}:\frac{n\theta_j^2}{\sigma^2}\geq u_n\right\}
			&&\text{(strong)},\\
			B_n(c_1\sigma^2,c_2\sigma^2)
			&=\left\{j\in\{2,\ldots,p_n\}:c_1\leq\frac{n\theta_j^2}{\sigma^2}\leq c_2\right\}
			&&\text{(moderate)},\\
			\mathcal V_n
			&=\left\{j\in\{2,\ldots,p_n\}:\frac{n\theta_j^2}{\sigma^2}\leq v_n\right\}
			&&\text{(weak)}.
		\end{aligned}
		\]
		The middle set uses the notation \eqref{Con 36}. Let $\mathcal T_n$ contain the remaining indices, whose normalized $\theta_j^2$'s lie in $(v_n,c_1)\cup(c_2,u_n)$. Figure~\ref{fig:signal-strength-partition} illustrates the three signal groups and these two transition intervals. The first component is treated separately because every candidate estimator includes it without shrinkage.
		
		By \eqref{Con 32}, the oracle MS risk decomposes as
		\begin{equation}\label{eq:four-threshold-ms}
			R_n(m_n^*\mid\mathcal M_N,\boldsymbol\mu)
			=\frac{\sigma^2}{n}(1+|\mathcal S_n|)
			+\sum_{j\in B_n(c_1\sigma^2,c_2\sigma^2)}d_j+\sum_{j\in\mathcal V_n}\theta_j^2
			+\sum_{j\in\mathcal T_n}d_j.
		\end{equation}
		Each strong component contributes $\sigma^2/n$, while each moderate component contributes between $c_1\sigma^2/n$ and $\sigma^2/n$. The weak components contribute their accumulated $\theta_j^2$'s. Thus, the first two groups approximately enter through their cardinalities, and the weak group enters through its total signal strength.
		
		To isolate these three contributions, suppose that the transition components satisfy
		\begin{equation}\label{eq:transition-risk-negligible}
			\sum_{j\in\mathcal T_n}d_j
			=o\!\left(R_n(m_n^*\mid\mathcal M_N,\boldsymbol\mu)\right).
		\end{equation}
		The condition requires the components in $\mathcal T_n$ to contribute a negligible proportion of the oracle MS risk; a small ratio $|\mathcal T_n|/p_n$ alone does not ensure this. Define
		\begin{equation*}
			D_n=\max\left\{
			1+|\mathcal S_n|,\;
			\frac{n}{\sigma^2}\sum_{j\in\mathcal V_n}\theta_j^2
			\right\}.
		\end{equation*}
		The constant $1$ accounts for the first component, and both terms use the noise level $\sigma^2/n$ as the risk unit. Under \eqref{eq:transition-risk-negligible}, the risk in \eqref{eq:four-threshold-ms} is of the same order as $(\sigma^2/n)\{D_n+|B_n(c_1\sigma^2,c_2\sigma^2)|\}$.
		
		For any fixed $c\geq\max\{c_2,1/c_1\}$, all strong and weak components eventually lie outside $B_n(c)$. Any components in $B_n(c)$ but outside $B_n(c_1\sigma^2,c_2\sigma^2)$ therefore belong to $\mathcal T_n$, so
		\[
		0\leq f_n(c)-f_n(c_1\sigma^2,c_2\sigma^2)
		\leq\frac{\sum_{j\in\mathcal T_n}d_j}
		{R_n(m_n^*\mid\mathcal M_N,\boldsymbol\mu)}
		\longrightarrow0.
		\]
		The moderate-group risk is comparable to its cardinality times $\sigma^2/n$. Consequently,
		\begin{align*}
			f_n(c)\to0
			&\quad\Longleftrightarrow\quad
			|B_n(c_1\sigma^2,c_2\sigma^2)|=o(D_n),\\
			\liminf_{n\to\infty}f_n(c)>0
			&\quad\Longleftrightarrow\quad
			D_n=O\!\left(|B_n(c_1\sigma^2,c_2\sigma^2)|\right).
		\end{align*}
		Since the preceding comparison holds for every fixed $c\geq\max\{c_2,1/c_1\}$, Theorem~\ref{Theorem 1} and Remark~\ref{Remark 1} imply that these same two cardinality conditions characterize $G_{n,N}\to0$ and $\liminf_{n\to\infty}G_{n,N}>0$, respectively.
		
		Under \eqref{eq:transition-risk-negligible}, transition components contribute a vanishing share of the oracle MS risk. To illustrate the roles of the three signal groups, we take $\mathcal T_n=\varnothing$ in the examples. For the first two examples, let $k_n,q_n,r_n\geq1$ denote the numbers of strong, moderate, and weak components, respectively, with $p_n=k_n+q_n+r_n\leq n$. Set
		\begin{equation}\label{eq:three-signal-profile}
			\theta_j^2=\frac{\sigma^2}{n}
			\begin{cases}
				u_{n,j},&1\leq j\leq k_n,\\
				1,&k_n<j\leq k_n+q_n,\\
				v_{n,j},&k_n+q_n<j\leq p_n,
			\end{cases}
		\end{equation}
		where $u_{n,1}\geq\cdots\geq u_{n,k_n}>1$ and $1>v_{n,k_n+q_n+1}\geq\cdots\geq v_{n,p_n}>0$. In the four-threshold partition, take $u_n=u_{n,k_n}\to\infty$ and $v_n=v_{n,k_n+q_n+1}\to0$. Let $\bar v_n=r_n^{-1}\sum_{j=k_n+q_n+1}^{p_n}v_{n,j}$ denote the average normalized strength of the weak components. Assume $\sum_{j=1}^{k_n}u_{n,j}+q_n+r_n\bar v_n=O(n)$ so that Assumption~\ref{Assump 1} holds.
		
		In units of $\sigma^2/n$, the strong, moderate, and weak groups contribute $k_n$, $q_n$, and $r_n\bar v_n$, respectively, to the oracle MS risk. For each fixed $c>1$, the band $B_n(c)$ contains exactly the $q_n$ moderate components for all sufficiently large $n$. Equations~\eqref{Con 32} and~\eqref{Con 22} then give
		\begin{equation}\label{eq:three-signal-share}
			f_n(c)=\frac{q_n}{k_n+q_n+r_n\bar v_n},
			\quad
			G_{n,N}=\frac12 f_n(c)+o(1).
		\end{equation}
		Oracle MA reduces the risk of each moderate component by $\sigma^2/(2n)$. The total risk reduction from the strong and weak components is of smaller order than the oracle MS risk.
	
	\begin{Ex}(A substantial proportion of risk from moderate signals)\label{ex:moderate-share}
		Local departures of order $n^{-1/2}$ from a smaller model are central to the model averaging analyses of \citet{hjort2003claeskens} and \citet{Hansen2014}. Here the moderate coefficients have this scale, while their number $q_n$ may increase with $n$. Set
		\[
		q_n=\lfloor n^{1/2}\rfloor,\quad r_n=\lfloor n^{3/4}\rfloor.
		\]
		Allow any non-increasing strong-signal sequence satisfying $n^{1/4}\leq u_{n,j}\leq2n^{1/4}$ for $j=1,\ldots,k_n$, and any non-increasing weak-signal sequence satisfying $n^{-1/2}\leq v_{n,j}\leq2n^{-1/2}$ for $k_n+q_n<j\leq p_n$. Assumption~\ref{Assump 1} holds for both choices of $k_n$ below. The weak components contribute $r_n\bar v_n\asymp n^{1/4}=o(q_n)$ in noise units. Equation~\eqref{eq:three-signal-share} gives, for every fixed $c>1$,
		\begin{itemize}
			\item If $k_n=q_n$, then $f_n(c)\to1/2$ and $G_{n,N}\to1/4$.
			\item If $k_n=\lfloor n^{1/4}\rfloor$, then $f_n(c)\to1$ and $G_{n,N}\to1/2$.
		\end{itemize}
		In both cases, $q_n/p_n\to0$, yet the moderate components account for a positive fraction of the oracle MS risk. Since $r_n\bar v_n=o(q_n)$, the limiting gain depends on the relative numbers of strong and moderate components.
	\end{Ex}
	
	\begin{Ex}(Many weak signals with different accumulated risks)\label{ex:weak-accumulation}
		Motivated by the evidence on dense predictive models in \citet{GiannoneLenzaPrimiceri2021}, we examine how the accumulated contribution of weak components affects $G_{n,N}$. We fix the numbers of strong, moderate, and weak components and vary the average strength of the weak group. For instance,
		\[
		k_n=\lfloor n^{1/4}\rfloor,\quad q_n=\lfloor n^{1/2}\rfloor,\quad r_n=\lfloor n^{3/4}\rfloor.
		\]
		Choose any non-increasing sequence with $n^{1/4}\leq u_{n,j}\leq2n^{1/4}$ for $j=1,\ldots,k_n$, and set
		\[
		v_{n,j}=\frac{2(p_n-j+1)}{r_n+1}\bar v_n,
		\quad k_n+q_n<j\leq p_n.
		\]
		The values $v_{n,j}$ decrease with $j$, have average $\bar v_n$, and are bounded above by $2\bar v_n$. The strong components are the same for all three choices of $\bar v_n$ below. Assumption~\ref{Assump 1} holds for all three choices. Since $k_n=o(q_n)$, \eqref{eq:three-signal-share} gives the following limits for every fixed $c>1$:
		\[
		\begin{array}{c|ccc}
			\bar v_n & \displaystyle\lim_{n\to\infty}\frac{r_n\bar v_n}{q_n}
			& \displaystyle\lim_{n\to\infty}f_n(c)
			& \displaystyle\lim_{n\to\infty}G_{n,N}\\[4pt]
			\hline
			n^{-1/2}&0&1&1/2\\
			n^{-1/4}&1&1/2&1/4\\
			n^{-1/8}&\infty&0&0
		\end{array}
		\]
		With the same counts $k_n$, $q_n$, and $r_n$ in all three cases, the limiting gain falls from $1/2$ to $1/4$ and then to zero as $r_n\bar v_n/q_n$ tends to zero, one, and infinity, respectively. In particular, the accumulated weak-signal contribution can make the moderate-signal risk share vanish even though $q_n\to\infty$.
	\end{Ex}
	
	\begin{Ex}(Weak signals in all components)\label{ex:weak-only}
		\citet{Baraud2002} uses equal-magnitude signals with unknown signs to establish detection limits in Gaussian sequence models. With coordinate noise variance $\sigma^2/n$, the dense construction at the detection boundary has $\theta_j^2\asymp\sigma^2/(n\sqrt{p_n})=o(\sigma^2/n)$. We now consider the case in which every component is weak, with strengths that may vary across components. Let $p_n\to\infty$ with $p_n\leq n$, and suppose that
		\[
		\theta_j^2=\frac{\sigma^2}{n}v_{n,j},\quad j=1,\ldots,p_n,
		\]
		where $1>v_{n,1}\geq\cdots\geq v_{n,p_n}>0$ and $v_{n,1}\to0$. The coefficients are completely ordered, and Assumption~\ref{Assump 1} holds because $\sum_{j=1}^{p_n}\theta_j^2\leq\sigma^2v_{n,1}\to0$.
		
		For $c_n=v_{n,1}^{-1/2}\to\infty$, we have $B_n(c_n)=\varnothing$ and $f_n(c_n)=0$, so Theorem~\ref{Theorem 1}(ii) gives $G_{n,N}\to0$. Thus, weak-signal accumulation alone cannot produce a substantial relative MA gain, even if $\sum_{j=2}^{p_n}v_{n,j}\to\infty$.
	\end{Ex}
	
	\section{Improvability under arbitrary ordering of  $\{|\theta_j|\}_{j=1}^{p_n}$}
	\label{Section 4}
	
	In this section, we extend the theoretical findings in Section \ref{Section 3} to the generic candidate set $\mathcal{M}_A$ instead of $\mathcal{M}_N$, where $\{|\theta_j|\}_{j=1}^{p_n}$ are no longer required to satisfy the monotonicity property in Assumption \ref{Assump 6}. This case is very important, because it is impossible to order the unknown $\{|\theta_j|\}_{j=1}^{p_n}$ in practice, and we normally construct a series of nested models $\mathcal{M}_A$ based on a prespecified sequence of covariates (or equivalently $\{\phi_j\}_{j=1}^{p_n}$).
	
	The main challenge in this case is to characterize the specific forms of $R_n(m_n^*\mid\mathcal M_A,\boldsymbol\mu)$ and $R_n(\boldsymbol w^*\mid\mathcal M_A,\boldsymbol\mu)$ and determine when their difference represents a substantial improvement over MS. We use the pool-adjacent-violators algorithm (PAVA, \citealp{Ayer1955}) to express the oracle MA risk through consecutive blocks, in accordance with the existing shrinkage representation of nested MA discussed below. We then show that the same partition also yields the oracle MS risk, which provides a common basis for the subsequent risk comparison.
	
	\subsection{PAVA representations of the oracle risks}\label{Section 4.1}
	First, recall that the oracle MS risk can be written as
	\[
	R_n(m_n^*\mid\mathcal M_A,\boldsymbol\mu)
	=
	\inf_{\boldsymbol\gamma\in\Gamma_n^{\mathrm{MS}}}
	R_n(\boldsymbol w\mid\mathcal M_A,\boldsymbol\mu),
	\]
	where $\Gamma_n^{\mathrm{MS}}=\{\boldsymbol\gamma^{(m)}:\gamma_j^{(m)}=\mathbf 1\{j\leq m\},\;m=1,\ldots,p_n\}$. Each element of $\Gamma_n^{\mathrm{MS}}$ therefore represents a model that uses the first $m$ projected components $\boldsymbol\phi_j$. Next, for any $\boldsymbol\gamma\in\Gamma_n$, the MA risk can be equivalently written as
	\begin{align}\label{Con 15}
		\begin{split}
			R_n(\boldsymbol w\mid\mathcal M_A,\boldsymbol\mu)
			&=
			\sum_{j=1}^{p_n}
			\left(\frac{\sigma^2}{n}+\theta_j^2\right)
			\left(
			\gamma_j-\frac{\theta_j^2}{\theta_j^2+\sigma^2/n}
			\right)^2
			+
			\sum_{j=1}^{p_n}
			\frac{\theta_j^2\sigma^2}{n\theta_j^2+\sigma^2}\\
			&=
			\sum_{j=1}^{p_n}(1-\gamma_j)^2\theta_j^2
			+
			\frac{\sigma^2}{n}\sum_{j=1}^{p_n}\gamma_j^2.
		\end{split}
	\end{align}
	
	Under Assumption~\ref{Assump 6}, the ratios $\theta_j^2/(\theta_j^2+\sigma^2/n)$ are non-increasing over $j=2,\ldots,p_n$. Thus, with $\gamma_1=1$ fixed, the coordinatewise minimizers for the remaining components satisfy the constraints in $\Gamma_n$. However, this is generally false under an arbitrary ordering of $\{|\theta_j|\}_{j=1}^{p_n}$. By the first representation in \eqref{Con 15}, the minimization of the MA risk over $\gamma_2,\ldots,\gamma_{p_n}$ is a weighted antitonic regression problem, with responses $\theta_j^2/(\theta_j^2+\sigma^2/n)$ and weights $\theta_j^2+\sigma^2/n$. This problem can be solved by the weighted PAVA \citep{Miles1959,Leeuw2009}. The general procedure is summarized as follows.
	\begin{enumerate}
		\item
		For $j=2,\ldots,p_n$, start with singleton blocks and let $\gamma_j=\theta_j^2/(\theta_j^2+\sigma^2/n)$.
		\item
		Whenever the fitted value of a block is smaller than that of the next block, merge them and assign a common value that minimizes the sum of their weighted squared errors. For instance, two adjacent singleton blocks with $\gamma_j<\gamma_{j+1}$ receive the common value
		\[
		\mathop{\arg\min}_{\gamma\in[\gamma_j,\gamma_{j+1}]}
		\sum_{k=j}^{j+1}
		\left(\frac{\sigma^2}{n}+\theta_k^2\right)
		\left(
		\gamma-\frac{\theta_k^2}{\theta_k^2+\sigma^2/n}
		\right)^2.
		\]
		For blocks formed by earlier merges, the same update uses the sum over all indices in their union.
		\item Repeat this procedure on adjacent blocks until all fitted values are non-increasing. Combine adjacent blocks with equal fitted values, and denote the resulting maximal constant blocks by $\mathcal J_{n,1},\ldots,\mathcal J_{n,L_n}$.
	\end{enumerate}
	The resulting blocks $\mathcal J_{n,1},\ldots,\mathcal J_{n,L_n}$ form a partition of $\{2,\ldots,p_n\}$. The first component is excluded from the PAVA because $\gamma_1=1$ is fixed. For each resulting block, define
	\[
	\bar\theta_{n,l}^2
	=
	\frac{1}{|\mathcal J_{n,l}|}
	\sum_{j\in\mathcal J_{n,l}}\theta_j^2,
	\quad l=1,\ldots,L_n.
	\]
	\begin{Rem}[Uniqueness of the PAVA solution]
		The choice of the violating adjacent pair in Step~2 does not affect the final solution. Since all weights $\theta_j^2+\sigma^2/n$ are positive, the objective in \eqref{Con 15} is strictly convex in $(\gamma_2,\ldots,\gamma_{p_n})$. Together with the compactness and convexity of $\Gamma_n$, this ensures that the global minimizer is unique.
		
		This global minimizer remains constant on every current block throughout the algorithm. The property holds initially because all blocks are singletons. Whenever two adjacent blocks violate the required order, the pooling property of weighted antitonic regression implies that the global minimizer must take the same value on both blocks \citep{Leeuw2009}. Their merger therefore preserves this property, so the global minimizer remains among the vectors that are constant on the updated blocks, regardless of which violating pair is chosen.
		
		At termination, each block is assigned the value that minimizes its weighted squared error. These values are non-increasing and lie in $[0,1)$, so, together with $\gamma_1=1$, they satisfy the constraints in $\Gamma_n$. The resulting vector therefore minimizes the risk among vectors that are constant on the final blocks. Since this class still contains the original global minimizer, the resulting vector also minimizes the original risk. Uniqueness then ensures that every valid merging order yields the same optimal transformed weights and oracle MA risk. The final combination of adjacent blocks with equal fitted values also uniquely determines the maximal constant blocks $\mathcal J_{n,1},\ldots,\mathcal J_{n,L_n}$.
	\end{Rem}
	
	The common fitted value on $\mathcal J_{n,l}$ is
	\[
	\frac{
		\sum_{j\in\mathcal J_{n,l}}
		(\theta_j^2+\sigma^2/n)
		\frac{\theta_j^2}{\theta_j^2+\sigma^2/n}
	}{
		\sum_{j\in\mathcal J_{n,l}}(\theta_j^2+\sigma^2/n)
	}
	=
	\frac{\bar\theta_{n,l}^2}{\bar\theta_{n,l}^2+\sigma^2/n}.
	\]
	The following proposition gives the resulting optimal MA weights and risk. Its risk representation specializes equation~(3.14) in Theorem~3.1 of \citet{Peng2024shrinkage} to the complete nested path $\mathcal M_A$ and the normalized risk used here.
	
	\begin{Pro}[PAVA representation of the optimal MA risk]
		\label{Proposition 1}
		Under Assumptions~\ref{Assump 1} and~\ref{Assump 5}, the optimal MA weights $\boldsymbol w^*\mid\mathcal M_A$ are unique and their transformed weights satisfy
		\begin{equation}\label{Con 21}
			\gamma_1^*=1
			\quad\text{and}\quad
			\gamma_j^*
			=
			\frac{\bar\theta_{n,l}^2}
			{\bar\theta_{n,l}^2+\sigma^2/n},
			\quad j\in\mathcal J_{n,l}.
		\end{equation}
		The optimal MA risk $R_n(\boldsymbol w^*\mid\mathcal M_A,\boldsymbol\mu)$ is
		\begin{equation}\label{Con 23}
			R_n(\boldsymbol w^*\mid\mathcal M_A,\boldsymbol\mu)
			=
			\frac{\sigma^2}{n}
			+
			\sum_{l=1}^{L_n}
			|\mathcal J_{n,l}|
			\frac{\bar\theta_{n,l}^2\sigma^2}
			{n\bar\theta_{n,l}^2+\sigma^2}.
		\end{equation}
	\end{Pro}
	
	The expressions \eqref{Con 21} and \eqref{Con 23} are the blockwise counterparts of the optimal transformed weights and MA risk under $\mathcal M_N$. Under $\mathcal M_N$, each ordered squared coefficient $\theta_{(j)}^2$, $j=2,\ldots,p_n$, is treated separately, with transformed optimal weight $\theta_{(j)}^2/(\theta_{(j)}^2+\sigma^2/n)$ and risk contribution $\theta_{(j)}^2\sigma^2/(n\theta_{(j)}^2+\sigma^2)$. Under $\mathcal M_A$, the monotonicity constraint may prevent these coordinatewise minimizers from being feasible. The PAVA instead assigns the common transformed weight $\bar\theta_{n,l}^2/(\bar\theta_{n,l}^2+\sigma^2/n)$ to every index in $\mathcal J_{n,l}$, whose risk contribution is $|\mathcal J_{n,l}|\bar\theta_{n,l}^2\sigma^2/(n\bar\theta_{n,l}^2+\sigma^2)$. Thus, in addition to the common part $\sigma^2/n$, $R_n(\boldsymbol{w}^*\mid\mathcal{M}_A,\boldsymbol{\mu})$ is obtained from $R_n(\boldsymbol{w}^*\mid\mathcal{M}_N,\boldsymbol{\mu})$ by replacing the individual signal strengths $\theta_{(j)}^2$ with the PAVA-fitted block averages $\bar{\theta}_{n,l}^2$ and replacing the coordinatewise summation with a blockwise summation weighted by the block sizes. In particular, when $\theta_j^2$ are already non-increasing, Proposition \ref{Proposition 1} reduces exactly to the corresponding results under $\mathcal{M}_N$. 
	
	The original model weights can be recovered by setting $\gamma_{p_n+1}^*=0$ and $w_m^*=\gamma_m^*-\gamma_{m+1}^*$ for $m=1,\ldots,p_n$. This shows that, apart from the first model, positive model weights can only occur at the right endpoints of $\mathcal J_{n,1},\ldots,\mathcal J_{n,L_n}$.
	
	\begin{Rem}
		Here, we discuss how our PAVA-based approach to computing the optimal MA risk and the corresponding weights relates to existing work.
		
		First, Theorem~3.1 of \cite{Peng2024shrinkage} characterizes the optimal MA risk in a similar setting without requiring $\{|\theta_j|\}_{j=1}^{p_n}$ to be ordered, and the accompanying proof identifies the corresponding optimal weights. The argument follows the same pooling principle as our PAVA approach. The distinction in the candidate sets is that his formulation allows a general nested collection $\mathcal M=\{k_1,\ldots,k_{M_n}\}$, where $1\leq k_1<\cdots<k_{M_n}\leq p_n$, whereas we consider the complete nested candidate collection $\mathcal{M}_A$. Accordingly, the oracle MA representation in Proposition~\ref{Proposition 1} is a special case of his Theorem~3.1. However, the two studies use this representation for different purposes. The analysis of \cite{Peng2024shrinkage} further compares this risk with the oracle MA risk under a relaxed weight constraint. In contrast, we use the PAVA partition to derive the optimal MS risk and identify an optimal model cutoff, which provides the basis for our subsequent characterization of the relative risk advantage of MA over MS when $\{|\theta_j|\}_{j=1}^{p_n}$ are unordered.
		
		\citet[Section~5]{ChenKlusowskiTan2024} also use weighted isotonic regression to compute combining weights for a series of nested estimators. The connection with Proposition~\ref{Proposition 1} lies in the same structural reduction: the nested models and nonnegative weights yield monotonicity constraints on the transformed weights $\boldsymbol{\gamma}$, which can also be handled through adjacent pooling. However, their stacking weights minimize a penalized loss without the sum-to-one constraint $\gamma_1=1$. Their isotonic fit depends on reductions in residual sums of squares and model-dimension increments, and a subsequent thresholding step determines the final weights. In contrast, Proposition~\ref{Proposition 1} minimizes the true MA risk $R_n(\boldsymbol{w}\mid\mathcal{M}_A,\boldsymbol{\mu})$ under the simplex weight space, with the PAVA partition determined by $\theta_j^2$. Thus, the two approaches share a monotone optimization principle but generally yield different partitions and weights. 
	\end{Rem}
		
	Next, we utilize the same PAVA partition to yield the optimal MS risk under $\mathcal{M}_A$, which is summarized in Proposition \ref{Proposition 2}.
	
	\begin{Pro}[PAVA representation of the optimal MS risk]
		\label{Proposition 2}
		Under Assumptions \ref{Assump 1} and \ref{Assump 5}, there exists an optimal MS cutoff at a PAVA block boundary. An optimal cutoff of $\{1,\ldots,p_n\}$ that determines the index of $m_n^*$ in $\mathcal{M}_A$ should include every block satisfying $\bar{\theta}^2_{n,l}>{\sigma^2}/{n}$ and exclude every block satisfying $\bar{\theta}^2_{n,l}<{\sigma^2}/{n}$.
		If a block average equals $\sigma^2/n$, including or excluding that entire
		block gives the same risk. The resulting oracle risk is
		\begin{equation}
			\label{Con 25}
			R_n(m_n^*\mid\mathcal M_A,\boldsymbol\mu)
			=
			\frac{\sigma^2}{n}
			+
			\sum_{l=1}^{L_n}
			|\mathcal J_{n,l}|
			\min\left(
			\bar{\theta}^2_{n,l},
			\frac{\sigma^2}{n}
			\right).
		\end{equation}
	\end{Pro}
	
	Equation~\eqref{Con 25} is the blockwise counterpart of the optimal MS risk under $\mathcal M_N$. Because $\theta_{(j)}^2$ are non-increasing under $\mathcal M_N$, coordinatewise thresholding at $\sigma^2/n$ is compatible with the nested model path, which gives
	\[
	R_n(m_n^*\mid\mathcal M_N,\boldsymbol\mu)
	=
	\frac{\sigma^2}{n}
	+
	\sum_{j=2}^{p_n}
	\min\left(
	\theta_{(j)}^2,
	\frac{\sigma^2}{n}
	\right).
	\]
	Under $\mathcal M_A$, such coordinatewise thresholding is generally incompatible with the given order. The PAVA instead produces blocks with non-increasing averages $\bar\theta_{n,l}^2$. For each block $\mathcal J_{n,l}$, the inclusion contributes $|\mathcal J_{n,l}|\sigma^2/n$ to $R_n(m_n^*\mid\mathcal M_A,\boldsymbol\mu)$, whereas the exclusion contributes $|\mathcal J_{n,l}|\bar{\theta}_{n,l}^2$. Taking the smaller contribution of the above two yields \eqref{Con 25}. 
	Thus, the unordered problem has the same oracle-risk form as the ordered problem after $\theta_2^2,\ldots,\theta_{p_n}^2$ are replaced by their PAVA block averages, each repeated according to its block size. The contribution of the first term remains $\sigma^2/n$.
	When $\{\theta_j^2\}_{j=1}^{p_n}$ are already non-increasing, the two oracle risks $R_n(m_n^*\mid\mathcal M_A,\boldsymbol\mu)$ and $R_n(m_n^*\mid\mathcal M_N,\boldsymbol\mu)$ coincide.
	
	\subsection{Asymptotic comparisons}
	From Propositions \ref{Proposition 1} and \ref{Proposition 2}, the exact difference between the optimal MS and MA risks is
	\begin{equation}\label{Con 26}
		R_n(m_n^*\mid\mathcal M_A,\boldsymbol\mu)
		-R_n(\boldsymbol w^*\mid\mathcal M_A,\boldsymbol\mu)
		=
		\sum_{l=1}^{L_n}
		|\mathcal J_{n,l}|
		\frac{
			\left\{\min\left(\bar{\theta}^2_{n,l},{\sigma^2}/{n}\right)\right\}^2
		}{
			\bar{\theta}^2_{n,l}+{\sigma^2}/{n}
		}.
	\end{equation}
	Define the relative gain of MA over MS as
	\begin{equation}\label{Con 20}
		G_{n,A}=\frac{R_n(m_n^*\mid\mathcal{M}_A,\boldsymbol{\mu})-			R_n(\boldsymbol{w}^*\mid\mathcal{M}_A,\boldsymbol{\mu})}{R_n(m_n^*\mid\mathcal{M}_A,\boldsymbol{\mu})},
	\end{equation}
	where the subscript $A$ shows that $G_{n,A}$ is constructed based on the candidate model set $\mathcal{M}_A$.
	Note that the relative gain has the following finite-sample bound.
	\begin{Pro}\label{Proposition 3}
		For the complete nested candidate set \(\mathcal M_A\), suppose the risk has the representation in \eqref{Con 15}. Then, for every \(p_n\geq1\),
		\begin{equation}\label{gain-range}
			0\leq G_{n,A}
			\leq
			\frac12\left\{
			1-\frac{\sigma^2}
			{nR_n(m_n^*\mid\mathcal M_A,\boldsymbol\mu)}
			\right\}
			\leq\frac{p_n-1}{2p_n}.
		\end{equation}
		For \(p_n\geq2\), the upper bound is attained when every PAVA block average equals \(\sigma^2/n\). 
	\end{Pro}
	This bound does not require an additional tail condition or an asymptotic growth restriction. It will also restrict the confidence intervals constructed below. The asymptotic behavior of \(G_{n,A}\) distinguishes the two cases in \eqref{Con 1} and \eqref{Con 2}: a value that tends to zero is negligible, whereas a value bounded away from zero represents a substantial relative improvement.
	
	For any $c>1$, define the proportion of $R_n(m_n^*\mid\mathcal M_A,\boldsymbol\mu)$ contributed by PAVA blocks whose $\bar\theta_{n,l}^2$ are within the levels of $\sigma^2/(cn)$ and $c\sigma^2/n$ as
	\begin{equation*}
		\tilde f_n(c)
		=
		\frac{
			\sum_{l:\,\sigma^2/(cn)\leq
				\bar{\theta}^2_{n,l}\leq c\sigma^2/n}
			|\mathcal J_{n,l}|
			\min(
			\bar{\theta}^2_{n,l},
			{\sigma^2}/{n}
			)
		}{
			R_n(m_n^*\mid\mathcal M_A,\boldsymbol\mu)
		}.
	\end{equation*}
	This quantity is the blockwise counterpart of $f_n(c)$ in Section~\ref{Section 3}. Both quantities measure the proportion of oracle MS risk contributed by components near the noise level $\sigma^2/n$, but the objects being compared with are now $\bar{\theta}_{n,l}^2$ instead of $\theta_{(j)}^2$. 
	
	\begin{Th}
		\label{Theorem 2}
		Under Assumptions \ref{Assump 1} and \ref{Assump 5}, (i) the MA improvement is asymptotically substantial,
		\[
		R_n(m_n^*\mid\mathcal M_A,\boldsymbol\mu)
		-R_n(\boldsymbol w^*\mid\mathcal M_A,\boldsymbol\mu)
		\asymp
		R_n(m_n^*\mid\mathcal M_A,\boldsymbol\mu),
		\]
		if and only if there exists a fixed $c>1$ such that $\liminf_{n\to\infty}\tilde f_n(c)>0$; (ii) the MA improvement is asymptotically negligible,
		\[
		R_n(m_n^*\mid\mathcal M_A,\boldsymbol\mu)
		-R_n(\boldsymbol w^*\mid\mathcal M_A,\boldsymbol\mu)
		=
		o\left(
		R_n(m_n^*\mid\mathcal M_A,\boldsymbol\mu)
		\right),
		\]
		if and only if there exists a sequence $c_n\to\infty$ such that $\tilde f_n(c_n)\to0$.
	\end{Th}
	
	The main theoretical contribution of Theorem~\ref{Theorem 2} is to extend the characterization of MA improvability beyond the non-increasing ordering of \(\{|\theta_j|\}_{j=1}^{p_n}\) imposed in \cite{Peng2022,xu2022,Lin2025nonstationary}. To the best of our knowledge, this is the first necessary-and-sufficient characterization of when the oracle MA improvement over oracle MS is asymptotically substantial or negligible under an arbitrary fixed ordering of $\{|\theta_j|\}_{j=1}^{p_n}$.
	
	The signal-composition interpretation in Section~\ref{sec:signal-composition} applies to these risks with each block average $\bar\theta_{n,l}^2$ repeated $|\mathcal J_{n,l}|$ times in \eqref{Con 23} and \eqref{Con 25}. We classify blocks according to $n\bar\theta_{n,l}^2/\sigma^2$ using the same thresholds $v_n,c_1,c_2,u_n$. Strong and moderate blocks contribute to the oracle MS risk at order $\sigma^2/n$ per component, whereas each weak block contributes $|\mathcal J_{n,l}|\bar\theta_{n,l}^2$. If the transition blocks contribute a negligible proportion of this risk, the same asymptotic comparisons apply to the total numbers of components in the strong and moderate blocks and the accumulated strength of the weak blocks in units of $\sigma^2/n$. The first-component contribution remains $\sigma^2/n$.
	
	\subsection{Effects of coefficient ordering on oracle risks}\label{sec:ordering-effects}
		
		The previous content in Section~\ref{Section 4} characterizes the relative advantage of MA under an arbitrary ordering. We now examine how the ordering itself affects the oracle risks. Throughout this subsection, we fix the orthonormal basis $\{\boldsymbol{\phi}_j\}_{j=1}^{p_n}$ and their corresponding coefficients $\{\theta_j\}_{j=1}^{p_n}$ and change only their order in the nested model sequence. As all components may be permuted, so the first component included in every candidate model may differ across orderings.
		
		\begin{Pro}[Effects of ordering on oracle risks]\label{prop:ordering-risk}
			Suppose that Assumption~\ref{Assump 5} holds with $\sigma^2>0$ and $p_n\geq2$.
			
			\textup{(i)} Among all permutations of the same coefficients $\{|\theta_j|\}_{j=1}^{p_n}$, the non-increasing ordering minimizes both oracle risks of MS and MA, whereas the non-decreasing ordering maximizes them. In particular,
			\begin{equation}\label{eq:ordering-risk-ratios}
				\begin{aligned}
					1
					\leq
					\frac{R_n(m_n^*\mid\mathcal M_A,\boldsymbol\mu)}
					{R_n(m_n^*\mid\mathcal M_N,\boldsymbol\mu)}
					\leq p_n,\quad
					1
					\leq
					\frac{R_n(\boldsymbol w^*\mid\mathcal M_A,\boldsymbol\mu)}
					{R_n(\boldsymbol w^*\mid\mathcal M_N,\boldsymbol\mu)}
					\leq p_n.
				\end{aligned}
			\end{equation}
			The upper bound $p_n$ is attainable for MS and can be approached arbitrarily closely for MA.
			
			\textup{(ii)} Let $\mathcal J_{n,1},\ldots,\mathcal J_{n,L_n}$ and $\bar\theta_{n,l}^2$ be the PAVA blocks and their average squared coefficients under $\mathcal M_A$, as defined in Section~\ref{Section 4.1}. The increases in the oracle risks relative to $\mathcal M_N$ satisfy
			\begin{equation}\label{eq:ordering-ms-excess}
				\begin{aligned}
					& R_n(m_n^*\mid\mathcal M_A,\boldsymbol\mu)
					-R_n(m_n^*\mid\mathcal M_N,\boldsymbol\mu)\\
					=&
					\min\left(\theta_{(1)}^2,\frac{\sigma^2}{n}\right)
					-\min\left(\theta_1^2,\frac{\sigma^2}{n}\right)+
					\sum_{l=1}^{L_n}
					\min\left\{
					\sum_{j\in\mathcal J_{n,l}}
					\left(\theta_j^2-\frac{\sigma^2}{n}\right)_+,\,
					\sum_{j\in\mathcal J_{n,l}}
					\left(\frac{\sigma^2}{n}-\theta_j^2\right)_+
					\right\},
				\end{aligned}
			\end{equation}
			and
			\begin{equation}\label{eq:ordering-ma-excess}
				\begin{aligned}
					& R_n(\boldsymbol w^*\mid\mathcal M_A,\boldsymbol\mu)
					-R_n(\boldsymbol w^*\mid\mathcal M_N,\boldsymbol\mu)\\
					=&
					\frac{\theta_{(1)}^2\sigma^2}{n\theta_{(1)}^2+\sigma^2}
					-\frac{\theta_1^2\sigma^2}{n\theta_1^2+\sigma^2}
					+
					\sum_{l=1}^{L_n}
					\frac{(\sigma^2/n)^2}{(\bar\theta_{n,l}^2+\sigma^2/n)^2}
					\sum_{j\in\mathcal J_{n,l}}
					\frac{(\theta_j^2-\bar\theta_{n,l}^2)^2}
					{\theta_j^2+\sigma^2/n},
				\end{aligned}
			\end{equation}
			where $(x)_+=\max\{x,0\}$.
		\end{Pro}
		
		The decompositions in \eqref{eq:ordering-ms-excess} and \eqref{eq:ordering-ma-excess} separate two sources of risk increase. The first difference accounts for which components are included, and it vanishes when the $\theta_{(1)}^2$ remains in the first position for both $\mathcal{M}_N$ and $\mathcal{M}_A$. The remaining terms measure the risk increase caused by the PAVA blocks.
		
		The two oracle risks respond differently to these PAVA blocks. For MS, a block $\mathcal J_{n,l}$ contributes a positive term in \eqref{eq:ordering-ms-excess} if and only if it contains $\theta_j^2$'s strictly above and strictly below $\sigma^2/n$. In particular, suppose that $\theta_{(1)}^2>\sigma^2/n$ and no $\theta_j^2$ equals $\sigma^2/n$. Any permutation that keeps all components with $\theta_j^2>\sigma^2/n$ before those with $\theta_j^2<\sigma^2/n$ leaves both the index of the optimal MS model and the corresponding oracle risk unchanged, regardless of the ordering within either group. Moreover, the MS risk decreases as components from the first group are included and increases as components from the second group are included. Consequently,
		\[
		m_n^*=\#\{j:\theta_j^2>\sigma^2/n\},
		\quad
		R_n(m_n^*\mid\mathcal M_A,\boldsymbol\mu)
		=
		R_n(m_n^*\mid\mathcal M_N,\boldsymbol\mu).
		\]
		Thus, $R_n(m_n^*\mid\mathcal M_N,\boldsymbol\mu)$ remains attainable under arbitrary permutations within the two groups, provided that their $\theta_j^2$'s are separated at $\sigma^2/n$.
		
		In contrast, a block $\mathcal J_{n,l}$ contributes a positive term in \eqref{eq:ordering-ma-excess} whenever its squared coefficients are not all equal, even if they lie on the same side of $\sigma^2/n$. Hence, permutations within either group can increase the oracle MA risk while leaving the optimal MS index and its risk unchanged.
		
		Note that neither $G_{n,A}\geq G_{n,N}$ nor $G_{n,A}\leq G_{n,N}$ holds in general. Compard to the benchmark $G_{n,N}$, the effect of ordering on $G_{n,A}$ depends on the relative changes in the two oracle risks.
		
		\begin{Ex}[Risk inflation caused by ordering]\label{ex:ordering-risk-inflation}
			Assume $p_n\to\infty$ and $p_n=o(n)$, and consider
			\[
			(\theta_1^2,\ldots,\theta_{p_n}^2)
			=
			\frac{\sigma^2}{n}
			\left(
			\underbrace{0,\ldots,0}_{p_n-1\text{ terms}},
			p_n-1
			\right).
			\]
			The total squared signal strength is $(p_n-1)\sigma^2/n$, so Assumption~\ref{Assump 1} holds. Under $\mathcal M_A$, the PAVA pools all components of $\{2,\ldots,p_n\}$ into one block with an average fitted value of $\sigma^2/n$. Propositions~\ref{Proposition 1} and~\ref{Proposition 2} give
			\[
			R_n(m_n^*\mid\mathcal M_A,\boldsymbol\mu)
			=\frac{p_n\sigma^2}{n},
			\quad
			R_n(\boldsymbol w^*\mid\mathcal M_A,\boldsymbol\mu)
			=\frac{(p_n+1)\sigma^2}{2n}.
			\]
			Under $\mathcal M_N$, the only nonzero coefficient becomes the first component, and
			\[
			R_n(m_n^*\mid\mathcal M_N,\boldsymbol\mu)
			=
			R_n(\boldsymbol w^*\mid\mathcal M_N,\boldsymbol\mu)
			=
			\frac{\sigma^2}{n}.
			\]
			Consequently, the MS and MA risks increase by factors $p_n$ and $(p_n+1)/2$, respectively. The ordering therefore changes their convergence rates, although both risks still tend to zero.
			
			At the same time,
			\[
			G_{n,N}=0,
			\quad
			G_{n,A}=\frac{p_n-1}{2p_n}\longrightarrow\frac12.
			\]
			The latter attains the upper bound in Proposition~\ref{Proposition 3}. Thus, a substantial relative advantage of MA can coexist with a much larger absolute MA risk than under the decreasing order.
		\end{Ex}

    \section{Feasible diagnostic for MA gains over MS}\label{Section 6}
		The preceding analysis characterizes the potential advantage of MA over MS through their oracle risks. These risks, however, depend on the unknown signal coefficients $\{|\theta_j|\}_{j=1}^{p_n}$ and hence cannot be evaluated directly from the observed data. This section constructs a data-based criterion and a confidence interval to compare this comparative advantage. 
		
		\subsection{Estimation consistency}
		Recall that in \eqref{Con 15} the MA risk can be equivalently written as
		\begin{equation*}
			R_n(\boldsymbol{w}\mid\mathcal{M}_A,\boldsymbol{\mu})
			=
			\sum_{j=1}^{p_n}(1-\gamma_j)^2\theta_j^2
			+
			\frac{\sigma^2}{n}\sum_{j=1}^{p_n}\gamma_j^2
		\end{equation*}
		for any $\boldsymbol{\gamma}\in\Gamma_n$.
		Direct estimation of this formula by replacing $\theta_j^2$ with $\hat\theta_j^2$ would introduce an additional variance term because
		\[
		\E(\hat\theta_j^2)
		=
		\theta_j^2+\frac{\sigma^2}{n}.
		\]
		To correct this bias, we define the projected Mallows criterion by
		\begin{equation}\label{Con 12}
			C_n(\boldsymbol{w}\mid\mathcal{M}_A,\boldsymbol{\mu})
			=
			\sum_{j=1}^{p_n}(1-\gamma_j)^2\hat\theta_j^2
			+
			\frac{2\sigma^2}{n}\sum_{j=1}^{p_n}\gamma_j
			-
			\frac{\sigma^2}{n}p_n.
		\end{equation}
		and for every fixed $\boldsymbol{\gamma}\in\Gamma_n$, $\E\left\{
		C_n(\boldsymbol{w}\mid\mathcal{M}_A,\boldsymbol{\mu})
		\right\}
		=
		R_n(\boldsymbol{w}\mid\mathcal{M}_A,\boldsymbol{\mu})$.
		We later utilize \eqref{Con 12} to construct an assessment criterion of the relative advantage of MA over MS.
		Define
		\[
		\begin{aligned}
			C_n(m_n^*\mid\mathcal{M}_A,\boldsymbol{\mu})
			&=
			\inf_{\boldsymbol{\gamma}\in\Gamma_n^{\mathrm{MS}}}
			C_n(\boldsymbol{w}\mid\mathcal{M}_A,\boldsymbol{\mu}),\\
			C_n(\boldsymbol{w}^*\mid\mathcal{M}_A,\boldsymbol{\mu})
			&=
			\inf_{\boldsymbol{\gamma}\in\Gamma_n}
			C_n(\boldsymbol{w}\mid\mathcal{M}_A,\boldsymbol{\mu}).
		\end{aligned}
		\]
		Recall that in \eqref{Con 20}, the relative gain of MA over MS based on $\mathcal{M}_A$ is defined as
		\[
		G_{n,A}=\frac{R_n(m_n^*\mid\mathcal{M}_A,\boldsymbol{\mu})-			R_n(\boldsymbol{w}^*\mid\mathcal{M}_A,\boldsymbol{\mu})}{R_n(m_n^*\mid\mathcal{M}_A,\boldsymbol{\mu})}.
		\]
		On this basis, define the practical gain diagnostic
		as
		\[
		\hat G_{n,A}
		=
		\frac{
			C_n(m_n^*\mid\mathcal{M}_A,\boldsymbol{\mu})
			-
			C_n(\boldsymbol{w}^*\mid\mathcal{M}_A,\boldsymbol{\mu})
		}{
			C_n(m_n^*\mid\mathcal{M}_A,\boldsymbol{\mu})
		}.
		\]
		The objective is to establish that $\hat G_{n,A}$ consistently approximates $G_{n,A}$, so that the potential gain from model averaging can be assessed from the data. 
		Although the denominator of $\hat G_{n,A}$ is not necessarily positive for every finite sample, the results below show that $C_n(m_n^*\mid\mathcal{M}_A,\boldsymbol{\mu})$ is asymptotically equivalent to the strict positive $R_n(m_n^*\mid\mathcal{M}_A,\boldsymbol{\mu})$ and $\hat G_{n,A}$ is therefore positive with probability tending to $1$.
		
		For the theoretical development, criterion \eqref{Con 12} is first written in terms of $\sigma^2$. When $\sigma^2$ is unknown, its feasible counterpart is obtained by replacing $\sigma^2$ with a consistent estimator. Next, we introduce the following sufficient conditions.
		
		\begin{Assump}[Sub-Gaussian errors]\label{Assump 4}
			The standardized errors $\{\varepsilon_i/\sigma\}_{i=1}^n$ are uniformly
			sub-Gaussian, i.e.,
			\[
			P\left(
			\left|
			\frac{\varepsilon_i}{\sigma}
			\right|
			\ge t
			\right)
			\le
			2\exp\left(
			-\frac{t^2}{K^2}
			\right)
			\]
			for every $t>0$ and some fixed constant $K>0$ independent of $i$ and $n$.
		\end{Assump}		
		\begin{Assump}\label{Assump 2}
			$s_{n,A}\to\infty$ and
			$p_n\log\,(e p_n)=o(s_{n,A}^2)$.
		\end{Assump}
		
		Assumption~\ref{Assump 4} specifies that the standardized errors $\{\varepsilon_i/\sigma\}_{i=1}^n$ have uniformly sub-Gaussian tails. Together with Assumption~\ref{Assump 5}, this condition permits the application of the Hanson--Wright inequality \citep{RudelsonVershynin2013} in the proof of Theorem~\ref{Theorem 3}.
		
		For Assumption \ref{Assump 2}, first recall that in Section \ref{Section 3},
		\[
		R_n(m_n^*\mid\mathcal{M}_A,\boldsymbol{\mu})
		=
		\frac{\sigma^2}{n}m_n^*+\sum_{j=m_n^*+1}^{p_n}\theta_j^2.
		\]
		with 
		\begin{equation}\label{Con 18}
			\sum_{j=m_n^*+1}^{k}\theta_j^2\leq \frac{\sigma^2(k-m_n^*)}{n}\;\;\text{for any}\;\;m_n^*+1\leq k\leq p_n.
		\end{equation}
		Consequently,
		\[
		s_{n,A}
		=
		\frac{nR_n(m_n^*\mid\mathcal{M}_A,\boldsymbol{\mu})}{\sigma^2}
		=
		m_n^*
		+
		\frac{n}{\sigma^2}
		\sum_{j=m_n^*+1}^{p_n}\theta_j^2,
		\]
		and \(s_{n,A}\) incorporates both the index \(m_n^*\) of the best single model and its normalized approximation error. Moreover, by taking $k=p_n$ in \eqref{Con 18}, it is easy to show $m_n^*\le s_{n,A}\leq p_n$. On this basis, the first requirement \(s_{n,A}\to\infty\) implies \(p_n\to\infty\), but it does not necessarily imply \(m_n^*\to\infty\).

		The second requirement in Assumption~\ref{Assump 2} restricts the divergence rate of $p_n$ with respect to $s_{n,A}$. In general, this cannot be expressed solely in terms of \(m_n^*\), because the normalized approximation error of \(s_{n,A}\) may be nonnegligible. If this component is \(O(m_n^*)\), then \(s_{n,A}\asymp m_n^*\), in which case the second requirement reduces to $p_n\log(e p_n)=o\{(m_n^*)^2\}$.
		
		The following theorem formalizes this uniform approximation and establishes the consistency of $\hat G_{n,A}$ for $G_{n,A}$.
	
	\begin{Th}
		\label{Theorem 3}
		Suppose Assumptions \ref{Assump 1}, \ref{Assump 5}, \ref{Assump 4} and \ref{Assump 2} hold. Then
		\begin{equation}\label{Con 3}
		C_n(m_n^*\mid\mathcal{M}_A,\boldsymbol{\mu})
		=
		R_n(m_n^*\mid\mathcal{M}_A,\boldsymbol{\mu})+o_p(R_n(m_n^*\mid\mathcal{M}_A,\boldsymbol{\mu})),
		\end{equation}
		and
		\begin{equation}\label{Con 4}
		C_n(\boldsymbol{w}^*\mid\mathcal{M}_A,\boldsymbol{\mu})
		=
		R_n(\boldsymbol{w}^*\mid\mathcal{M}_A,\boldsymbol{\mu})+o_p(R_n(m_n^*\mid\mathcal{M}_A,\boldsymbol{\mu})).
	\end{equation}
		Consequently,
		\[
		\hat G_{n,A}-G_{n,A}=o_p(1).
		\]
	\end{Th}
	Under the stated assumptions, Theorem~\ref{Theorem 3} establishes the consistency of $\hat G_{n,A}$ to $G_{n,A}$. 
	A key practical advantage of Theorem~\ref{Theorem 3} is that it does not require the unknown sequence \(\{|\theta_j|\}_{j=1}^{p_n}\) to satisfy the monotonicity condition in Assumption~\ref{Assump 6}. For any ordering specified in advance, the corresponding practical diagnostic \(\hat G_{n,A}\) can be computed from the data when \(\sigma^2\) is known. 
	
	Theorem~\ref{Theorem 3} does not guarantee that $\hat G_{n,A}$ lies in $[0,(p_n-1)/(2p_n)]$ for every finite sample. Although the inclusion of the MS class in the MA class ensures $C_n(\boldsymbol w^*\mid\mathcal M_A,\boldsymbol\mu)\leq C_n(m_n^*\mid\mathcal M_A,\boldsymbol\mu)$, the oracle gain bound does not apply to the ratio of the optimized criteria. The possible cases are as follows.
		\begin{enumerate}
			\item If $0<C_n(\boldsymbol w^*\mid\mathcal M_A,\boldsymbol\mu)\leq C_n(m_n^*\mid\mathcal M_A,\boldsymbol\mu)$, then $\hat G_{n,A}\in[0,1)$, but it may exceed $(p_n-1)/(2p_n)$.
			\item If $C_n(\boldsymbol w^*\mid\mathcal M_A,\boldsymbol\mu)\leq0<C_n(m_n^*\mid\mathcal M_A,\boldsymbol\mu)$, then $\hat G_{n,A}\geq1$.
			\item If $C_n(m_n^*\mid\mathcal M_A,\boldsymbol\mu)=0$, the ratio defining $\hat G_{n,A}$ is undefined.
			\item If $C_n(\boldsymbol w^*\mid\mathcal M_A,\boldsymbol\mu)\leq C_n(m_n^*\mid\mathcal M_A,\boldsymbol\mu)<0$, then $\hat G_{n,A}\leq0$.
		\end{enumerate}
		Since the oracle MS risk $R_n(m_n^*\mid\mathcal M_A,\boldsymbol\mu)$ is strictly positive, \eqref{Con 3} implies
		\[
		\Pr\left\{
		C_n(m_n^*\mid\mathcal M_A,\boldsymbol\mu)>0
		\right\}\to1.
		\]
		The probability of a zero denominator tends to zero as $n$ increases, so the consistency result in Theorem~\ref{Theorem 3} still holds if we define $\hat G_{n,A}=0$ in this case. In particular,
	\[
	G_{n,A}\to0
	\quad\Longleftrightarrow\quad
	\hat G_{n,A}\to_p0,
	\]
	whereas
	\[
	\liminf_{n\to\infty}G_{n,A}>0
	\quad\Longleftrightarrow\quad
	\text{there exists a constant \(c>0\) such that }
	\Pr(\hat G_{n,A}>c)\to1.
	\]
	
	Therefore, Theorem~\ref{Theorem 3} provides asymptotic justification for using \(\hat G_{n,A}\) to distinguish a negligible oracle gain from a substantial one under any prespecified ordering, without estimating or recovering an oracle ordering of \(\{|\theta_j|\}_{j=1}^{p_n}\). For a certain finite sample, however, a positive point estimate $\hat G_{n,A}$ alone does not quantify its reliability. The finite-sample procedure developed next addresses the latter issue.
	
	\subsection{A finite-sample confidence interval}\label{Section 5.2}
		
		We construct a confidence interval for $G_{n,A}$ under Gaussian errors, with the design and candidate ordering fixed. For $p_n\geq2$, define
		\begin{equation*}
			\Lambda_m=\frac{n}{\sigma^2}\sum_{j=m}^{p_n}\theta_j^2,
			\quad m=2,\ldots,p_n,
			\quad \Lambda_{p_n+1}=0.
		\end{equation*}
		It can be shown that
		\begin{equation}\label{Con 31}
			\begin{aligned}
			\frac{n}{\sigma^2}R_n(m\mid\mathcal M_A,\boldsymbol\mu)&={m}+{\Lambda_{m+1}},\\
			\frac{n}{\sigma^2}
			R_n(\boldsymbol w\mid\mathcal M_A,\boldsymbol\mu)
			&=\sum_{j=1}^{p_n}\gamma_j^2+\sum_{m=2}^{p_n}
			\left\{(1-\gamma_m)^2-(1-\gamma_{m-1})^2\right\}\Lambda_m.
			\end{aligned}
		\end{equation}
		For $\boldsymbol\lambda=(\lambda_2,\ldots,\lambda_{p_n+1})$ satisfying $\lambda_2\geq\cdots\geq\lambda_{p_n+1}=0$, let
		\begin{equation}\label{eq:ci-oracle-risks}
			\begin{aligned}
				R_{n,\mathrm{MS}}(\boldsymbol\lambda)
				&=\min_{1\leq m\leq p_n}\{m+\lambda_{m+1}\},\\
				R_{n,\mathrm{MA}}(\boldsymbol\lambda)
				&=\min_{\boldsymbol\gamma\in\Gamma_n}
				\left[
				\sum_{j=1}^{p_n}\gamma_j^2+
				\sum_{m=2}^{p_n}
				\left\{(1-\gamma_m)^2-(1-\gamma_{m-1})^2\right\}\lambda_m
				\right].
			\end{aligned}
		\end{equation}
		By comparing \eqref{Con 31} and \eqref{eq:ci-oracle-risks}, we find that, at $\boldsymbol\Lambda=(\Lambda_2,\ldots,\Lambda_{p_n+1})$, $G_{n,A}=1-R_{n,\mathrm{MA}}(\boldsymbol\Lambda)/R_{n,\mathrm{MS}}(\boldsymbol\Lambda)$.
		
		For $\lambda\geq0$, define
		\begin{equation*}
			F_m(\lambda)
			=
			\Pr\left\{
			\chi^2_{p_n-m+1}(\lambda)
			\leq\frac{n}{\sigma^2}\sum_{j=m}^{p_n}\hat\theta_j^2
			\right\},
			\quad m=2,\ldots,p_n,
		\end{equation*}
		where $\chi_d^2(\lambda)$ denotes a noncentral chi-square random variable with $d$ degrees of freedom and noncentrality parameter $\lambda\geq0$. For $0<\alpha<1$, the simultaneous tail bounds are
		\begin{align}
			\underline\Lambda_m(\alpha)
			&=\max_{m\leq j\leq p_n}
			\inf\left\{\lambda\geq0:
			F_j(\lambda)\leq1-\frac{\alpha}{2(p_n-1)}\right\},
			\label{tail-lower}\\
			\overline\Lambda_m(\alpha)
			&=\min_{2\leq j\leq m}
			\inf\left\{\lambda\geq0:
			F_j(\lambda)\leq\frac{\alpha}{2(p_n-1)}\right\}.
			\label{tail-upper}
		\end{align}
		Set $\underline\Lambda_{p_n+1}(\alpha)=\overline\Lambda_{p_n+1}(\alpha)=0$, and write $\underline{\boldsymbol\Lambda}=(\underline\Lambda_2(\alpha),\cdots,\underline\Lambda_{p_n}(\alpha))$ and $\overline{\boldsymbol\Lambda}=(\overline\Lambda_2(\alpha),\cdots,\overline\Lambda_{p_n}(\alpha))$. Both are nonnegative and nonincreasing, i.e., 
		\[
		\begin{aligned}
			\underline\Lambda_2(\alpha)\geq\cdots\geq\underline\Lambda_{p_n}(\alpha)\geq0,\quad\text{and}\quad
			\overline\Lambda_2(\alpha)\geq\cdots\geq\overline\Lambda_{p_n}(\alpha)\geq0.
		\end{aligned}
		\]
		Define
		\begin{equation}\label{eq:ci-tail-region}
			\mathcal C_{n,\alpha}
			=
			\left\{
			\boldsymbol\lambda:
			\lambda_2\geq\cdots\geq\lambda_{p_n+1}=0,\ 
			\underline{\boldsymbol\Lambda}\leq\boldsymbol\lambda
			\leq\overline{\boldsymbol\Lambda}
			\right\},
		\end{equation}
		where vector inequalities are componentwise.
		
		If $\mathcal C_{n,\alpha}\neq\varnothing$, define
		\begin{equation}\label{eq:ci-endpoints}
			\begin{aligned}
				\underline G_{n,\alpha}
				&=
				1-\max_{\boldsymbol\lambda\in\mathcal C_{n,\alpha}}
				\frac{R_{n,\mathrm{MA}}(\boldsymbol\lambda)}
				{R_{n,\mathrm{MS}}(\boldsymbol\lambda)},\\
				\overline G_{n,\alpha}
				&=
				\min\left\{
				1-\frac{R_{n,\mathrm{MA}}(\underline{\boldsymbol\Lambda})}
				{R_{n,\mathrm{MS}}(\overline{\boldsymbol\Lambda})},
				\frac12\left(
				1-\frac1{R_{n,\mathrm{MS}}(\overline{\boldsymbol\Lambda})}
				\right)
				\right\},
			\end{aligned}
		\end{equation}
		and
		\begin{equation}\label{Con 30}
			\mathcal I_{n,\alpha}^{\mathrm{tail}}
			=
			[\underline G_{n,\alpha},\overline G_{n,\alpha}].
		\end{equation}
		The lower endpoint minimizes the gain over the same tail vector for both oracle risks. The upper endpoint follows from coordinatewise monotonicity of the risks and Proposition~\ref{Proposition 3}. If the lower and upper bounds are not compatible, that is, the following event is violated for certain $m$, 
		\begin{equation}\label{Con 33}
		\{\underline\Lambda_m(\alpha)\leq \overline\Lambda_m(\alpha),\;\text{for}\;m=2,\ldots,p_n\},
		\end{equation}
		then $\mathcal C_{n,\alpha}=\varnothing$, and we set $\mathcal I_{n,\alpha}^{\mathrm{tail}}=[0,(p_n-1)/(2p_n)]$. For $p_n=1$, set $\mathcal I_{n,\alpha}^{\mathrm{tail}}=\{0\}$.
		
		\begin{Th}[Finite-sample coverage]\label{Theorem 4}
			For any prespecified candidate ordering, $\mathcal I_{n,\alpha}^{\mathrm{tail}}$ is nonempty and contained in $[0,(p_n-1)/(2p_n)]$. If $\boldsymbol\epsilon\sim \text{Normal}\,(\boldsymbol0,\sigma^2\boldsymbol I_n)$ with known $\sigma^2>0$, then
			\[
			\Pr\left\{G_{n,A}\in\mathcal I_{n,\alpha}^{\mathrm{tail}}\right\}
			\geq1-\alpha,
			\quad 0<\alpha<1.
			\]
		\end{Th}
		
		Theorem~\ref{Theorem 4} requires neither Assumption~\ref{Assump 2} nor positivity of the practical MS criterion. Both endpoints of $\mathcal I_{n,\alpha}^{\mathrm{tail}}$ are nonnegative, so the interval excludes zero exactly when its lower endpoint is positive. This provides evidence that oracle MA has strictly smaller risk than oracle MS. If the interval contains zero, a positive gain is not established, but this does not imply equality of the two oracle risks.
			
			More generally, for $p_n\geq2$ and a prespecified $0\leq\delta<(p_n-1)/(2p_n)$, reject
			\[
			H_0:G_{n,A}\leq\delta
			\quad\text{against}\quad
			H_1:G_{n,A}>\delta
			\]
			when $\inf\mathcal I_{n,\alpha}^{\mathrm{tail}}>\delta$. This test has size at most $\alpha$, since rejection under the null implies failure of coverage. The choice $\delta=0$ gives the zero-exclusion rule above, whereas $\delta>0$ assesses whether the relative risk reduction exceeds a specified magnitude. These conclusions concern the oracle gain at the current sample size and do not establish its asymptotic behavior.
		
		\begin{Rem}[Computation of the confidence interval]\label{remark:ci-computation}
			The upper endpoint in \eqref{eq:ci-endpoints} requires only the oracle risks evaluated at the tail bounds, which are available from the MS risk formula and PAVA. The lower endpoint additionally requires optimization of the risk ratio over $\mathcal C_{n,\alpha}$. We describe a reduction to one-dimensional problems and explain how numerical approximation can preserve the coverage guarantee.
			
			Suppose $p_n\geq2$ and $\mathcal C_{n,\alpha}\neq\varnothing$. For $m=1,\ldots,p_n$ and $t\in[\underline\Lambda_{m+1}(\alpha),\overline\Lambda_{m+1}(\alpha)]$, define
			\[
			\lambda_j^{(m)}(t)=
			\begin{cases}
				\overline\Lambda_j(\alpha),&2\leq j\leq m,\\
				\min\{\overline\Lambda_j(\alpha),t\},&m+1\leq j\leq p_n+1.
			\end{cases}
			\]
			Then
			\begin{equation}\label{eq:ci-scalar}
				1-\underline G_{n,\alpha}
				=
				\max_{1\leq m\leq p_n}
				\max_{\underline\Lambda_{m+1}(\alpha)\leq t\leq\overline\Lambda_{m+1}(\alpha)}
				\frac{R_{n,\mathrm{MA}}(\boldsymbol\lambda^{(m)}(t))}{m+t}.
			\end{equation}				
			Since $R_{n,\mathrm{MS}}(\boldsymbol\lambda)=\min_{1\leq m\leq p_n}\{m+\lambda_{m+1}\}$ and all candidate risks are positive, we have
			\[
			\frac{R_{n,\mathrm{MA}}(\boldsymbol\lambda)}
			{R_{n,\mathrm{MS}}(\boldsymbol\lambda)}
			=
			\max_{1\leq m\leq p_n}
			\frac{R_{n,\mathrm{MA}}(\boldsymbol\lambda)}
			{m+\lambda_{m+1}}.
			\]
			For fixed $m$ and $t\in[\underline\Lambda_{m+1}(\alpha),\overline\Lambda_{m+1}(\alpha)]$, consider any $\boldsymbol\lambda\in\mathcal C_{n,\alpha}$ satisfying $\lambda_{m+1}=t$. The denominator $m+\lambda_{m+1}$ then equals $m+t$. The constraints defining $\mathcal C_{n,\alpha}$ imply $\lambda_j\leq\overline\Lambda_j(\alpha)$ for every $j$, while the monotonicity of $\boldsymbol\lambda$ also gives $\lambda_j\leq t$ for $j\geq m+1$. Hence, $\lambda_j\leq\lambda_j^{(m)}(t)$ for every $j=2,\ldots,p_n+1$. Moreover, the nonincreasing bounds $\underline\Lambda_j(\alpha)$ and $\overline\Lambda_j(\alpha)$, together with the specified range of $t$, ensure that $\boldsymbol\lambda^{(m)}(t)\in\mathcal C_{n,\alpha}$ and $\lambda_{m+1}^{(m)}(t)=t$. Since $R_{n,\mathrm{MA}}(\boldsymbol\lambda)$ is nondecreasing in each coordinate $\lambda_j$, we obtain $R_{n,\mathrm{MA}}(\boldsymbol\lambda)\leq R_{n,\mathrm{MA}}(\boldsymbol\lambda^{(m)}(t))$. Therefore,
			\[
			\max_{{\boldsymbol\lambda\in\mathcal C_{n,\alpha},
					\lambda_{m+1}=t}}
			\frac{R_{n,\mathrm{MA}}(\boldsymbol\lambda)}
			{m+\lambda_{m+1}}
			=
			\frac{R_{n,\mathrm{MA}}(\boldsymbol\lambda^{(m)}(t))}
			{m+t}.
			\]
			The maximum over $m=1,\ldots,p_n$ and $t\in[\underline\Lambda_{m+1}(\alpha),\overline\Lambda_{m+1}(\alpha)]$ then gives \eqref{eq:ci-scalar}.
			
			For numerical implementation, suppose the optimization procedure returns valid bounds $\underline M\leq1-\underline G_{n,\alpha}\leq\overline M$ with $\overline M-\underline M\leq\varepsilon$, where $\varepsilon>0$ is a prescribed tolerance. Define
			\[
			\underline G_{n,\alpha}^{\mathrm{num}}
			=
			\max\left\{
			0,\,
			1-\frac{R_{n,\mathrm{MA}}(\overline{\boldsymbol\Lambda})}
			{R_{n,\mathrm{MS}}(\underline{\boldsymbol\Lambda})},\,
			1-\overline M
			\right\}.
			\]
			The first two terms are lower bounds for the gain at every feasible tail vector $\boldsymbol{\lambda}$, and the third is a lower bound for its minimum over $\mathcal C_{n,\alpha}$. Thus, all three terms are no greater than $\underline G_{n,\alpha}$. The optimization gap also gives
			\[
			\underline G_{n,\alpha}-\varepsilon
			\leq1-\overline M
			\leq\underline G_{n,\alpha}^{\mathrm{num}}
			\leq\underline G_{n,\alpha}.
			\]
			Consequently, the interval $[\underline G_{n,\alpha}^{\mathrm{num}},\overline G_{n,\alpha}]$ contains the theoretical interval and extends it by at most $\varepsilon$ at the lower endpoint, thereby preserving its coverage guarantee. The same calculation applies to $\mathcal I_{n,\alpha}^{\mathrm{plug}}$ after the tail bounds are replaced by $\underline\Lambda_m^{\mathrm{plug}}(\alpha)$ and $\overline\Lambda_m^{\mathrm{plug}}(\alpha)$ obtained from the noncentral $F$ construction in Section~\ref{Section 5.3}. The target $\boldsymbol\Lambda$ and the oracle-risk functions remain unchanged. We use this procedure to compute the confidence intervals in Experiment~4 of Section~\ref{Section 6.4}.
		\end{Rem}
	
	\subsection{Feasible diagnostic}\label{Section 5.3}
	As $\sigma^2$ is unknown in the projected Mallows criterion $C_n(\boldsymbol{w}\mid\mathcal{M}_A,\boldsymbol{\mu})$, we define $\hat\sigma_n^2$ as an estimator of $\sigma^2$, and give the feasible projected Mallows criterion as
	\begin{align}\label{Con 16}
		\hat C_n(\boldsymbol w\mid\mathcal M_A,\boldsymbol\mu)
		=
		\sum_{j=1}^{p_n}(1-\gamma_j)^2\hat\theta_j^2
		+
		2\frac{\hat\sigma_n^2}{n}\sum_{j=1}^{p_n}\gamma_j
		-
		p_n\frac{\hat\sigma_n^2}{n}.
	\end{align}
	Accordingly, define
	\[
	\hat C_n(m^*_{n,\mathrm{plug}}\mid\mathcal M_A,\boldsymbol\mu)
	=
	\inf_{\boldsymbol\gamma\in\Gamma_n^{\mathrm{MS}}}
	\hat C_n(\boldsymbol w\mid\mathcal M_A,\boldsymbol\mu),
	\]
	\[
	\hat C_n(\boldsymbol w^*_{\mathrm{plug}}\mid\mathcal M_A,\boldsymbol\mu)
	=
	\inf_{\boldsymbol\gamma\in\Gamma_n}
	\hat C_n(\boldsymbol w\mid\mathcal M_A,\boldsymbol\mu),
	\]
	and
	\[
	\hat G_{n,A}^{\mathrm{plug}}
	=
	1-
	\frac{
		\hat C_n(\boldsymbol w^*_{\mathrm{plug}}\mid\mathcal M_A,\boldsymbol\mu)
	}{
		\hat C_n(m^*_{n,\mathrm{plug}}\mid\mathcal M_A,\boldsymbol\mu)
	}.
	\]
	On this basis, we establish the consistency property of $\hat G_{n,A}^{\mathrm{plug}}$ to $G_{n,A}$.
	
	\begin{Cor}\label{Corollary 3}
		Suppose the conditions of Theorem~\ref{Theorem 3} hold. The estimator $\hat\sigma_n^2$ need not be independent of the data used to construct $\hat C_n(\boldsymbol w\mid\mathcal M_A,\boldsymbol\mu)$, but it satisfies
		\begin{equation}\label{Con 13}
			\frac{p_n}{n}
			\left|\hat\sigma_n^2-\sigma^2\right|
			=
			o_p\left\{R_n(m_n^*\mid\mathcal M_A,\boldsymbol\mu)\right\}.
		\end{equation}
		Then
		\[
		\hat C_n(m^*_{n,\mathrm{plug}}\mid\mathcal M_A,\boldsymbol\mu)
		=
		R_n(m_n^*\mid\mathcal M_A,\boldsymbol\mu)
		+
		o_p\left\{R_n(m_n^*\mid\mathcal M_A,\boldsymbol\mu)\right\},
		\]
		\[
		\hat C_n(\boldsymbol w^*_{\mathrm{plug}}\mid\mathcal M_A,\boldsymbol\mu)
		=
		R_n(\boldsymbol w^*\mid\mathcal M_A,\boldsymbol\mu)
		+
		o_p\left\{R_n(m_n^*\mid\mathcal M_A,\boldsymbol\mu)\right\},
		\]
		and
		\[
		\hat G_{n,A}^{\mathrm{plug}}-G_{n,A}=o_p(1).
		\]
	\end{Cor}
	
	Similar to $\hat G_{n,A}$, this feasible diagnostic $\hat G_{n,A}^{\mathrm{plug}}$ does not necessarily fall into $[0,(p_n-1)/(2p_n)]$, and this result depends on the values of $\hat C_n(\boldsymbol w^*_{\mathrm{plug}}\mid\mathcal M_A,\boldsymbol\mu)$ and $\hat C_n(m^*_{n,\mathrm{plug}}\mid\mathcal M_A,\boldsymbol\mu)$. 
	Likewise, Corollary~\ref{Corollary 3} implies
	\[
	\Pr\left\{
	\hat C_n(m^*_{n,\mathrm{plug}}\mid\mathcal M_A,\boldsymbol\mu)>0
	\right\}\to1,
	\]
	so the finite-sample behavior of $\hat G_{n,A}^{\mathrm{plug}}$ discussed above does not affect its asymptotic validity.
	
	For finite-sample inference with unknown variance, suppose $p_n<n$, $\boldsymbol P_{p_n}\boldsymbol\mu=\boldsymbol\mu$, and $\operatorname{rank}(\boldsymbol P_{p_n})=p_n$. Define
		\begin{equation}\label{tail-full-variance}
			\hat\sigma_{\mathrm f}^2
			=
			\frac{\boldsymbol y^{\mathsf T}(\boldsymbol I_n-\boldsymbol P_{p_n})\boldsymbol y}
			{n-p_n}.
		\end{equation}
		For $p_n\geq2$ and $\hat\sigma_{\mathrm f}^2>0$, replace $F_m$ in \eqref{tail-lower}--\eqref{tail-upper} by
		\begin{equation}\label{tail-cdf-unknown}
			F_m^{\mathrm{plug}}(\lambda)
			=
			\Pr\left\{
			F_{p_n-m+1,n-p_n}(\lambda)
			\leq
			\frac{n\sum_{j=m}^{p_n}\hat\theta_j^2}
			{(p_n-m+1)\hat\sigma_{\mathrm f}^2}
			\right\},
		\end{equation}
		where $F_{d,r}(\lambda)$ denotes a noncentral $F$ variable with degrees of freedom $(d,r)$ and noncentrality parameter $\lambda$. Denote the resulting bounds by $\underline\Lambda_m^{\mathrm{plug}}(\alpha)$ and $\overline\Lambda_m^{\mathrm{plug}}(\alpha)$.
		
		The target $\boldsymbol\Lambda$ and the risk functions in \eqref{eq:ci-oracle-risks} remain unchanged. We substitute the plug-in tail bounds into \eqref{eq:ci-tail-region}--\eqref{Con 30} and denote the resulting interval by $\mathcal I_{n,\alpha}^{\mathrm{plug}}$. If the bounds are incompatible or $\hat\sigma_{\mathrm f}^2=0$, set $\mathcal I_{n,\alpha}^{\mathrm{plug}}=[0,(p_n-1)/(2p_n)]$; for $p_n=1$, set $\mathcal I_{n,\alpha}^{\mathrm{plug}}=\{0\}$.
		
		\begin{Cor}[Finite-sample coverage with unknown variance]\label{Corollary 4}
			Suppose $\boldsymbol\epsilon\sim \text{Normal}\,(\boldsymbol0,\sigma^2\boldsymbol I_n)$, $p_n<n$, $\boldsymbol P_{p_n}\boldsymbol\mu=\boldsymbol\mu$, and $\operatorname{rank}(\boldsymbol P_{p_n})=p_n$. For any prespecified candidate ordering, $\mathcal I_{n,\alpha}^{\mathrm{plug}}$ is nonempty, lies in $[0,(p_n-1)/(2p_n)]$, and satisfies
			\[
			\Pr\left\{G_{n,A}\in\mathcal I_{n,\alpha}^{\mathrm{plug}}\right\}
			\geq1-\alpha,
			\quad 0<\alpha<1.
			\]
	\end{Cor}
	Under the conditions of Corollary~\ref{Corollary 4}, the same interpretation and tests apply to $\mathcal I_{n,\alpha}^{\mathrm{plug}}$: exclusion of zero supports a positive oracle MA gain, and rejection when $\inf\mathcal I_{n,\alpha}^{\mathrm{plug}}>\delta$ has size at most $\alpha$.
	
	\begin{Ex}\label{Example 4}
		The full-model residual estimator in \eqref{tail-full-variance} is used in Corollary~\ref{Corollary 4} for finite-sample inference. We now demonstrate when it also satisfies \eqref{Con 13}, so that this estimator can be used in Corollary~\ref{Corollary 3}.
		
		Suppose the conditions of Theorem~\ref{Theorem 3} hold, $p_n<n$, $\boldsymbol P_{p_n}\boldsymbol\mu=\boldsymbol\mu$, and $\operatorname{rank}(\boldsymbol P_{p_n})=p_n$. Since $(\boldsymbol I_n-\boldsymbol P_{p_n})\boldsymbol\mu=0$, the Hanson--Wright inequality \citep{RudelsonVershynin2013} applied to the residual projection gives
		\[
		\hat\sigma_{\mathrm f}^2-\sigma^2
		=
		\frac{
			\boldsymbol\epsilon^{\T}(\boldsymbol I_n-\boldsymbol P_{p_n})\boldsymbol\epsilon
			-\sigma^2(n-p_n)
		}{n-p_n}
		=
		O_p\left(\frac{\sigma^2}{\sqrt{n-p_n}}\right).
		\]
		Consequently, 
		\[
		\frac{
			(p_n/n)|\hat\sigma_{\mathrm f}^2-\sigma^2|
		}{
			R_n(m_n^*\mid\mathcal M_A,\boldsymbol\mu)
		}
		=
		\frac{p_n}{s_{n,A}}
		\left|\frac{\hat\sigma_{\mathrm f}^2}{\sigma^2}-1\right|
		=
		O_p\left(\frac{p_n}{s_{n,A}\sqrt{n-p_n}}\right).
		\]
		This shows that the consistency of $\hat\sigma_{\mathrm{f}}^2$ alone need not suffice for \eqref{Con 13}.
		
		Under Assumption~\ref{Assump 2}, a sufficient additional condition is
		\begin{equation}\label{Con 17}
			\frac{p_n}{(n-p_n)\log(e p_n)}=O(1),
		\end{equation}
		because
		\[
		\frac{p_n}{s_{n,A}\sqrt{n-p_n}}
		=
		\left\{
		\frac{p_n}{(n-p_n)\log(e p_n)}
		\frac{p_n\log(e p_n)}{s_{n,A}^2}
		\right\}^{1/2}
		\to0.
		\]
		Thus, \eqref{Con 13} holds with $\hat\sigma_n^2=\hat\sigma_{\mathrm f}^2$, and Corollary~\ref{Corollary 3} yields $\hat G_{n,A}^{\mathrm{plug}}-G_{n,A}=o_p(1)$. 
		
		These additional rate conditions ensure consistency of the point diagnostic $\hat G_{n,A}^{\mathrm{plug}}$. Under Gaussian errors, the finite-sample coverage in Corollary~\ref{Corollary 4} does not require \eqref{Con 17} or asymptotic consistency of $\hat\sigma_{\mathrm f}^2$.
	\end{Ex}
	
	\section{Rashomon sets of candidate models}\label{sec:rashomon}
		
		Sections~\ref{Section 3} and~\ref{Section 4} compare the optimal risks of MS and MA. We now characterize the individual candidate models whose risks are close to the MS optimum. According to the definition of the Rashomon-set \citep{breiman2001statistical,FisherRudinDominici2019}, for $\varepsilon\geq0$, define
		\begin{equation}
			\label{eq:rashomon-definition}
			\mathcal R_{n,A}(\varepsilon)
			=
			\left\{
			m\in\{1,\ldots,p_n\}:
			R_n(m\mid\mathcal M_A,\boldsymbol\mu)
			\leq
			(1+\varepsilon)
			R_n(m_n^*\mid\mathcal M_A,\boldsymbol\mu)
			\right\}.
		\end{equation}
		The same definition with $\mathcal M_A$ replaced by $\mathcal M_N$ gives $\mathcal R_{n,N}(\varepsilon)$. Each element $m$ identifies one of the $p_n$ nested candidate models, and $\varepsilon$ specifies the permissible proportional increase over the optimal MS risk. The tolerance $\varepsilon$ may be chosen to reflect a practically acceptable relative increase in risk \citep{RudinEtAl2022}, and statistical approaches to tolerance selection based on model distinguishability are discussed by \citet{PaesCruzCalmonDiaz2023}. Throughout this section, we retain the model setup and Assumption~\ref{Assump 5}, with $\sigma^2>0$.
		
		\subsection{Complete ordering}
		\label{subsec:rashomon-ordered}
		
		Under Assumption~\ref{Assump 6}, the successive risk increments are nondecreasing along the nested candidate path. This property determines the interval structure of the Rashomon set and yields bounds on its cardinality.
		
		\begin{Pro}[Rashomon set under complete ordering]
			\label{prop:rashomon-ordered}
			Suppose Assumption~\ref{Assump 6} holds, and let $m_n^*$ be any oracle MS cutoff under $\mathcal M_N$. For $\varepsilon\geq0$, define
			\begin{align}
				m_{n,-}(\varepsilon)
				&=\min\left\{
				m\in\{1,\ldots,m_n^*\}:
				\sum_{j=m+1}^{m_n^*}
				\left(\theta_{(j)}^2-\frac{\sigma^2}{n}\right)
				\leq\varepsilon R_n(m_n^*\mid\mathcal M_N,\boldsymbol\mu)
				\right\},
				\label{eq:rashomon-left}\\
				m_{n,+}(\varepsilon)
				&=\max\left\{
				m\in\{m_n^*,\ldots,p_n\}:
				\sum_{j=m_n^*+1}^{m}
				\left(\frac{\sigma^2}{n}-\theta_{(j)}^2\right)
				\leq\varepsilon R_n(m_n^*\mid\mathcal M_N,\boldsymbol\mu)
				\right\}.
				\label{eq:rashomon-right}
			\end{align}
			Then
			\begin{equation*}
				\begin{aligned}
					\mathcal R_{n,N}(\varepsilon)
					&=\{m_{n,-}(\varepsilon),\ldots,m_{n,+}(\varepsilon)\},\\
					|\mathcal R_{n,N}(\varepsilon)|
					&=m_{n,+}(\varepsilon)-m_{n,-}(\varepsilon)+1.
				\end{aligned}
			\end{equation*}
			Moreover,
			\begin{equation}
				\label{eq:rashomon-count-upper}
				|\mathcal R_{n,N}(\varepsilon)|
				\leq
				\min\left\{
				p_n,\,
				\left\lfloor
				\frac{(1+\varepsilon)n}{\sigma^2}
				R_n(m_n^*\mid\mathcal M_N,\boldsymbol\mu)
				\right\rfloor
				\right\}.
			\end{equation}
			If $R_n(m_n^*\mid\mathcal M_N,\boldsymbol\mu)<p_n\sigma^2/n$, then
			\begin{equation}
				\label{eq:rashomon-count-lower}
				|\mathcal R_{n,N}(\varepsilon)|
				\geq
				1+\min\left\{
				p_n-m_n^*,\,
				\left\lfloor
				\frac{
					\varepsilon R_n(m_n^*\mid\mathcal M_N,\boldsymbol\mu)(p_n-m_n^*)
				}{
					p_n\sigma^2/n-R_n(m_n^*\mid\mathcal M_N,\boldsymbol\mu)
				}
				\right\rfloor
				\right\}.
			\end{equation}
			If $R_n(m_n^*\mid\mathcal M_N,\boldsymbol\mu)=p_n\sigma^2/n$, then every cutoff from $m_n^*$ to $p_n$ is optimal, and $|\mathcal R_{n,N}(\varepsilon)|\geq p_n-m_n^*+1$.
		\end{Pro}
		
		Under complete ordering, the risk increments $\sigma^2/n-\theta_{(j)}^2$ are nondecreasing in $j$. Hence, $\mathcal R_{n,N}(\varepsilon)$ is a single integer interval $\{m_{n,-}(\varepsilon),\ldots,m_{n,+}(\varepsilon)\}$ containing every oracle cutoff, and its cardinality is $m_{n,+}(\varepsilon)-m_{n,-}(\varepsilon)+1$. The ordering of the squared coefficients $\theta_{(j)}^2$ therefore ensures the interval structure, while their magnitudes determine its endpoints.
		
		Specifically, each deletion to the left of $m_n^*$ increases the risk by $\theta_{(j)}^2-\sigma^2/n$, whereas each addition to its right increases the risk by $\sigma^2/n-\theta_{(j)}^2$. The endpoints in \eqref{eq:rashomon-left}--\eqref{eq:rashomon-right} extend as far as these cumulative increases remain within the tolerance $\varepsilon R_n(m_n^*\mid\mathcal M_N,\boldsymbol\mu)$. For a fixed tolerance, smaller differences between the relevant squared coefficients $\theta_{(j)}^2$ and the noise threshold $\sigma^2/n$ permit more cutoffs on the corresponding side. Different cumulative increases on the two sides can produce an interval asymmetric about $m_n^*$. For fixed $n$, $\sigma^2$, and squared coefficients $\theta_{(j)}^2$, a larger $\varepsilon$ gives a weakly larger interval and cardinality.
		
		\subsection{Arbitrary ordering}
		\label{subsec:rashomon-unordered}
		
		Under $\mathcal M_A$, the risk increments need not be monotone, and the Rashomon set may contain gaps along the nested candidate path. The PAVA blocks from Section~\ref{Section 4} determine an enclosing interval, while the signal distribution within each block determines which cutoffs remain in the set. To describe this distinction, define
		\begin{equation}
			\label{eq:rashomon-block-discrepancy}
			H_n(1)=0,\quad
			H_n(m)
			=\sum_{j\in\mathcal J_{n,l},\,j\leq m}
			\left(\bar\theta_{n,l}^2-\theta_j^2\right),
			\quad m\in\mathcal J_{n,l}.
		\end{equation}
		Thus, $H_n(m)$ is the difference between the signal accumulated up to $m$ within its block $\mathcal J_{n,l}$ and the amount prescribed by the block average.
		
		\begin{Pro}[Rashomon set under arbitrary ordering]
			\label{prop:rashomon-unordered}
			Let $m_n^*$ be any oracle MS cutoff under $\mathcal M_A$, and let $\varepsilon\geq0$. Then $H_n(m)\geq0$ for every $m=1,\ldots,p_n$, with equality at $m=1$ and at the right endpoint of every PAVA block. For this ordering, let $m_{n,-}(\varepsilon)$ and $m_{n,+}(\varepsilon)$ be the smallest and largest cutoffs satisfying
			\begin{equation}
				\label{eq:rashomon-enclosing-interval}
				\frac{m\sigma^2}{n}
				+\sum_{l=1}^{L_n}
				\left|\mathcal J_{n,l}\cap\{m+1,\ldots,p_n\}\right|
				\bar\theta_{n,l}^2
				\leq
				(1+\varepsilon)R_n(m_n^*\mid\mathcal M_A,\boldsymbol\mu).
			\end{equation}
			The cutoffs satisfying \eqref{eq:rashomon-enclosing-interval} form a nonempty integer interval, and the Rashomon set is exactly
			\begin{equation}
				\label{eq:rashomon-unordered-set}
				\begin{aligned}
					\mathcal R_{n,A}(\varepsilon)
					&=\Biggl\{m\in\{m_{n,-}(\varepsilon),\ldots,m_{n,+}(\varepsilon)\}:
					H_n(m)+\frac{m\sigma^2}{n}\\
					&\quad+\sum_{l=1}^{L_n}
					\left|\mathcal J_{n,l}\cap\{m+1,\ldots,p_n\}\right|
					\bar\theta_{n,l}^2
					\leq(1+\varepsilon)R_n(m_n^*\mid\mathcal M_A,\boldsymbol\mu)
					\Biggr\}.
				\end{aligned}
			\end{equation}
			Its cardinality satisfies
			\begin{equation}
				\label{eq:rashomon-unordered-count}
				\begin{aligned}
					&\mathbf 1\{m_{n,-}(\varepsilon)=1\}+
					\sum_{l=1}^{L_n}
					\mathbf 1\left\{
					m_{n,-}(\varepsilon)
					\leq\max\mathcal J_{n,l}
					\leq m_{n,+}(\varepsilon)
					\right\}\\
					&\quad+
					\sum_{l=1}^{L_n}
					\sum_{j=1}^{|\mathcal J_{n,l}|-1}
					\mathbf 1\left\{
						R_n(\min\mathcal J_{n,l}-1\mid\mathcal M_A,\boldsymbol\mu)
						+\frac{j\sigma^2}{n}\leq
						(1+\varepsilon)R_n(m_n^*\mid\mathcal M_A,\boldsymbol\mu)
					\right\}\\
					&\leq
					|\mathcal R_{n,A}(\varepsilon)|
					\leq
					m_{n,+}(\varepsilon)-m_{n,-}(\varepsilon)+1,
				\end{aligned}
			\end{equation}
			where $\min\mathcal J_{n,l}$ represents the smallest index of elements in the block $\mathcal J_{n,l}$.
		\end{Pro}
		The PAVA blocks $\mathcal J_{n,l}$, their averages $\bar\theta_{n,l}^2$, and the risk tolerance $\varepsilon R_n(m_n^*\mid\mathcal M_A,\boldsymbol\mu)$ determine the enclosing interval $\{m_{n,-}(\varepsilon),\ldots,m_{n,+}(\varepsilon)\}$. Models outside this interval cannot belong to $\mathcal R_{n,A}(\varepsilon)$. For models within the interval, membership additionally depends on $H_n(m)$ through \eqref{eq:rashomon-unordered-set}: a sufficiently large $H_n(m)$ raises the risk above the allowed bound and excludes the model. Block endpoints satisfy $H_n(m)=0$, so those within the interval always belong to the set. Interior cutoffs may nevertheless be excluded, leaving gaps between included block endpoints. Thus, under arbitrary ordering, the Rashomon set need not be a single integer interval.
		
		The cardinality therefore depends on the enclosing interval's length and the number of models within it that are excluded because of $H_n(m)$. In \eqref{eq:rashomon-unordered-count}, the upper bound is the interval length, while the lower bound counts models whose inclusion is guaranteed by the PAVA blocks $\mathcal J_{n,l}$ and averages $\bar\theta_{n,l}^2$. Both bounds are attainable with these quantities fixed, so they give the smallest and largest possible cardinalities compatible with the same block information. The transformed coefficients $\theta_j^2$ determine $H_n(m)$ and thereby determine the exact set and its cardinality.
		
		\begin{Rem}[Rashomon cardinality and the oracle MA gain]\label{rem:rashomon-oracle-gain}
			Under the nested framework, the cardinality of the Rashomon set can provide a lower bound on $G_{n,A}$. Recall that $s_{n,A}=nR_n(m_n^*\mid\mathcal M_A,\boldsymbol\mu)/\sigma^2$. For any fixed ordering of $\{|\theta_j|\}_{j=1}^{p_n}$ and $\varepsilon\geq0$, we have
			\begin{equation}\label{eq:rashomon-gain-lower}
				G_{n,A}
				\geq
				\left[\frac{\max\mathcal R_{n,A}(\varepsilon)-\min\mathcal R_{n,A}(\varepsilon)}{2s_{n,A}}-\varepsilon\right]_+
				\geq
				\left[\frac{|\mathcal R_{n,A}(\varepsilon)|-1}{2s_{n,A}}-\varepsilon\right]_+,
			\end{equation}
			where $\min\mathcal R_{n,A}(\varepsilon)$ and $\max\mathcal R_{n,A}(\varepsilon)$ denote the indices of the smallest and largest models in $\mathcal R_{n,A}(\varepsilon)$, respectively, and $[x]_+=\max\{x,0\}$. In particular, for a fixed $\varepsilon$, the condition $\liminf_{n\to\infty}\{|\mathcal R_{n,A}(\varepsilon)|-1\}/(2s_{n,A})>\varepsilon$ implies $\liminf_{n\to\infty}G_{n,A}>0$. A zero lower bound does not imply that the oracle gain is negligible.
			
			To establish \eqref{eq:rashomon-gain-lower}, consider the average of models $\min\mathcal R_{n,A}(\varepsilon)$ and $\max\mathcal R_{n,A}(\varepsilon)$ with equal weights. The nested structure gives
			\[
			\begin{aligned}
				&\frac1n\E\left\|
				\frac{\hat{\boldsymbol\mu}_{\min\mathcal R_{n,A}(\varepsilon)\mid\mathcal M_A}+\hat{\boldsymbol\mu}_{\max\mathcal R_{n,A}(\varepsilon)\mid\mathcal M_A}}2
				-\boldsymbol\mu
				\right\|^2\\
				&=
				\frac{R_n(\min\mathcal R_{n,A}(\varepsilon)\mid\mathcal M_A,\boldsymbol\mu)+3R_n(\max\mathcal R_{n,A}(\varepsilon)\mid\mathcal M_A,\boldsymbol\mu)}4
				-\frac{\{\max\mathcal R_{n,A}(\varepsilon)-\min\mathcal R_{n,A}(\varepsilon)\}\sigma^2}{2n}\\
				&\leq
				(1+\varepsilon)R_n(m_n^*\mid\mathcal M_A,\boldsymbol\mu)
				-\frac{\{\max\mathcal R_{n,A}(\varepsilon)-\min\mathcal R_{n,A}(\varepsilon)\}\sigma^2}{2n}.
			\end{aligned}
			\]
			The oracle MA risk $R_n(\boldsymbol{w}^*\mid\mathcal M_A,\boldsymbol\mu)$ cannot exceed the risk of this risk. Together with $G_{n,A}\geq0$ and $\max\mathcal R_{n,A}(\varepsilon)-\min\mathcal R_{n,A}(\varepsilon)\geq|\mathcal R_{n,A}(\varepsilon)|-1$, this proves the stated bounds. Under complete ordering, the same argument and proof apply to $\mathcal M_N$.
			
			A plug-in version of the lower bound in \eqref{eq:rashomon-gain-lower} can be obtained from the projected Mallows criterion in \eqref{Con 12}. Its statistical properties require separate analysis. The purpose of this remark is to establish the theoretical relationship between $G_{n,A}$ and the Rashomon set cardinality $|\mathcal R_{n,A}(\varepsilon)|$.
		\end{Rem}
		
		\subsection{Non-nested models in an orthogonal basis}
		\label{subsec:rashomon-subsets}
		
		For the fixed orthonormal basis $\{\boldsymbol\phi_j\}_{j=1}^{p_n}$, let $S_1,\ldots,S_{2^{p_n}-1}$ denote all distinct nonempty subsets of $\{1,\ldots,p_n\}$, and define
		\[
		\mathcal M_{\mathrm{AS}}
		=
		\{S_m\subset \{1,\ldots,p_n\}:m=1,\ldots,2^{p_n}-1\}.
		\] 
		Here $m$ identifies a model, while $|S_m|$ is its number of components. The corresponding least squares estimator projects $\boldsymbol y$ onto $\operatorname{span}\{\boldsymbol\phi_j:j\in S_m\}$. Note that, after the decomposition in Section \ref{Section 2}, the subspaces of $\{\boldsymbol\phi_j\}_{j=1}^{p_n}$ need not coincide with those generated by subsets of the original design matrix.
		
		Let $m_n^*$ denote any oracle MS index under $\mathcal M_{\mathrm{AS}}$. For $\varepsilon\geq0$, define
		\begin{equation}
			\label{eq:rashomon-as-definition}
			\mathcal R_{n,\mathrm{AS}}(\varepsilon)
			=
			\left\{
			m\in\{1,\ldots,2^{p_n}-1\}:
			R_n(m\mid\mathcal M_{\mathrm{AS}},\boldsymbol\mu)
			\leq
			(1+\varepsilon)
			R_n(m_n^*\mid\mathcal M_{\mathrm{AS}},\boldsymbol\mu)
			\right\}.
		\end{equation}
		According to \cite{Peng2024}, for every model $m$,
		\begin{equation}
			\label{eq:rashomon-as-risk}
			R_n(m\mid\mathcal M_{\mathrm{AS}},\boldsymbol\mu)
			=
			\frac{|S_m|\sigma^2}{n}
			+
			\sum_{j\notin S_m}\theta_j^2.
		\end{equation}
		If $\max_j\theta_j^2>\sigma^2/n$, an oracle model is given by $S_{m_n^*}=\{j:\theta_j^2>\sigma^2/n\}$. Otherwise, any singleton corresponding to a largest $\theta_j^2$ is optimal. In both cases,
		\begin{equation}
			\label{eq:rashomon-as-oracle}
			R_n(m_n^*\mid\mathcal M_{\mathrm{AS}},\boldsymbol\mu)
			=
			\sum_{j=1}^{p_n}
			\min\left(\theta_j^2,\frac{\sigma^2}{n}\right)
			+
			\left(
			\frac{\sigma^2}{n}-\max_{1\leq j\leq p_n}\theta_j^2
			\right)_+,
		\end{equation}
		where $x_+=\max(x,0)$.

		\begin{Pro}[Rashomon set for non-nested models]
			\label{prop:rashomon-subsets}
			The Rashomon set is exactly
			\begin{equation}
				\label{eq:rashomon-as-set}
				\begin{aligned}
					\mathcal R_{n,\mathrm{AS}}(\varepsilon)
					=
					\Biggl\{
					m\in\{1,\ldots,2^{p_n}-1\}:\;
					&\sum_{j\in S_m}
					\left(\frac{\sigma^2}{n}-\theta_j^2\right)_+
					+
					\sum_{j\notin S_m}
					\left(\theta_j^2-\frac{\sigma^2}{n}\right)_+\\
					&\leq
					\varepsilon R_n(m_n^*\mid\mathcal M_{\mathrm{AS}},\boldsymbol\mu)
					+
					\left(\frac{\sigma^2}{n}-\max_j\theta_j^2\right)_+
					\Biggr\}.
				\end{aligned}
			\end{equation}
			At $\epsilon=0$, $\mathcal R_{n,\mathrm{AS}}(0)$ contains models with the lowest oracle MS risk, so
			\begin{equation}
				\label{eq:rashomon-as-oracle-count}
				|\mathcal R_{n,\mathrm{AS}}(0)|
				=
				\begin{cases}
					2^{|\{j:\theta_j^2=\sigma^2/n\}|},
					& \max_j\theta_j^2>\sigma^2/n,\\[3pt]
					2^{|\{j:\theta_j^2=\sigma^2/n\}|}-1,
					& \max_j\theta_j^2=\sigma^2/n,\\[3pt]
					|\{j:\theta_j^2=\max_k\theta_k^2\}|,
					& \max_j\theta_j^2<\sigma^2/n.
				\end{cases}
			\end{equation}
			For every $\varepsilon\geq0$,
			\begin{equation}
				\label{eq:rashomon-as-count-bounds}
				\begin{aligned}
					&\max\left\{
					|\mathcal R_{n,\mathrm{AS}}(0)|,\,
					\sum_{k=0}^{\min\left\{
						p_n-|S_{m_n^*}|,\,
						\left\lfloor
						\frac{n\varepsilon R_n(m_n^*\mid\mathcal M_{\mathrm{AS}},\boldsymbol\mu)}{\sigma^2}
						\right\rfloor
						\right\}}
					\binom{p_n-|S_{m_n^*}|}{k}
					\right\}\\
					&\quad\leq
					|\mathcal R_{n,\mathrm{AS}}(\varepsilon)|
					\leq
					\sum_{k=1}^{\min\left\{
						p_n,\,
						\left\lfloor
						\frac{n(1+\varepsilon)R_n(m_n^*\mid\mathcal M_{\mathrm{AS}},\boldsymbol\mu)}{\sigma^2}
						\right\rfloor
						\right\}}
					\binom{p_n}{k}.
				\end{aligned}
			\end{equation}
		\end{Pro}
		
		The left-hand side of \eqref{eq:rashomon-as-set} adds $\sigma^2/n-\theta_j^2$ for each included component with $\theta_j^2<\sigma^2/n$ and $\theta_j^2-\sigma^2/n$ for each omitted component with $\theta_j^2>\sigma^2/n$. Its minimum over all nonempty models $S_m$ is $(\sigma^2/n-\max_j\theta_j^2)_+$; when every $\theta_j^2<\sigma^2/n$, this minimum is attained by a singleton with the largest $\theta_j^2$. The difference between the left-hand side of \eqref{eq:rashomon-as-set} and its minimum is the excess risk of model $m$ over $m_n^*$. Thus, \eqref{eq:rashomon-as-set} includes exactly those models for which this difference does not exceed $\varepsilon R_n(m_n^*\mid\mathcal M_{\mathrm{AS}},\boldsymbol\mu)$.
		
		The intuition behind \eqref{eq:rashomon-as-count-bounds} is as follows. Each component added to $S_{m_n^*}$ increases the risk by at most $\sigma^2/n$. Hence, whenever $k\sigma^2/n\leq\varepsilon R_n(m_n^*\mid\mathcal M_{\mathrm{AS}},\boldsymbol\mu)$, all $\binom{p_n-|S_{m_n^*}|}{k}$ models obtained by adding $k$ components belong to $\mathcal R_{n,\mathrm{AS}}(\varepsilon)$. The binomial sum in \eqref{eq:rashomon-as-count-bounds} counts these models over all permitted values of $k$, while the maximum with $|\mathcal R_{n,\mathrm{AS}}(0)|$ accounts for the fact that every optimal model also belongs to the Rashomon set.
		
		Next, for the upper bound in \eqref{eq:rashomon-as-count-bounds}, every $m\in\mathcal R_{n,\mathrm{AS}}(\varepsilon)$ satisfies
		\[
		\frac{|S_m|\sigma^2}{n}
		\leq R_n(m\mid\mathcal M_{\mathrm{AS}},\boldsymbol\mu)
		\leq (1+\varepsilon)R_n(m_n^*\mid\mathcal M_{\mathrm{AS}},\boldsymbol\mu),
		\]
		where the first equation is because \eqref{eq:rashomon-as-risk}.
		Consequently, its size $|S_m|$ cannot exceed $\min\{p_n,\lfloor n(1+\varepsilon)R_n(m_n^*\mid\mathcal M_{\mathrm{AS}},\boldsymbol\mu)/\sigma^2\rfloor\}$. There are $\binom{p_n}{k}$ candidate models of size $k$, so the sum over all permitted sizes gives the upper bound in \eqref{eq:rashomon-as-count-bounds}.
		
		\subsection{Inference for the Rashomon set}
		\label{subsec:rashomon-inference}
		
		The Rashomon sets characterized above depend on the unknown squared coefficients $\theta_j^2$. We now construct inner and outer sets from the observed data, with the aim that, at a prescribed confidence level, the inner set contains only members of the oracle Rashomon set and the outer set contains all its members, simultaneously for all $\varepsilon\geq0$. Their cardinalities then provide lower and upper confidence bounds for the number of models satisfying the Rashomon risk constraint. The following framework applies to both the nested and non-nested candidate families considered above.
		
		Let $\mathcal M_K$ be a prespecified candidate family, with models indexed by $m=1,\ldots,|\mathcal M_K|$. The choices $K\in\{N,A,\mathrm{AS}\}$ correspond to the different model sets considered above. Let $m_n^*\mid \mathcal{M}_K$ be an oracle MS index under $\mathcal M_K$, and define $\mathcal R_{n,K}(\varepsilon)$ by the same risk inequality as in \eqref{eq:rashomon-definition}.
		
		The common requirement for inference is a pair of simultaneous bounds on
		\[
		R_n(m\mid\mathcal M_K,\boldsymbol\mu)
		-
		(1+\varepsilon)R_n(m_n^*\mid\mathcal M_K,\boldsymbol\mu).
		\]
		For data-based lower and upper bounds $\underline D_{n,\alpha,K}(m,\varepsilon)$ and $\overline D_{n,\alpha,K}(m,\varepsilon)$, define
		\begin{align}
			\hat{\mathcal R}^{\,\mathrm{in}}_{n,\alpha,K}(\varepsilon)
			&=
			\left\{
			m\in\{1,\ldots,|\mathcal M_K|\}:
			\overline D_{n,\alpha,K}(m,\varepsilon)\leq0
			\right\},
			\label{eq:rashomon-general-inner}\\
			\hat{\mathcal R}^{\,\mathrm{out}}_{n,\alpha,K}(\varepsilon)
			&=
			\left\{
			m\in\{1,\ldots,|\mathcal M_K|\}:
			\underline D_{n,\alpha,K}(m,\varepsilon)\leq0
			\right\}.
			\label{eq:rashomon-general-outer}
		\end{align}
		An upper bound below zero certifies the Rashomon risk inequality, while a lower bound above zero excludes the model.
		
		\begin{Th}[Simultaneous confidence bounds for Rashomon sets]
			\label{thm:rashomon-confidence}
			Fix $0<\alpha<1$ and a candidate family $\mathcal M_K$. Suppose that, with probability at least $1-\alpha$,
			\begin{equation}
				\label{eq:rashomon-difference-bounds}
				\begin{aligned}
					\underline D_{n,\alpha,K}(m,\varepsilon)
					\leq
					R_n(m\mid\mathcal M_K,\boldsymbol\mu)
					-
					(1+\varepsilon)R_n(m_n^*\mid\mathcal M_K,\boldsymbol\mu)\leq
					\overline D_{n,\alpha,K}(m,\varepsilon)
				\end{aligned}
			\end{equation}
			simultaneously for all $m$ and all $\varepsilon\geq0$. Then, with probability at least $1-\alpha$, simultaneously for all $\varepsilon\geq0$,
			\begin{align}
				\hat{\mathcal R}^{\,\mathrm{in}}_{n,\alpha,K}(\varepsilon)
				&\subseteq
				\mathcal R_{n,K}(\varepsilon)
				\subseteq
				\hat{\mathcal R}^{\,\mathrm{out}}_{n,\alpha,K}(\varepsilon),
				\label{eq:rashomon-coverage}\\
				\left|\hat{\mathcal R}^{\,\mathrm{in}}_{n,\alpha,K}(\varepsilon)\right|
				&\leq
				|\mathcal R_{n,K}(\varepsilon)|
				\leq
				\left|\hat{\mathcal R}^{\,\mathrm{out}}_{n,\alpha,K}(\varepsilon)\right|.
				\label{eq:rashomon-count-bounds}
			\end{align}
		\end{Th}
		
			For the nested family $\mathcal M_A$, retain the Gaussian assumptions and tail quantities of Section~\ref{Section 5.2}. For known $\sigma^2$ and $p_n\geq2$, let $\mathcal B_n\subseteq\{(a,b):1\leq a<b<p_n\}$ be a prespecified family of intervals. Let $\underline\Lambda_{a,b}(\alpha/2)$ and $\overline\Lambda_{a,b}(\alpha/2)$ be bounds for $\Lambda_{a+1}-\Lambda_{b+1}$ that hold simultaneously over $\mathcal B_n$ with probability at least $1-\alpha/2$. For $\mathcal B_n\neq\varnothing$, choose prespecified $\alpha_{ab}>0$ such that $\sum_{(a,b)\in\mathcal B_n}\alpha_{ab}\leq\alpha/2$. For each $(a,b)\in\mathcal B_n$, define
			\[
			F_{ab}(v)=\Pr\left\{\chi^2_{b-a}(v)\leq\frac{n}{\sigma^2}\sum_{j=a+1}^{b}\hat\theta_j^2\right\},
			\quad v\geq0,
			\]
			and set
			\[
			\begin{aligned}
				\underline\Lambda_{a,b}(\alpha/2)
				&=\inf\{v\geq0:F_{ab}(v)\leq1-\alpha_{ab}/2\},\\
				\overline\Lambda_{a,b}(\alpha/2)
				&=\inf\{v\geq0:F_{ab}(v)\leq\alpha_{ab}/2\}.
			\end{aligned}
			\]
			The noncentrality parameter is $\Lambda_{a+1}-\Lambda_{b+1}$, and the union bound gives the stated simultaneous coverage.
			
			We combine these bounds with the tail region in \eqref{eq:ci-tail-region}:
			\begin{equation}\label{eq:rashomon-local-region}
				\mathcal C^{\mathrm R}_{n,\alpha}
				=
				\left\{\boldsymbol\lambda\in\mathcal C_{n,\alpha/2}:
				\underline\Lambda_{a,b}(\alpha/2)
				\leq\lambda_{a+1}-\lambda_{b+1}
				\leq\overline\Lambda_{a,b}(\alpha/2),\;
				(a,b)\in\mathcal B_n
				\right\}.
			\end{equation}
			The union bound gives $\Pr\{\boldsymbol\Lambda\in\mathcal C^{\mathrm R}_{n,\alpha}\}\geq1-\alpha$. If $\mathcal B_n=\varnothing$, take $\mathcal C^{\mathrm R}_{n,\alpha}=\mathcal C_{n,\alpha}$.
			
			For $\mathcal C^{\mathrm R}_{n,\alpha}\neq\varnothing$, define
			\begin{align}
				\underline D_{n,\alpha,A}(m,\varepsilon)
				&=\frac{\sigma^2}{n}
				\min_{\boldsymbol\lambda\in\mathcal C^{\mathrm R}_{n,\alpha}}
				\left\{m+\lambda_{m+1}-(1+\varepsilon)R_{n,\mathrm{MS}}(\boldsymbol\lambda)\right\},
				\label{eq:rashomon-nested-difference-lower}\\
				\overline D_{n,\alpha,A}(m,\varepsilon)
				&=\frac{\sigma^2}{n}
				\max_{\boldsymbol\lambda\in\mathcal C^{\mathrm R}_{n,\alpha}}
				\left\{m+\lambda_{m+1}-(1+\varepsilon)R_{n,\mathrm{MS}}(\boldsymbol\lambda)\right\}.
				\label{eq:rashomon-nested-difference-upper}
			\end{align}
			Both risks are evaluated at the same $\boldsymbol\lambda$. These bounds satisfy \eqref{eq:rashomon-difference-bounds} simultaneously for all $m$ and $\varepsilon\geq0$, so the sets in \eqref{eq:rashomon-general-inner}--\eqref{eq:rashomon-general-outer} satisfy Theorem~\ref{thm:rashomon-confidence}. If the region is empty, take $\underline D_{n,\alpha,A}=-\infty$ and $\overline D_{n,\alpha,A}=+\infty$, giving an empty inner set and the full candidate family as the outer set. For $p_n=1$, both sets equal $\{1\}$. The same construction applies to $\mathcal M_N$ under a prespecified ordering satisfying Assumption~\ref{Assump 6}.
			
			For unknown $\sigma^2$, under the conditions of Corollary~\ref{Corollary 4}, construct the tail bounds using \eqref{tail-cdf-unknown} and replace $F_{ab}(v)$ in the local-bound construction above by
			\[
			\Pr\left\{
			F_{b-a,n-p_n}(v)
			\leq
			\frac{n\sum_{j=a+1}^{b}\hat\theta_j^2}{(b-a)\hat\sigma_{\mathrm f}^2}
			\right\}.
			\]
			The error allocations $\alpha_{ab}$ remain unchanged. The target $\boldsymbol\Lambda$ remains unchanged. The signs of the same two extrema determine the inner and outer sets, since $\sigma^2/n>0$, and the simultaneous coverage guarantee remains valid. For $p_n\geq2$, use an empty inner set and the full outer set if $\hat\sigma_{\mathrm f}^2\leq0$ or the resulting region is empty.
		
		For the non-nested family $\mathcal M_{\mathrm{AS}}$, assume $\boldsymbol\epsilon\sim \text{Normal}\,(\boldsymbol0,\sigma^2\boldsymbol I_n)$ with known $\sigma^2>0$. We construct simultaneous confidence intervals for $\{\theta_j^2\}_{j=1}^{p_n}$. Let $F_{1,v}(x)$ denote the distribution function of $\chi_1^2(v)$. Since $n\hat\theta_j^2/\sigma^2\sim\chi_1^2(n\theta_j^2/\sigma^2)$, define
		\begin{align}
			\ell_j
			&=
			\frac{\sigma^2}{n}
			\inf\left\{
			v\geq0:
			F_{1,v}\left(\frac{n\hat\theta_j^2}{\sigma^2}\right)
			\leq1-\frac{\alpha}{2p_n}
			\right\},
			\label{eq:rashomon-as-ci-lower}\\
			u_j
			&=
			\frac{\sigma^2}{n}
			\inf\left\{
			v\geq0:
			F_{1,v}\left(\frac{n\hat\theta_j^2}{\sigma^2}\right)
			\leq\frac{\alpha}{2p_n}
			\right\},
			\label{eq:rashomon-as-ci-upper}
		\end{align}
		for $j=1,\ldots,p_n$.
		Then $0\leq\ell_j\leq u_j$ and it can be shown that
		\begin{equation}
			\label{eq:rashomon-as-ci-coverage}
			\Pr\{\ell_j\leq\theta_j^2\leq u_j,\ j=1,\ldots,p_n\}
			\geq1-\alpha.
		\end{equation}
		Let $\mathcal C^{\mathrm{AS}}_{n,\alpha}=\prod_{j=1}^{p_n}[\ell_j,u_j]$. For $\boldsymbol t=(t_1,\ldots,t_{p_n})\in\mathcal C^{\mathrm{AS}}_{n,\alpha}$, define
		\begin{equation}
			\label{eq:rashomon-as-difference}
			\begin{aligned}
				D_{n,\varepsilon}(m;\boldsymbol t)
				=
				\frac{|S_m|\sigma^2}{n}
				+
				\sum_{j\notin S_m}t_j-
				(1+\varepsilon)
				\left\{
				\sum_{j=1}^{p_n}
				\min\left(t_j,\frac{\sigma^2}{n}\right)
				+
				\left(\frac{\sigma^2}{n}-\max_jt_j\right)_+
				\right\}.
			\end{aligned}
		\end{equation}
		At $\boldsymbol t=(\theta_1^2,\ldots,\theta_{p_n}^2)$, this is the risk difference in \eqref{eq:rashomon-difference-bounds}. Therefore, valid bounds for \eqref{eq:rashomon-as-difference} are
		\begin{align}
			\underline D_{n,\alpha,\mathrm{AS}}(m,\varepsilon)
			&=
			\inf_{\boldsymbol t\in\mathcal C^{\mathrm{AS}}_{n,\alpha}}
			D_{n,\varepsilon}(m;\boldsymbol t),
			\label{eq:rashomon-as-difference-lower}\\
			\overline D_{n,\alpha,\mathrm{AS}}(m,\varepsilon)
			&=
			\sup_{\boldsymbol t\in\mathcal C^{\mathrm{AS}}_{n,\alpha}}
			D_{n,\varepsilon}(m;\boldsymbol t).
			\label{eq:rashomon-as-difference-upper}
		\end{align}
		Together with \eqref{eq:rashomon-general-inner}--\eqref{eq:rashomon-general-outer}, these equations define the non-nested inner and outer sets $\hat{\mathcal R}^{\,\mathrm{in}}_{n,\alpha,\mathrm{AS}}(\varepsilon)$ and $\hat{\mathcal R}^{\,\mathrm{out}}_{n,\alpha,\mathrm{AS}}(\varepsilon)$. The inner set contains models satisfying the risk inequality for every $\boldsymbol t\in\mathcal C^{\mathrm{AS}}_{n,\alpha}$; the outer set contains models satisfying it for at least one such $\boldsymbol t$. 
		
		If all $t_j<\sigma^2/n$, the term $\sum_{j=1}^{p_n}\min(t_j,\sigma^2/n)$ equals the risk $\sum_{j=1}^{p_n}t_j$ of the empty model. Since every model in $\mathcal M_{\mathrm{AS}}$ must contain at least one component, the minimum risk is instead $\sum_{j=1}^{p_n}t_j+\sigma^2/n-\max_jt_j$, attained by a singleton with the largest $t_j$. This gives the correction $(\sigma^2/n-\max_jt_j)_+$ in \eqref{eq:rashomon-as-difference}, which is zero when $\max_jt_j\geq\sigma^2/n$.
		
		Both constructions obtain the lower and upper bounds by optimizing the risk difference over a common confidence region. Their validity follows from simultaneous coverage of the true parameter vector and Theorem~\ref{thm:rashomon-confidence}.
		
		\begin{Rem}[Precision of the inner and outer sets]
			\label{rem:rashomon-precision}
			To assess the precision of the inner and outer sets, define
			\[
			\Delta_{n,\alpha,K}(\varepsilon)
			=
			\max_{1\leq m\leq|\mathcal M_K|}
			\left\{
			\overline D_{n,\alpha,K}(m,\varepsilon)
			-
			\underline D_{n,\alpha,K}(m,\varepsilon)
			\right\}.
			\]
			On the event in \eqref{eq:rashomon-difference-bounds}, simultaneously for all $\varepsilon\geq0$,
			\begin{equation}
				\label{eq:rashomon-general-undetermined}
				\begin{aligned}
					&\hat{\mathcal R}^{\,\mathrm{out}}_{n,\alpha,K}(\varepsilon)
					\backslash
					\hat{\mathcal R}^{\,\mathrm{in}}_{n,\alpha,K}(\varepsilon)\\
					\subseteq&
					\left\{
					m\in\{1,\ldots,|\mathcal M_K|\}:
					\begin{aligned}
						-\Delta_{n,\alpha,K}(\varepsilon)
						&<
						R_n(m\mid\mathcal M_K,\boldsymbol\mu)\\
						&\quad-(1+\varepsilon)
						R_n(m_n^*\mid\mathcal M_K,\boldsymbol\mu)\leq\Delta_{n,\alpha,K}(\varepsilon)
					\end{aligned}
					\right\}.
				\end{aligned}
			\end{equation}
			Consequently, the difference between $|\hat{\mathcal R}^{\,\mathrm{out}}_{n,\alpha,K}(\varepsilon)|$ and $|\hat{\mathcal R}^{\,\mathrm{in}}_{n,\alpha,K}(\varepsilon)|$ is at most the number of models in the set on the right-hand side of \eqref{eq:rashomon-general-undetermined}. For the non-nested construction,
			\begin{equation}
				\label{eq:rashomon-as-width-bound}
				\Delta_{n,\alpha,\mathrm{AS}}(\varepsilon)
				\leq
				(1+\varepsilon)\sum_{j=1}^{p_n}(u_j-\ell_j).
			\end{equation}
			These bounds relate the difference between $|\hat{\mathcal R}^{\,\mathrm{out}}_{n,\alpha,\mathrm{AS}}(\varepsilon)|$ and $|\hat{\mathcal R}^{\,\mathrm{in}}_{n,\alpha,\mathrm{AS}}(\varepsilon)|$ to the interval widths $u_j-\ell_j$ and the number of models whose risks are close to the allowed risk threshold in \eqref{eq:rashomon-general-undetermined}. 
		\end{Rem}
		
		For fixed $\alpha$ and a component satisfying $\theta_j^2=O(\sigma^2/n)$, the intervals in \eqref{eq:rashomon-as-ci-lower}--\eqref{eq:rashomon-as-ci-upper} satisfy
		\[
		u_j-\ell_j
		=
		O_p\left\{
		\frac{\sigma^2}{n}\log\left(\frac{p_n}{\alpha}\right)
		\right\}.
		\]
		When $\theta_j^2=o(\sigma^2/n)$, we have $n\hat\theta_j^2/\sigma^2\Rightarrow\chi_1^2$. The inversion in \eqref{eq:rashomon-as-ci-lower}--\eqref{eq:rashomon-as-ci-upper} yields a width at least a fixed positive multiple of $\sigma^2/n$ with probability bounded away from zero. Thus, $u_j-\ell_j$ is not $o_p(\sigma^2/n)$, and the interval need not shrink at the scale of $\theta_j^2$.
		If the intervals $[\ell_j,u_j]$ are enlarged, the region $\mathcal C^{\mathrm{AS}}_{n,\alpha}$ expands, which can make the inner set smaller and the outer set larger. Recall that, for $\boldsymbol t=(\theta_1^2,\ldots,\theta_{p_n}^2)$, the inequality $D_{n,\varepsilon}(m;\boldsymbol t)\leq0$ means that
		\[
		R_n(m\mid\mathcal M_{\mathrm{AS}},\boldsymbol\mu)\leq (1+\varepsilon)R_n(m_n^*\mid\mathcal M_{\mathrm{AS}},\boldsymbol\mu)
		\]
		The inner set requires this inequality to hold for every $\boldsymbol{t}\in\mathcal C^{\mathrm{AS}}_{n,\alpha}$, whereas the outer set requires it to hold for at least one such $\boldsymbol t$. This can produce a small lower bound and a large upper bound for $|\mathcal R_{n,\mathrm{AS}}(\varepsilon)|$, reducing the precision of the inference despite the valid coverage guarantee.
		
	\section{Simulation studies}\label{sec:simulations}
	
	We use the orthogonal Gaussian model
	\begin{equation}\label{eq:sim-model}
		\boldsymbol y=\boldsymbol\mu+\boldsymbol\epsilon,\quad \boldsymbol\mu=\sqrt n\sum_{j=1}^{p_n}\theta_j\boldsymbol\phi_j,\quad \boldsymbol\epsilon\sim \text{Normal}\,(\boldsymbol0,\sigma^2\boldsymbol I_n),
	\end{equation}
	where $\boldsymbol\phi_j^{\mathsf T}\boldsymbol\phi_k=\mathbf1\{j=k\}$, $p_n=\lfloor n^{2/3}\rfloor$, and $\sigma^2=1$. Candidate model $m$ includes the first $m$ components, $m=1,\ldots,p_n$. The coefficients satisfy $\theta_j\geq0$ and Assumption~\ref{Assump 1}.
	
	Let $\mathcal N_0=\{200,500,1000,2000,5000\}$. Experiments~1 and~2 calculate $(G_{n,N},f_n(2))$ and $(G_{n,A},\tilde f_n(2))$ under ordered and arbitrarily ordered coefficients, respectively, at
	\[
	n\in\mathcal N_0\cup
	\left\{
	\left\lfloor 200(5000)^{r/54}+\frac12\right\rfloor:
	r=0,\ldots,54
	\right\}.
	\]
	This grid contains 59 distinct sample sizes between $200$ and $10^6$, combining logarithmic spacing with the sample sizes in $\mathcal N_0$. All oracle quantities are evaluated from the known coefficients and variance, without random replications.
	
	Experiments~3 and~4 reuse the five settings of Experiment~2 at
	\[
	n\in\mathcal N_0\cup\{10000,20000,100000\}.
	\]
	The three larger sample sizes extend the assessment of estimation accuracy and interval precision. Experiment~3 examines $|\hat G_{n,A}-G_{n,A}|$ and $|\hat G_{n,A}^{\mathrm{plug}}-G_{n,A}|$. Experiment~4 evaluates the practical coverage and mean relative length of $\mathcal I_{n,0.05}^{\mathrm{tail}}$ and $\mathcal I_{n,0.05}^{\mathrm{plug}}$. The two implementations use the known variance $\sigma^2$ and the full-model estimate $\hat\sigma_{\mathrm f}^2$, respectively.
	
	The supplementary comparisons use the settings of Experiments~1 and~2 at $n\in\mathcal N_0$. They compare Mallows $C_p$ selection and MMA through the mean loss $n^{-1}\|\hat{\boldsymbol\mu}-\boldsymbol\mu\|^2$, to assess whether the oracle gain patterns also appear for fitted procedures. Experiments~3 and~4 use 500 independent Gaussian replications for each setting and sample size. All procedures compared within a replication use the same simulated data.
	
	\subsection{Experiment 1: Ordered signals}\label{sec:sim-ordered}
	
	This experiment examines the critical-risk characterization in Theorem~\ref{Theorem 1} under six non-increasing coefficient sequences. The first four settings use the following decay patterns: polynomial decay, $\theta_j=j^{-\alpha_1}$ with $\alpha_1=1$ or $1.25$; exponential decay, $\theta_j=\exp(-j^{\alpha_2})$ with $\alpha_2=1$; and mixed decay, $\theta_j=j^{-1}$ for $j\leq k$ and $\theta_j=\exp(-j)$ thereafter, where $k=\lfloor\sqrt n\rfloor$. Two additional settings use
	\begin{equation}\label{eq:sim-ordered-critical}
		\theta_j^2=
		\begin{cases}
			\sigma^2/\sqrt n,&1\leq j\leq k,\\
			\sigma^2/n,&k<j\leq k+q,\\
			0,&k+q<j\leq p_n,
		\end{cases}
	\end{equation}
	with $q=k$ or $q=\lfloor\sqrt k\rfloor$, to separate the number of critical components from their contribution to the oracle MS risk.
	
	\begin{figure}[t]
		\centering
		\includegraphics[width=\textwidth]{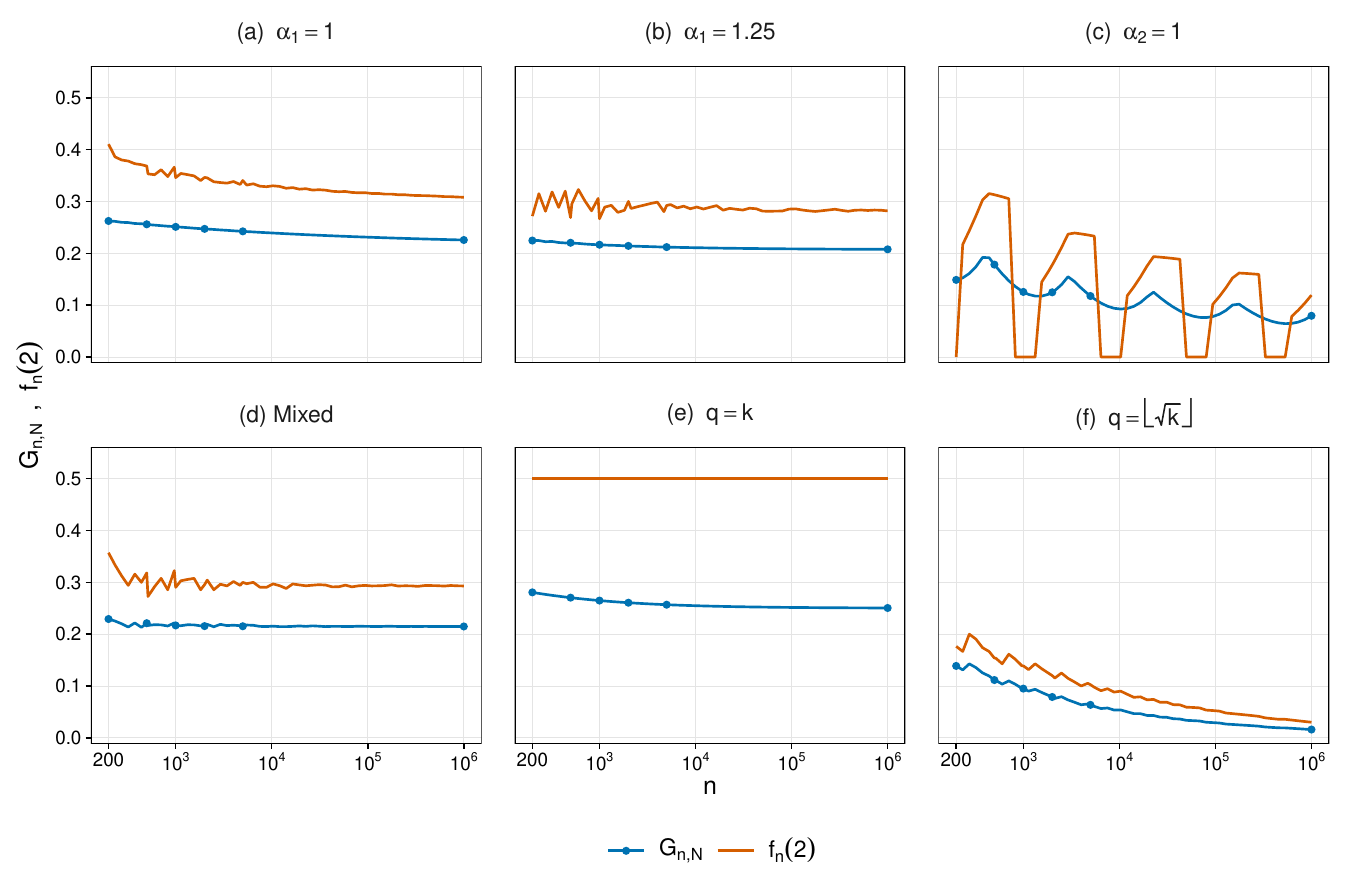}
		\caption{Exact oracle gain $G_{n,N}$ (blue) and critical-risk proportion $f_n(2)$ (orange) in Experiment~1. Panels (a)--(d) use polynomial decay with $\alpha_1=1$ and $1.25$, exponential decay, and mixed decay. Panels (e)--(f) use \eqref{eq:sim-ordered-critical} with $q=k$ and $q=\lfloor\sqrt k\rfloor$. Both curves connect exact calculations at 59 sample sizes between $200$ and $10^6$ on a logarithmic horizontal axis, without smoothing. Blue circles mark only $n=200,500,1000,2000,5000$ and $10^6$; orange curves have no markers. The curves contain no Monte Carlo error.}
		\label{fig:sim-exp1-oracle}
	\end{figure}
	
	Figure~\ref{fig:sim-exp1-oracle}(a), (b), and (d) display sustained gains together with critical-risk proportions that remain positive as $n$ increases. Under polynomial decay, the components $d_j$'s near the threshold $\sigma^2/n$ contribute a non-vanishing fraction of the oracle MS risk, in accordance with Theorem~\ref{Theorem 1}(i). Panel (d) shows that this behavior persists under mixed decay: the polynomial segment retains a substantial critical-risk contribution even though the remaining coefficients decay exponentially. These patterns agree with Theorem~\ref{Theorem 1}(i), which identifies the risk contribution near the noise level as the relevant feature. The small fluctuations in $f_n(2)$ occur as indices enter or leave $B_n(2)$. In the mixed setting, each unit increase in $k=\lfloor\sqrt n\rfloor$ also changes the coefficient at the new cutoff from $e^{-k}$ to $k^{-1}$, which can produce small jumps in both $f_n(2)$ and $G_{n,N}$.
	
	Panel (c) exhibits obvious oscillations within an overall declining pattern. The orange curve repeatedly rises and returns to zero because successive coefficients enter and leave the critical band. Specifically, under $\theta_j=\exp(-j)$ and $\sigma^2=1$, a component belongs to $B_n(2)$ when $e^{2j}/2\leq n\leq2e^{2j}$. These intervals are separated by gaps, during which the displayed band is empty. The blue curve also fluctuates as successive components pass through the noise threshold, where shrinkage offers the greatest relative benefit for an individual component. However, $G_{n,N}$ includes contributions from all optional components and can remain positive when $f_n(2)=0$. The connecting line segments provide visual guides between calculated values and do not imply continuous transitions across changes in band membership.
	
	The declining scale of the curves in panel (c) is consistent with Theorem~\ref{Theorem 1}(ii). For every fixed $c>1$, the exponential sequence has only a bounded number of components in $B_n(c)$, each contributing at most $\sigma^2/n$, while the oracle MS risk is of order $(\log n)/n$. Consequently, $f_n(c)\to0$ for every fixed $c>1$, which satisfies the equivalent criterion in Remark~\ref{Remark 1}. This theoretical argument establishes the vanishing gain beyond the displayed choice $c=2$. 
	
	Panels (e)--(f) isolate the distinction between the number of critical components and their risk share. Panel (e) shows a constant critical-risk proportion and a gain that approaches a positive level, whereas both quantities decline in panel (f). In both settings, $q$ diverges, but $f_n(c)=q/(k+q)$ eventually holds for every fixed $c>1$. Thus, $q=k$ gives a critical-risk proportion of $1/2$ and satisfies Theorem~\ref{Theorem 1}(i), while $q=\lfloor\sqrt k\rfloor$ gives a vanishing proportion and satisfies part~(ii). The exact gain formula yields limits of $1/4$ and zero, respectively, in agreement with the plotted trends. The small jumps in panel (f) follow from integer changes in $k$ and $q$. Although $q\to\infty$ in both settings, the choice $q=\lfloor\sqrt k\rfloor$ gives $f_n(2)=q/(k+q)\to0$ and hence $G_{n,N}\to0$. Thus, the divergence of $q$ alone does not ensure a non-vanishing gain.

	\subsection{Experiment 2: Arbitrary fixed orderings}\label{sec:sim-arbitrary}
	
	We first consider
	\begin{equation}\label{eq:sim-smooth}
		\theta_1^2=\frac{4\sigma^2}{n},\quad \theta_j^2=\frac{\sigma^2}{n}\left[0.5+3.5\left\{1-\frac{j-2}{p_n-2}\right\}^2\right],\quad j=2,\ldots,p_n.
	\end{equation}
	For $\{|\theta_j|\}_{j=2}^{p_n}$, we compare the descending order, permutations within consecutive groups of length $\max\{2,\lfloor0.1(p_n-1)\rfloor\}$, and a global permutation across all $p_n-1$ positions. The final shorter group is also permuted in the second setting. Each permutation is fixed for a given $n$ before data generation. The three orders leave $\theta_1^2$ and $\sum_{j=1}^{p_n}\theta_j^2$ unchanged.
	
\begin{figure}[t]
		\centering
		\includegraphics[width=\textwidth]{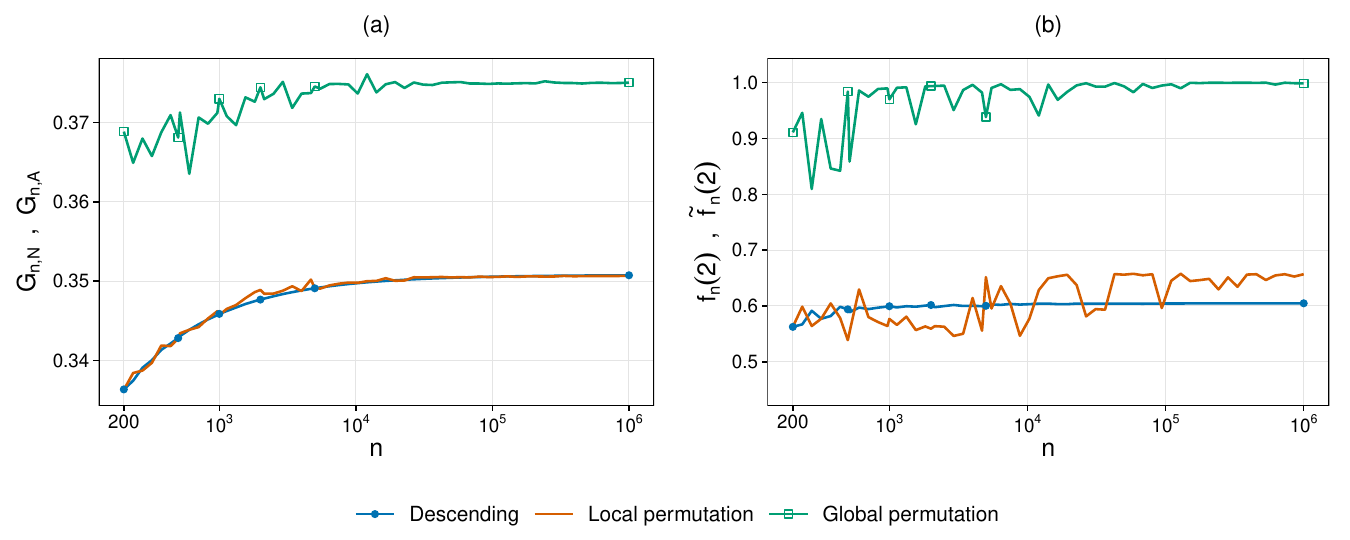}
		\caption{Comparison of the three orders of \eqref{eq:sim-smooth}. The descending order uses $(G_{n,N},f_n(2))$, and the permuted orders use $(G_{n,A},\tilde f_n(2))$. Lines connect all calculated values, and markers identify $n\in\mathcal N_0\cup\{10^6\}$. The horizontal axis is logarithmic.}
		\label{fig:sim-exp2-order}
	\end{figure}

	Figure~\ref{fig:sim-exp2-order} shows nearly coincident gains under descending and local orders, while global permutation gives a larger gain and a larger $\tilde f_n(2)$. Under global permutation, PAVA pools coefficients of different magnitudes into blocks, with a larger fraction of the oracle MS risk contributed by blocks satisfying $\sigma^2/(2n)\leq\bar\theta_{n,l}^2\leq2\sigma^2/n$. The difference between the descending and locally permuted risk proportions remains visible because $\tilde f_n(2)$ changes discontinuously when a block mean crosses a band boundary. A larger relative gain need not imply a lower oracle MA risk, because the gain is measured relative to the oracle MS risk under the same ordering. Proposition~\ref{prop:ordering-risk} shows that the descending order minimizes both oracle risks.
	
\begin{figure}[t]
		\centering
		\includegraphics[width=\textwidth]{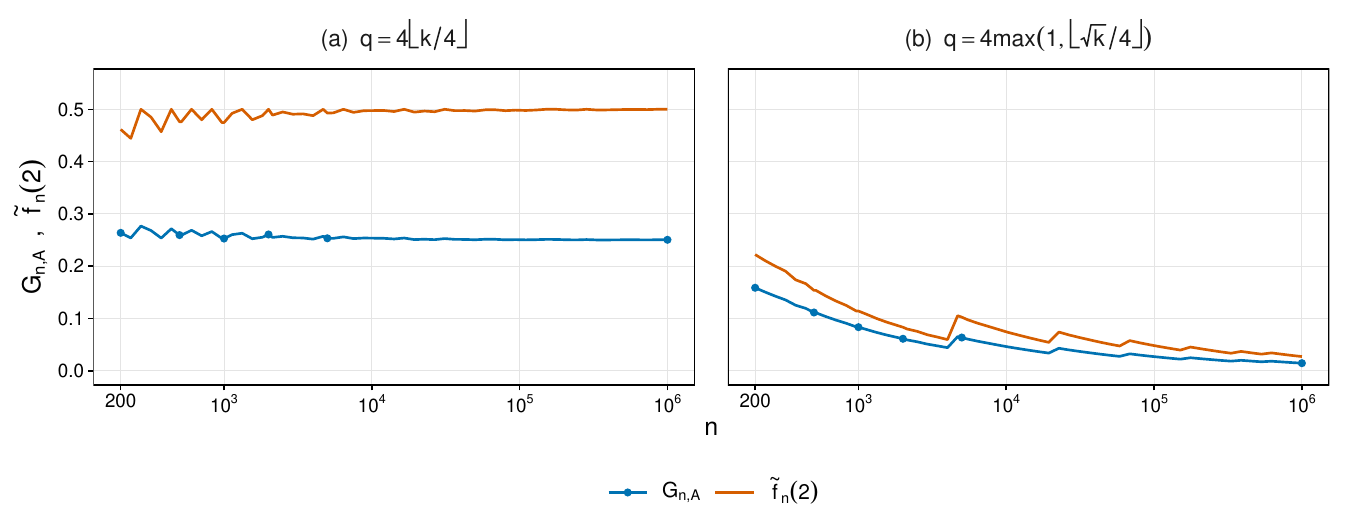}
		\caption{Exact $G_{n,A}$ (blue) and $\tilde f_n(2)$ (orange) for $q=4\lfloor k/4\rfloor$ and $q=4\max\{1,\lfloor\sqrt k/4\rfloor\}$. Lines connect all calculated values, and blue circles mark $n\in\mathcal N_0\cup\{10^6\}$. The horizontal axis is logarithmic.}
		\label{fig:sim-exp2-block}
	\end{figure}

	The remaining two settings use $k=\lfloor\sqrt n\rfloor$ and
	\[
	\theta_j^2=
	\begin{cases}
		\sigma^2/\sqrt n,&1\leq j\leq k,\\
		4\sigma^2/n,&k<j\leq k+q,\quad j-k\equiv0\pmod4,\\
		0,&\text{otherwise},
	\end{cases}
	\]
	with $q=4\lfloor k/4\rfloor$ or $q=4\max\{1,\lfloor\sqrt k/4\rfloor\}$. Thus, the segment $j=k+1,\ldots,k+q$ repeats $(0,0,0,4)\sigma^2/n$. PAVA combines this segment into one maximal block of length $q$ and mean $\sigma^2/n$. For sufficiently large $n$,
	\[
	\tilde f_n(2)=\tilde f_n(n^{1/4})=\frac{q}{k+q}.
	\]
	The first choice has $q/k\to1$ and satisfies Theorem~\ref{Theorem 2}(i). The second has $q/k\to0$ and satisfies part~(ii) with $c_n=n^{1/4}$. Together with the exact gain
	\[
	G_{n,A}=\frac{(k-1)/(1+\sqrt n)+q/2}{k+q},
	\]
	these relations give
	\[
	(\tilde f_n(2),G_{n,A})\to(1/2,1/4)
	\quad\text{or}\quad
	(0,0),
	\]
	respectively, in agreement with Figure~\ref{fig:sim-exp2-block}. Both settings contain one block with mean $\sigma^2/n$, but its contribution $q/(k+q)$ determines whether the gain persists. The small jumps follow from integer changes in $k$ and $q$.
	
	\subsection{Experiment 3: Accuracy of the practical gain diagnostic}\label{sec:sim-diagnostic}
	
	We compare $\hat G_{n,A}$ and $\hat G_{n,A}^{\mathrm{plug}}$ under the five settings of Experiment~2: the descending, locally permuted, and globally permuted orders of \eqref{eq:sim-smooth}, and the two block settings. For each setting and sample size, the coefficients and their ordering remain fixed across replications. The reference value $G_{n,A}$ is computed from the true coefficients and variance, with $G_{n,A}=G_{n,N}$ for the descending order. The two estimators use $\sigma^2$ and $\hat\sigma_{\mathrm f}^2$, respectively, and are evaluated on the same simulated observations. Each replication recomputes the minima defining the estimators, and values outside $[0,(p_n-1)/(2p_n)]$ are retained.

\begin{figure}[t]
		\centering
		\includegraphics[width=\textwidth]{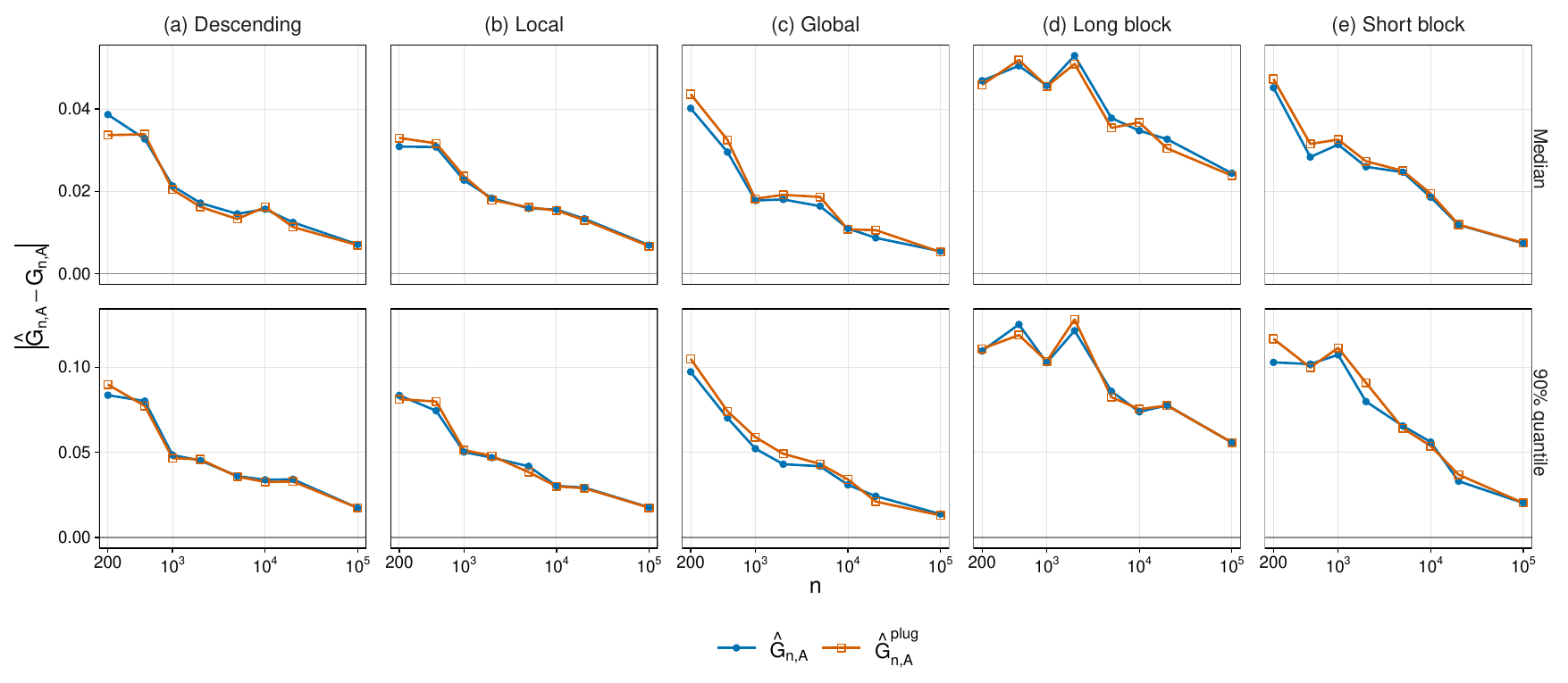}
		\caption{Median (top row) and 90th percentile (bottom row) of $|\hat G_{n,A}-G_{n,A}|$ and $|\hat G_{n,A}^{\mathrm{plug}}-G_{n,A}|$ under the five settings of Experiment~2. Blue circles and orange open squares correspond to $\hat G_{n,A}$ and $\hat G_{n,A}^{\mathrm{plug}}$, respectively. The horizontal axis is logarithmic.}
		\label{fig:exp3-error}
	\end{figure}

	Figure~\ref{fig:exp3-error} shows that the median and 90th percentile of both absolute errors generally decrease with $n$, consistently with the theoretical convergence results. The curves for $\hat G_{n,A}$ and $\hat G_{n,A}^{\mathrm{plug}}$ are close across most sample sizes, indicating similar accuracy under known and estimated variance.
	
	The setting $q=4\lfloor k/4\rfloor$ generally produces larger errors, with a visible difference from the other settings at larger $n$. This pattern also occurs for $\hat G_{n,A}$, so it cannot be attributed solely to variance estimation. The consistency results therefore coexist with differences in finite-sample accuracy across the coefficient sequences.
	
	\subsection{Experiment 4: Confidence intervals}\label{Section 6.4}
	
\begin{figure}[t]
		\centering
		\includegraphics[width=\textwidth]{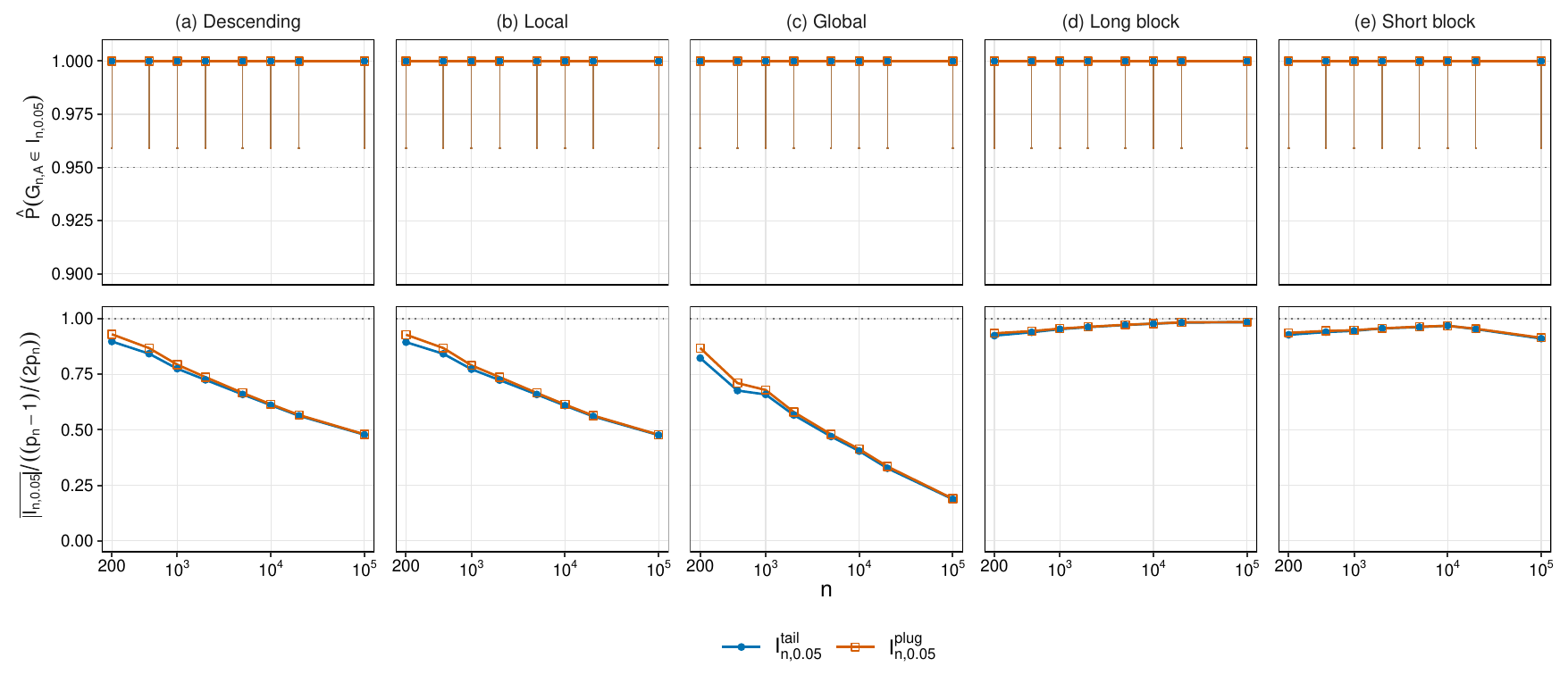}
		\caption{Coverage (top row) and mean relative length (bottom row) of $\mathcal I_{n,0.05}^{\mathrm{tail}}$ (blue circles) and $\mathcal I_{n,0.05}^{\mathrm{plug}}$ (orange open squares) under the five settings of Experiment~2. Relative length is interval length divided by $(p_n-1)/(2p_n)$. Error bars are 95\% Wilson intervals for Monte Carlo uncertainty based on 500 replications. Dotted lines indicate nominal coverage and the full parameter-range length. Coverage curves and error bars coincide for the two implementations. The horizontal axis is logarithmic.}
		\label{fig:exp4-coverage-length}
	\end{figure}

	We evaluate the coverage and precision of the nominal $95\%$ confidence intervals $\mathcal I_{n,0.05}^{\mathrm{tail}}$ and $\mathcal I_{n,0.05}^{\mathrm{plug}}$ under the same five settings and sample sizes as Experiment~3. The two intervals use $\sigma^2$ and $\hat\sigma_{\mathrm f}^2$, with tail bounds obtained by inverting noncentral chi-square and noncentral $F$ distributions, respectively. Both intervals are computed from the same simulated observations using the procedure in Remark~\ref{remark:ci-computation}. Figure~\ref{fig:exp4-coverage-length} reports practical coverage and mean interval length divided by $(p_n-1)/(2p_n)$.
	
	The coverage curves coincide above the nominal level throughout the grid. This agrees with Theorem~\ref{Theorem 4} and Corollary~\ref{Corollary 4}, which provide coverage lower bounds rather than equality with the nominal level. Known and estimated variance also give similar interval lengths.
	
	Under the three orders of \eqref{eq:sim-smooth}, mean relative length decreases with $n$. The descending and locally permuted orders produce nearly identical curves, whereas global permutation yields shorter intervals. In contrast, the long-block intervals remain close to the full parameter range, and the short-block intervals also remain wide. Their relative lengths do not decrease monotonically over the grid. Thus, the increasing accuracy of the point diagnostics in Experiment~3 does not imply comparably precise finite-sample intervals.
	
	\subsection{Experiment 5: Structure and inference of Rashomon sets}\label{sec:sim-rashomon}
		
		We examine $\mathcal R_{n,K}(\varepsilon)$ under the five settings of Experiment~2, with $K=N$ for the descending order and $K=A$ otherwise. Exact calculations use the oracle grid of Experiments~1 and~2. Inference uses the eight sample sizes of Experiments~3 and~4, with 500 Gaussian replications per setting and sample size. We set $\alpha=\varepsilon=0.05$, and known- and estimated-variance procedures use the same observations. To construct $\mathcal B_n$, we start from $[1,p_n]$ and recursively bisect each interval $[a,b]$ at $\lfloor(a+b)/2\rfloor$ until $b-a=1$, retaining the resulting intervals with $b<p_n$. The tail bounds use $\mathcal C_{n,\alpha/2}$. For the local bounds, we divide the remaining $\alpha/2$ equally among tree depths containing intervals in $\mathcal B_n$, then equally among those intervals at each depth to obtain $\alpha_{ab}$. The risk-difference extrema are evaluated jointly over $\mathcal C^{\mathrm R}_{n,\alpha}$.
		
\begin{figure}[t]
			\centering
			\includegraphics[width=\textwidth]{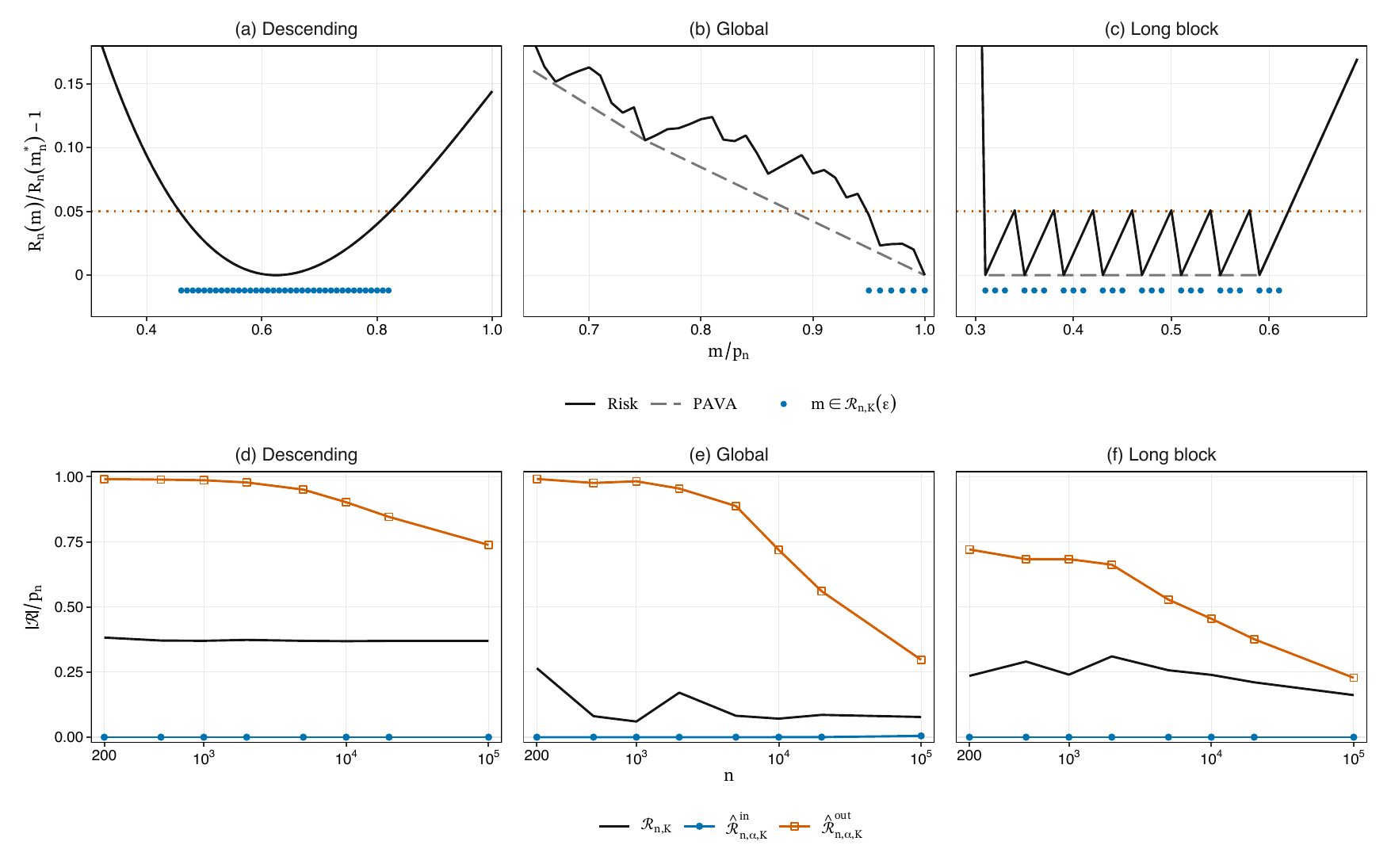}
			\caption{Rashomon sets for the descending order, global permutation, and long critical block. Upper panels show $R_n(m)/R_n(m_n^*)-1$ at $n=1000$, where $R_n(m)=R_n(m\mid\mathcal M_K,\boldsymbol\mu)$, together with the corresponding PAVA curve. The horizontal line is $\varepsilon=0.05$; points below zero mark members of $\mathcal R_{n,K}(0.05)$, not negative excess risks. Lower panels show $|\mathcal R_{n,K}(0.05)|/p_n$ and mean inner and outer cardinalities divided by $p_n$, using $\hat\sigma_{\mathrm f}^2$, $\alpha=0.05$, and 500 replications. Their horizontal axes are logarithmic.}
			\label{fig:exp5-main}
	\end{figure}

		The upper panels of Figure~\ref{fig:exp5-main} illustrate Propositions~\ref{prop:rashomon-ordered} and~\ref{prop:rashomon-unordered}. The descending order produces a contiguous set of cutoffs. Under global permutation, the PAVA risk curve determines an enclosing interval, while $H_n(m)$ excludes some of its cutoffs; the displayed Rashomon set remains contiguous. For the long block, positive within-block discrepancies exclude interior cutoffs and produce separated groups. Thus, the PAVA curve alone need not determine membership.
		
		The lower panels show that the normalized difference between outer and inner cardinalities generally decreases with $n$, particularly under global permutation and for the long block. However, the inner sets remain empty for the descending order and long block, and contain only a small fraction of the true members under global permutation. The long-block oracle cardinality relative to $p_n$ also decreases, so a smaller normalized difference does not establish recovery of the set. These results illustrate the distinction between simultaneous coverage and precise membership determination.
		
		The block settings also connect Rashomon geometry to the oracle gains in Experiment~2. At $\varepsilon=0$, we have $\mathcal R_{n,A}(0)=\{k,k+4,\ldots,k+q\}$ and $s_{n,A}=k+q$. The span and cardinality bounds in \eqref{eq:rashomon-gain-lower} therefore equal $q/\{2(k+q)\}$ and $q/\{8(k+q)\}$, respectively. For the long block, these bounds converge to $1/4$ and $1/16$, while $G_{n,A}\to1/4$. For the short block, both bounds and $G_{n,A}$ converge to zero despite $|\mathcal R_{n,A}(0)|\to\infty$. Thus, an increasing number of optimal models alone does not ensure a nonvanishing relative MA gain.
		
	\section{An application to diamond prices}\label{sec:diamonds}
	
	We apply the proposed diagnostic to the public \texttt{diamonds} data from \texttt{ggplot2}. This dataset contains prices in US dollars and nine attributes describing the weight, dimensions, and quality of 53,940 round-cut diamonds. A subset of these data was considered by \citet{welandawe2024reliability}. We use log price as the response and include log-transformed carat weight and physical dimensions, depth and table percentages, and indicators for cut, color, and clarity as predictors. We remove 20 observations with nonpositive physical dimensions and 344 repeated predictor vectors, retaining the first occurrence without reference to price. This leaves 53,576 observations. Treatment coding uses Fair, D, and I1 as the reference levels for cut, color, and clarity, respectively, giving 23 covariate columns and $p_n=24$ after inclusion of the intercept. Centers and scales are estimated from each training sample only.
	
	We randomly divide the observations into four disjoint subsets: 15\% for variable ordering, 15\% for an independent check of that ordering, 20\% for testing, and the remainder for the training pool. On the ordering sample, we rank the covariate columns by their absolute Pearson correlations with log price. The candidate models start with an intercept and successively add one column in this order. We fix the ordering and test set across 500 repetitions. In each repetition, we randomly permute the training pool and use its first $n$ observations for $n\in\{500,1000,2000,5000,10000,20000,25000\}$, so the training samples are nested. For each sample, we compute $\hat G_{n,A}^{\mathrm{plug}}$ using the full-model residual variance estimate with $n-24$ degrees of freedom, and evaluate MMA and $C_p$ selection on the same test set. We summarize their predictive comparison by the percentage reduction in mean test MSE, with each mean taken over the 500 repetitions.
	
\begin{figure}[t]
		\centering
		\includegraphics[width=\textwidth]{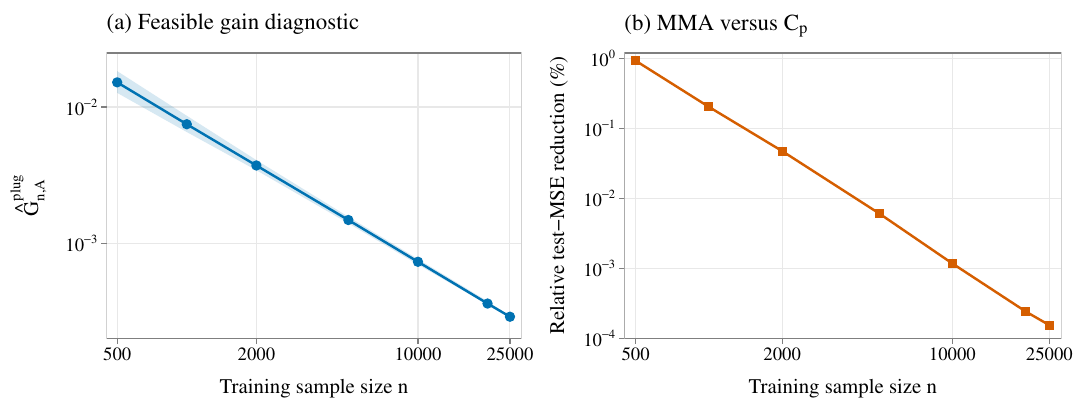}
		\caption{Diamonds application with a fixed candidate ordering and test set over 500 repetitions. Panel (a) reports the median $\hat G_{n,A}^{\mathrm{plug}}$ and its 10th--90th percentile band across repetitions. The band describes variability across training subsamples and is not a confidence interval for $G_{n,A}$. Panel (b) reports the percentage reduction in mean test MSE, $100(1-\overline{\mathrm{MSE}}_{\mathrm{MMA}}/\overline{\mathrm{MSE}}_{C_p})$, where the bars denote averages over repetitions. Both axes in each panel are logarithmic.}
		\label{fig:diamonds-main}
	\end{figure}

	Figure~\ref{fig:diamonds-main}(a) shows that the median $\hat G_{n,A}^{\mathrm{plug}}$ decreases steadily as $n$ increases; the shaded region represents the 10th--90th percentiles across repetitions. Panel~(b) shows a corresponding decline in the percentage reduction in mean test MSE. Although MMA has a lower mean test MSE throughout the displayed range, its advantage becomes negligible at the larger sample sizes. The two panels thus provide qualitatively consistent evidence that a small estimated gain accompanies little predictive benefit from averaging within this candidate family. Their common downward trend does not, however, establish a common convergence rate.
	
	\section{Conclusion}\label{Section 8}
	
	This paper characterizes the relative oracle risk improvement of MA over MS for nested linear regression models with homoscedastic errors. We establish necessary and sufficient conditions for this improvement to be asymptotically substantial or negligible without imposing a specific coefficient decay pattern. Under complete ordering, the comparison is determined by the proportion of oracle MS risk contributed by optional components whose squared coefficients are near the noise level $\sigma^2/n$. Under an arbitrary fixed ordering, a common PAVA partition provides representations of both oracle risks, and the corresponding criterion depends on the risk contribution of blocks whose average squared coefficients are near $\sigma^2/n$. We further derive a sharp upper bound on $G_{n,A}$, develop a consistent practical diagnostic under suitable conditions, and construct confidence intervals and associated tests with finite-sample guarantees under Gaussian errors. Feasible versions accommodate unknown error variance. The simulation studies illustrate the theoretical characterization and assess the diagnostic and inference procedures, while the diamond-price application demonstrates how the diagnostic can help assess the potential benefit of MA within a given candidate family.
	
	Two directions merit further investigation. First, the oracle comparison could be extended to non-nested candidate models. The common PAVA representation of the two oracle risks relies on the nested structure. Without this structure, the comparison must account for the overlap among candidate model spaces and the associated covariance terms. A central question is whether an analogous characterization of substantial and negligible gains can be established through the approximation errors and covariance structure of the candidate estimators. Second, the confidence intervals can remain wide in the critical-block settings despite their finite-sample coverage guarantee. Further work could sharpen the simultaneous tail bounds through the dependence among the observed tail strengths and optimize the upper endpoint over the same feasible set.
	
	\clearpage

	\baselineskip=16pt
	\bibliographystyle{abbrvnat}
	\bibliography{ref}
\end{document}